\documentclass[a4paper, 11pt]{article}
\pdfoutput=1
\usepackage{jcappub}
\usepackage[utf8]{inputenc}
\usepackage{physics}
\usepackage{amsmath, amssymb, amsthm}
\usepackage{siunitx}
\usepackage{xcolor}
\usepackage{graphicx}
\usepackage{cancel}
\usepackage{caption}
\usepackage{subcaption}
\usepackage{mathrsfs}
\usepackage{numprint}
\usepackage{multirow,multicol, makecell, booktabs}
\usepackage{framed}
\usepackage{mdframed}

\usepackage{tikz}
\usetikzlibrary{positioning,fit}

\renewcommand{\tilde}{\widetilde}
\newcommand{\mpl}{M_{\rm pl}}
\newcommand{\cint}{\operatorname{Ci}}

\newcommand{\closedopeninterval}[2]{[#1,#2)}

\numberwithin{equation}{section}

\newcommand{\decay}{f} 
\newcommand{\dimps}{P} 
\newcommand{\dimlessps}{\mathcal{P}}
\newcommand{\trap}{\ast} 
\newcommand{\present}{0} 
\newcommand{\osc}{\rm osc} 
\newcommand{\eq}{\rm eq} 
\newcommand{\eps}{\varrho} 
\newcommand{\ellipticE}{\textrm{E}} 

\title{ALP Dark Matter Mini-Clusters from Kinetic Fragmentation}

\author[a]{Cem Eröncel\footnote{Current e-mail address: \texttt{cem.eroncel@itu.edu.tr}},}
\author[a,b]{Géraldine Servant,}

\affiliation[a]{Deutsches Elektronen-Synchrotron DESY, Notkestr. 85, 22607 Hamburg, Germany}
\affiliation[b]{II. Institute of Theoretical Physics, Universit\"{a}t Hamburg D-22761, Germany}

\emailAdd{cem.eroncel@desy.de}
\emailAdd{geraldine.servant@desy.de}

\abstract{We show that very compact axion mini-clusters can form
in models where axion-like-particle (ALP) dark matter is produced via the kinetic misalignment mechanism, which is well-motivated in pre-inflationary $U(1)$ symmetry breaking scenarios.
This is due to ALP fragmentation.
We predict denser halos than what has been obtained so far in the literature 
from standard misalignment  in post-inflationary $U(1)$ breaking scenarios or from large misalignment.
The main reason is that adiabatic fluctuations are significant at early times, therefore, even if amplification from parametric resonance effects is moderate, the final size of ALP fluctuations is larger in kinetic misalignment.
We compare  halo mass functions and halo spectra obtained in kinetic misalignment, large misalignment and   standard misalignment  respectively. Our analysis does not depend on the specific model realization of the kinetic misalignment mechanism. We present our results generally as a function of the ALP mass and the ALP decay constant only. We show that a sizable region of this ALP parameter space can be tested by future experiments that probe  small-scale structures.}


\arxivnumber{2207.10111}

\begin{document}

\begin{flushright}
\footnotesize
DESY-22-115 \\
\end{flushright}
\color{black}

\maketitle
\flushbottom

\section{Introduction}
\label{sec:introduction}

Axion-like-particles (ALPs) are prime candidates for dark matter, and they have been subject to extensive theoretical research in recent years (see \cite{Marsh:2015xka,Irastorza:2018dyq,Grin:2019mub,Choi:2020rgn} for recent reviews). Typically, they are pseudo-Nambu-Goldstone bosons of a global $U(1)$ symmetry that is spontaneously broken at some high scale $f$ which is referred as the ALP decay constant. In this picture, ALPs are massless after symmetry breaking due to the continuous shift symmetry. They obtain a mass at a lower scale $\Lambda_{b}\ll f$ via their interactions with a strong gauge sector that breaks the continuous shift symmetry explicitly. This explicit breaking creates a periodic potential $V(\theta,T)$ for the ALP field $\theta$ that is minimized at $\theta=0$ and leads to an ALP mass $m^{2}(T)=\decay^{-2}\eval{V''(\theta)}_{\theta=0}$. At high temperatures the potential can well be approximated by $V(\theta,T)=\Lambda_{b}^{4}(T)\qty[1-\cos(\theta)]$ where $\Lambda_{b}^{4}(T)$ represents the height of the potential barrier. By far the best-known ALP model is the QCD axion, see \cite{DiLuzio:2020wdo} for a recent review, that arises as a by-product of the Peccei-Quinn solution to the Strong CP problem \cite{Peccei:1977hh,Weinberg:1977ma,Wilczek:1977pj}. In this case, the strong sector is QCD so that $\Lambda_{b}\sim \Lambda_{\rm QCD}$, thus the QCD axion mass $m_{0}\equiv m(T=0)$ and $f$ are related to each other. In a generic ALP model, both $m_{0}$ and $f$ are free parameters.


The couplings of ALPs to Standard Model particles depend on the UV completion of a particular ALP model, however they typically scale as $f^{-1}$. The weak coupling and the small mass make ALPs cosmologically stable, hence they are well-motivated dark matter candidates \cite{Preskill:1982cy,Dine:1982ah,Abbott:1982af}. The most widely studied production mechanism goes by the name \emph{misalignment mechanism}. After the spontaneous breaking of $U(1)$, the ALP field selects a random initial angle $\theta_{i}=\closedopeninterval{-\pi}{\pi}$ on each Hubble patch. The later evolution is governed by $\ddot{\theta}+3H(t)\dot{\theta}+m^{2}(t)\sin\theta=0$ which is derived from the Klein-Gordon equation with the Friedmann-Lemaitre-Robertson-Walker (FLRW) metric by neglecting the spatial gradient terms. Here $H(t)$ is the Hubble scale, and $m(t)$ is the axion mass which is in general temperature-dependent. The initial velocity is usually assumed to be vanishing. In this case, the ALP field is initially frozen due to the Hubble friction. Later it starts to oscillate around the minimum $\theta=0$ once the Hubble scale drops below the ALP mass. After this point, the ALP field behaves effectively as pressureless cold matter, and contributes to the dark matter relic density.

Recently, some works modified the usual assumption that the ALP field is static in the early universe, and studied the consequences of a large initial kinetic energy for the ALP field \cite{Co:2019jts,Chang:2019tvx}. The kinetic energy causes a delay in the onset of oscillations so that the ALP dark matter parameter space can be expanded to lower values of the ALP decay constant, offering very promising prospects for upcoming experiments, see in particular the recent overview in \cite{Eroncel:2022vjg}. This variation is referred as the \emph{kinetic misalignment mechanism (KMM)}, whereas the original case is called the \emph{standard misalignment mechanism (SMM)}. Such a kinetic energy can arise if the $U(1)$ symmetry is explicitly broken at a very high scale by the high-dimensional operators which induces a kick for the angular (axion) mode once the radial mode (saxion) starts oscillating \cite{Co:2019jts,Co:2019wyp,Co:2020dya,Co:2020jtv,Co:2021lkc,Gouttenoire:2021jhk}. Another possibility, named \emph{trapped misalignment}, is that the ALP potential has a non-trivial temperature dependence such that the ALP field first starts oscillating around the high-temperature minimum $\theta=\pm\pi$, then move to the low-temperature minimum $\theta=0$ after some critical temperature $T_{c}$ \cite{DiLuzio:2021gos,DiLuzio:2021pxd}. A large kinetic energy can also be realized in models with more than one ALP field \cite{Daido:2015bva,Daido:2015cba}.

In \cite{Eroncel:2022vjg}, we pointed out that ALP fluctuations play a crucial role in the KMM setup. When the ALP field is rolling over the potential barriers and when it is oscillating around the minimum, the fluctuations can grow exponentially via parametric resonance. This effect is called axion fragmentation \cite{Fonseca:2019ypl}. It was demonstrated in \cite{Eroncel:2022vjg} that in large part of the KMM parameter space all the energy in the homogeneous mode is converted into ALP fluctuations. This means that ALP dark matter consists of particles that are mildly relativistic at the end of fragmentation, and cool down later via redshift. This is in contrast with the SMM case where the dark matter is an extremely cold coherently oscillating scalar field. In this work, we continue our study of ALP fluctuations in the KMM setup, and discuss the observational consequences of fragmentation for the ALP miniclusters. We calculate the halo spectrum, typical densities of the ALP halos for a given halo mass, and show that a sizable region of the $(m,f)$-ALP parameter space can be tested by future experiments that probe  small-scale structures.

Most of the literature on ALP miniclusters assumes the scenario where the $U(1)$ symmetry breaking, Peccei-Quinn (PQ) in the case of the QCD axion, takes place after inflation, referred as the \emph{post-inflationary scenario}. In this case, the  ALP field takes uncorrelated initial values in causally disconnected Hubble patches. Since our universe contains many of these Hubble patches, the ALP field does have $\mathcal{O}(1)$ density fluctuations in the post-inflationary scenario. These fluctuations gravitationally collapse very early, and form dense objects called ALP miniclusters \cite{Hogan:1988mp,Kolb:1993zz,Kolb:1993hw,Kolb:1994fi}. In recent years, this process has been studied in detail both semi-analytically \cite{Hardy:2016mns,Enander:2017ogx,Fairbairn:2017dmf,Fairbairn:2017sil}, and numerically \cite{Bertschinger:1985pd,Zurek:2006sy,Ricotti:2009bs,PhysRevD.96.123519,Delos:2017thv,Vaquero:2018tib,Buschmann:2019icd,Eggemeier:2019khm,OHare:2021zrq}. For a recent review see \cite{Niemeyer:2019aqm}. A recent discussion of the observability of the QCD axion miniclusters in the post-inflationary scenario can be found in \cite{Ellis:2022grh}, while \cite{Barman:2021rdr} studies how the properties of the miniclusters are modified in the presence of kinetic misalignment.


In the case when $U(1)$ breaking happens before or during inflation, all Hubble patches are inflated to huge sizes so that the initial angle is the same in all the observable universe. This is referred as the \emph{pre-inflationary scenario}. Even in this scenario the ALP field has some fluctuations. Unavoidable ones are the adiabatic fluctuations that arise due to the temperature fluctuations of the radiation bath. If ALPs are present during inflation, and they are massless or light compared to the inflation scale $H_{I}$, they also pick up quantum fluctuations given by $\delta \theta\simeq H_{I}/(2\pi f_{I})$ where $f_{I}$ is the effective decay constant during inflation which might be different from its value today \cite{Linde:1991km}. The ratio of the isocurvature fluctuations to the adiabatic ones are highly constrained by the Planck 2018 data \cite{Aghanim:2018eyx,Planck:2018jri}. Throughout this work we will consider the pre-inflationary scenario, and assume that the inflationary isocurvature fluctuations are negligible.

Even though the initial size of the fluctuations is small in the pre-inflationary scenario, in some cases fluctuations can be enhanced significantly due to various instabilities \cite{Johnson:2008se}, or as a result of an early matter dominated era \cite{Nelson:2018via}.  For the former, a well-known example related to the periodic potential is when the initial angle $\theta_{i}$ is very close to the top of the potential\footnote{Such an initial condition can be realized with dynamical mechanisms without tuning \cite{Co:2018mho,Takahashi:2019pqf,Huang:2020etx}.}, i.e. $\abs{\pi-\theta_{i}}\ll 1$, referred as the \emph{large misalignment mechanism (LMM)} \cite{Arvanitaki:2019rax} or \emph{extreme axion} \cite{Zhang:2017dpp}. In this case, the onset of oscillations is delayed from the conventional condition $m\sim 3H$ due to the tiny potential gradient at the top. As we will see in detail in Section \ref{sec:comp-with-stand}, this delay causes the fluctuations to grow exponentially via tachyonic instability and parametric resonance \cite{PhysRevD.50.7690,Greene:1998pb,Zhang:2017flu,Zhang:2017dpp,Cedeno:2017sou,Arvanitaki:2019rax}. This feature also arises in ALP models with non-periodic potentials for sufficiently large initial angles \cite{Jaeckel:2016qjp,Alonso-Alvarez:2018tus,Olle:2019kbo,Berges:2019dgr,Chatrchyan:2020pzh,Fukunaga:2020mvq,Chatrchyan:2022askdja}, and via the interaction between the QCD axion and QCD matters during the QCD phase transition \cite{Sikivie:2021trt,Kitajima:2021inh}. The common conditions in all these cases are the delay of the onset of oscillations and a large initial field value. The latter causes the ALP field to probe the non-quadratic parts of its potential which causes instabilities, while the former ensures that the field amplitude decays slowly so that the fragmentation becomes efficient. We expect that the effects studied in this work are relevant for any ALP model which predict a delay in the onset of oscillations, such as the recently proposed \emph{frictional misalignment} \cite{Papageorgiou:2022prc}.

The enhancement of fluctuations does also occur in KMM \cite{Eroncel:2022vjg,Fonseca:2019ypl}. The main reason is the delayed onset of oscillations due to the initial kinetic energy. As a result, the physics responsible for fragmentation in KMM and LMM models is quite similar. However, there are subtle differences which we study thoughtfully in Section \ref{sec:comp-with-stand}. We demonstrate that in the LMM scenario, some modes exhibit significant growth even before the homogeneous mode starts oscillating since they become tachyonic. Due to this effect, the relative amplification is stronger in LMM compared to KMM. However, the initial size of the fluctuations is much smaller in LMM  due to the fact that the homogeneous mode behaves as dark energy $w\approx 0$ in the early time limit, and dark energy has no adiabatic perturbations \cite{Marsh:2015xka}. On the other hand, the equation of state of the homogeneous mode in KMM is $w\approx 1$, thus adiabatic perturbations are not zero at early times. We shall see that because of this difference in the initial conditions, the final size of the fluctuations are larger in KMM compared to LMM, even though the amount of amplification is stronger in the latter.

The fragmentation process modifies the matter power spectrum significantly at the scales $k \sim m_{\rm osc}a_{\rm osc}$ that get amplified most strongly. Here $m_{\rm osc}$ and $a_{\rm osc}$ are the ALP mass and the scale factor at the onset of oscillations. Therefore, fragmentation hypothesis can be tested via experiments that probe the matter power spectrum \cite{Brito:2022lmd}. An important consequence of fragmentation is that the peak of the matter power spectrum can reach  $\mathcal{O}(1)$ values right after the fragmentation. These fluctuations gravitationally collapse very early in the matter era, and can create very dense and compact ALP miniclusters like in the post-inflationary scenario \cite{Arvanitaki:2019rax}. Our primary goal in this work is to calculate the properties of these miniclusters semi-analytically, and comment on their discovery prospects. We compare the predictions of all the production mechanisms that we mentioned so far: SMM, LMM, KMM, post-inflationary scenario. Such a comparison can be the first step towards an ambitious goal of inferring the ALP production mechanism from the small-scale power spectrum observations. Our analysis is model-independent. The link to specific model implementations of KMM will be presented in \cite{Eroncel:2022a}.

In \cite{Eroncel:2022vjg} we demonstrated that fragmentation occurs both in temperature-dependent and independent potentials. In this work, we concentrate on ALP models with temperature-independent potentials. The constant ALP mass is denoted by $m$. A discussion of the temperature-dependent potential including a precise analysis for the QCD axion will be presented in a future publication.

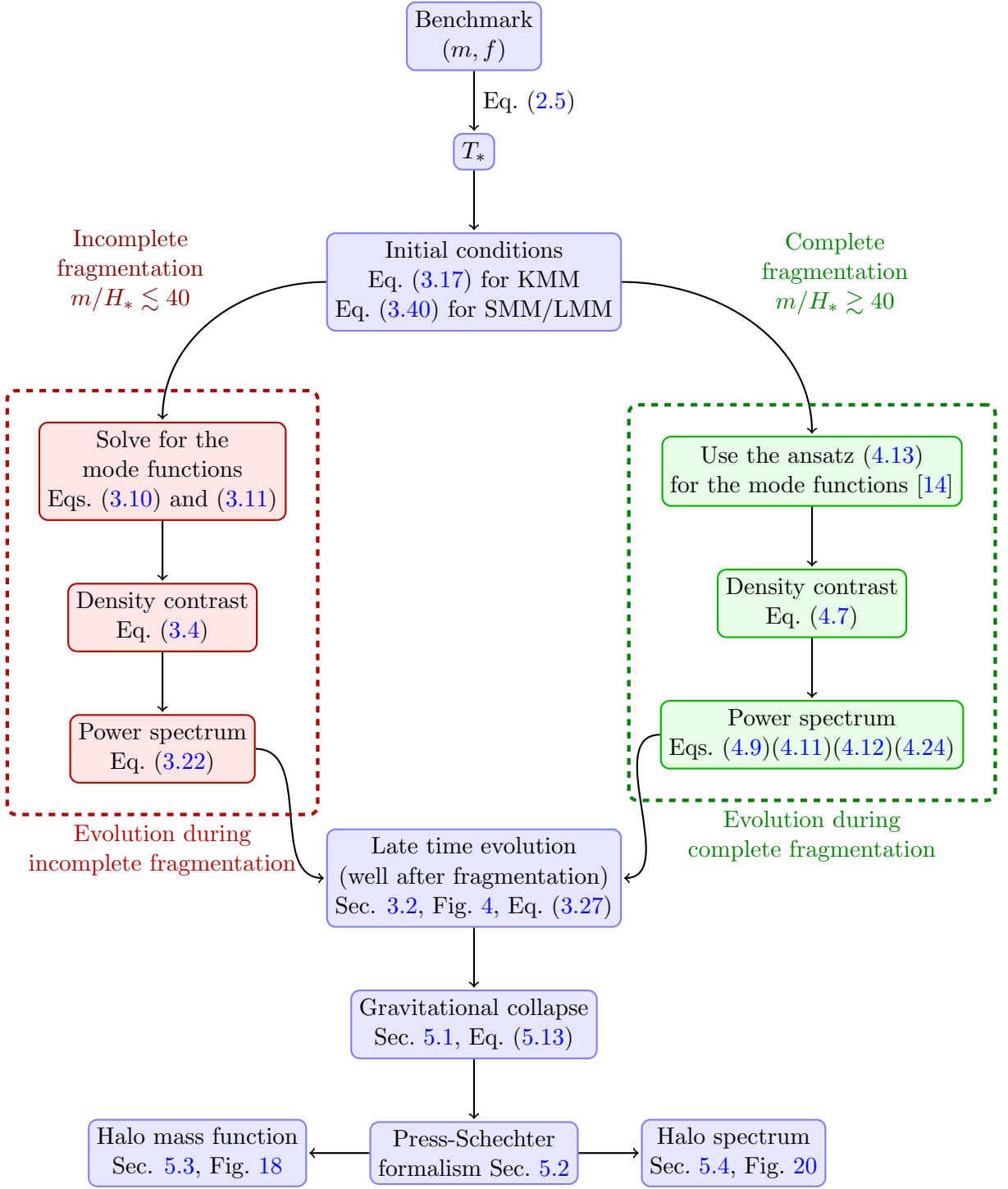
\begin{figure}[tbp]
  \centering
  \begin{tikzpicture}
    [default/.style={rounded corners,draw=blue!50,fill=blue!10,thick,align=center},
    hidden/.style={draw=white,fill=white,minimum width=4cm,minimum height=2.5cm},
    nonlinear/.style={rounded corners,draw=green!70!black,fill=green!10,thick,align=center},
    linear/.style={rounded corners,draw=red!70!black,fill=red!10,thick,align=center},
    arrow/.style={->,shorten >=1pt,thick}]
    \node (benchmark) [default] {Benchmark \\ $(m,f)$};
    \node (Tast) [default] [below=of benchmark] {$T_{\trap}$};
    \node (initial) [default] [below=of Tast] {Initial conditions\\Eq.~\eqref{eq:228} for KMM\\Eq.~\eqref{eq:307} for SMM/LMM};
    \node (empty) [hidden] [below=of initial] {};
    \node (empty2) [hidden] [below=of empty] {};
    \node (linear mode) [linear] [left=of empty] {Solve for the \\mode functions\\Eqs.~\eqref{eq:6} and \eqref{eq:219}};
    \node (linear density) [linear] [below=of linear mode] {Density contrast\\Eq.~\eqref{eq:199}};
    \node (linear ps) [linear] [below=of linear density] {Power spectrum\\Eq.~\eqref{eq:22}};
    \node (nonlinear mode) [nonlinear] [right=of empty] {Use the ansatz \eqref{eq:209}\\for the mode functions \cite{Eroncel:2022vjg}};
    \node (nonlinear density) [nonlinear] [below=of nonlinear mode] {Density contrast\\Eq.~\eqref{eq:215}};
    \node (nonlinear ps) [nonlinear] [below=of nonlinear density] {Power spectrum\\Eqs.~\eqref{eq:267}\eqref{eq:206}\eqref{eq:207}\eqref{eq:271}};
    \node (latetime) [default] [below=of empty2] {Late time evolution \\ (well after fragmentation) \\ Sec. \ref{sec:evol-at-latet}, Fig.~\ref{fig:cm-mod-function-eval-full}, Eq.~\eqref{eq:237}};
    \node (collapse) [default] [below=of latetime] {Gravitational collapse\\Sec.~\ref{sec:crit-dens-coll}, Eq.~\eqref{eq:252}};
    \node (ps) [default] [below=of collapse] {Press-Schechter \\ formalism Sec.~\ref{sec:press-schecht-form}};
    \node (hmf) [default] [left=of ps] {Halo mass function \\Sec. \ref{sec:mass-distr-halos:}, Fig.~\ref{fig:grand-money-plot-hmf}};
    \node (halo spectrum) [default] [right=of ps] {Halo spectrum\\Sec.~\ref{sec:halo-spectrum}, Fig.~\ref{fig:grand-money-plot}};
    \draw [arrow] (benchmark) to node[auto] {Eq.~\eqref{eq:4}} (Tast);
    \draw [arrow] (Tast) to (initial);
    \draw [arrow] (initial) to [out=180,in=90] node[auto,align=center,swap,red!50!black] {Incomplete\\fragmentation\\$m/H_{\trap}\lesssim 40$} (linear mode);
    \draw [arrow] (initial) to [out=0,in=90] node[auto,align=center,green!50!black] {Complete\\fragmentation\\$m/H_{\trap}\gtrsim 40$} (nonlinear mode);
    \draw [arrow] (linear mode) to (linear density);
    \draw [arrow] (linear density) to (linear ps);
    \draw [arrow] (linear ps) to [out=0,in=180] (latetime);
    \draw [arrow] (nonlinear mode) to (nonlinear density);
    \draw [arrow] (nonlinear density) to (nonlinear ps);
    \draw [arrow] (nonlinear ps) to [out=180,in=0] (latetime);
    \draw [arrow] (latetime) to (collapse);
    \draw [arrow] (collapse) to (ps);
    \draw [arrow] (ps) to (halo spectrum);
    \draw [arrow] (ps) to (hmf);
    \node (early linear) [draw=red!70!black, ultra thick, dashed, inner sep=0.5cm, rounded corners, fit={(linear mode) (linear density) (linear ps)}, label={[align=center,red!70!black]below:Evolution during \\ incomplete fragmentation}] {};
    \node (early nonlinear) [draw=green!50!black, ultra thick, dashed, rounded corners, inner sep=0.5cm, fit={(nonlinear mode) (nonlinear ps)}, label={[align=center,green!50!black]below:Evolution during \\ complete fragmentation}] {};
  \end{tikzpicture}
  \caption{\small \it Summary of the procedure followed in this paper.}
  \label{fig:pipeline}
\end{figure}

The main programme of the paper is sketched in Figure \ref{fig:pipeline}.
The ultimate goal  is to predict the observable ALP DM small-scale structures that can form at late times from gravitational collapse as signatures of the kinetic misalignment production mechanism.
 In Section \ref{sec:brief-review-alp}, we briefly review the key results of \cite{Eroncel:2022vjg} which are relevant for this work. The most important quantity for this paper is the trapping temperature $T_{\trap}$ that can be calculated for any benchmark point $(m,f)$. 
From this, one can derive the initial conditions for the ALP fluctuations. This is exposed in detail in Appendix \ref{sec:deriv-adiab-init-1}. The main quantity that is needed to determine the late-time evolution of the fluctuations is their power spectrum. There are two approaches for this calculation one can follow depending on the relative values  of the ALP mass $m$ and the Hubble rate at trapping $H_*$ that determine whether fragmentation is complete or incomplete.
 In Section \ref{sec:axion-dens-pert}, we calculate the power spectrum corresponding to the ALP density contrast in KMM with fragmentation using the cosmological perturbation theory. We also do the same exercise for LMM, and compare the two mechanisms. In Section \ref{sec:alp-dens-pert-nl}, we describe a semi-analytical estimate of the power spectrum in the case of complete fragmentation where the cosmological perturbation theory breaks down. In this section, we also make a comparison with the post-inflationary scenario. 
 Next,  In Section \ref{sec:grav-coll-fluct}, the formation of the dark matter halos is studied analytically via the Press-Schechter (PS) formalism.
We use the results obtained for the power spectrum to derive the halo mass function (HMF), and the halo spectrum. We briefly describe the observational prospects in Section \ref{sec:observ-prosp}, and finally conclude in Section \ref{sec:conclusion-1}. More technical details are presented in appendices.
Important equations are inside frames. Those equations which are new are in addition in blue background.

\paragraph{Notation:}

We use the metric convention $\operatorname{diag}\qty(-,+,+,+)$, and $M_{Pl}\approx  2.435 \times 10^{18}$ GeV denotes the reduced Planck mass.


\section{Brief review of ALP fragmentation in kinetic misalignment}
\label{sec:brief-review-alp}

In this section, we summarize the main findings of \cite{Eroncel:2022vjg}. We consider an ALP field $\theta$ with the Lagrangian\footnote{In realistic ALP models, the potential does not need to have the simple periodic form assumed in \eqref{eq:1}. The primary example is the potential of the QCD axion at low energies. However, even in the QCD axion case, most of the dynamics determining the cosmological evolution occurs at high temperatures $T\gg \Lambda_{\rm QCD}$ when the potential in \eqref{eq:1} provides a good description.}
\begin{equation}
  \label{eq:1}
  \mathcal{L}=-\frac{f^{2}}{2}g^{\mu\nu}\partial_{\mu}\theta\partial_{\nu}\theta-V(\theta)=-\frac{f^{2}}{2}g^{\mu\nu}\partial_{\mu}\theta\partial_{\nu}\theta-m^{2}f^{2}\qty(1-\cos \theta).
\end{equation}
In KMM, the evolution of the ALP homogeneous mode $\Theta$ can be divided into two regimes. When the ALP kinetic energy $f^{2}\dot{\Theta}^{2}/2$ dominates over the size of the potential barrier $2m^{2}f^{2}$, then the ALP field rolls with the velocity $\dot{\Theta}\propto a^{-3}$ so that its energy density redshifts as $\rho_{\Theta}\propto a^{-6}$. Shortly after its kinetic energy becomes sub-dominant compared to barrier height, it starts to oscillate around one of the minima, and behaves as Cold dark matter (CDM).

For the early evolution, one can introduce the quantity called the \emph{yield} defined by 
\begin{equation}
  \label{eq:yield}
Y=f^{2}\dot{\Theta}(T)/s(T)
\end{equation}
 where $s(T)$ is the entropy density of the universe. Assuming entropy conservation, this quantity is conserved at early times when $\rho_{\Theta}\propto a^{-6}$. The ALP relic density today can be expressed as
\begin{equation}
  \label{eq:2}
  h^{2}\Omega_{\Theta,0}\approx h^{2}\Omega_{\rm DM}\qty(\frac{m}{5\times 10^{-3}\,\si{\electronvolt}})\qty(\frac{Y}{40}),
\end{equation}
where we took $\Omega_{\rm DM}=h^{2}0.12$ \cite{Aghanim:2018eyx}. This result is also valid for a temperature-dependent potential with $m$ replaced by $m_{0}$.

The ALP field gets \emph{trapped} once its kinetic energy becomes comparable to the barrier height. This defines the \emph{trapping temperature} $T_{\trap}$:
\begin{equation}
  \label{eq:3}
  \frac{1}{2}f^{2}\dot{\Theta}^{2}(T_{\trap})=2f^{2}m^{2}(T_{\trap}).
\end{equation}
This depends on the barrier height at zero temperature, and on the high-temperature scaling of the potential. For a temperature-independent potential it simplifies to  \cite{Eroncel:2022vjg}
\begin{equation}
\boxed{
  \label{eq:4}
  \frac{T_{\trap}}{\Lambda_{b}}\approx\qty(6\times 10^{2})\qty(\frac{g_{s}(T_{\trap})}{72})^{-1/3}\qty(\frac{\Lambda_{b}}{\si{\giga\electronvolt}})^{1/3}\qty(\frac{h^{2}\Omega_{\Theta,0}}{h^{2}\Omega_{\rm DM}})^{-1/3}} \  ,
\end{equation}
where $g_{s}$ denotes the number of effective degrees of freedom in the entropy.

The different fragmentation regimes in KMM depend strongly on the hierarchy between the ALP mass and Hubble scale at trapping, i.e. $m/H_{\trap}$, which can be obtained after solving for $T_{\trap}$ in \eqref{eq:4}. These regimes can be summarized as follows:
\begin{enumerate}
  \item \textbf{Standard misalignment ($m/H_{\trap}<3$): }The onset of oscillations is not delayed from its conventional value $m(T_{\osc})=3 H(T_{\osc})$, and there is no kinetic misalignment.
  \item \textbf{Kinetic misalignment with incomplete fragmentation ($3< m/H_{\trap}\lesssim 4\times 10^{1}$): } The onset of oscillations is delayed due to the non-zero initial velocity, but the fragmentation is weak such that the energy density in fluctuations is always sub-dominant.
  \item \textbf{Complete fragmentation after trapping ($4\times 10^{1}\lesssim m/H_{\trap}\lesssim 5\times 10^{2}$): }The ALP field is completely fragmented, and the fragmentation ends after it would have been trapped by the potential in the absence of fragmentation, i.e. $T_{\rm end}<T_{\trap}$.
  \item \textbf{Complete fragmentation before trapping ($5\times 10^{2}\lesssim m/H_{\trap}\lesssim 5\times 10^{3}$): }The ALP field is completely fragmented, and the fragmentation ends before it would have been trapped by the potential in the absence of fragmentation, i.e. $T_{\rm end}>T_{\trap}$.
  \item \textbf{Non-perturbative ($5\times 10^{3}\lesssim m/H_{\trap}$): }The ALP field is non-perturbative already at the beginning of fragmentation due to the large adiabatic perturbations, so our analytical solution breaks down.
\end{enumerate}
In this work, we will denote the second region as the \emph{linear regime},  corresponding to incomplete fragmentation, while the third and fourth regions will be called the \emph{non-linear regime}, corresponding to complete fragmentation.

Our primary goal in this paper is to study the observational consequences of fragmentation. For this, we need to calculate the power spectrum of the ALP density perturbations at late times. Our pipeline to calculate the power spectrum depends whether the corresponding benchmark point is in the linear or non-linear regime.
\begin{itemize}
  \item For the benchmark points in the linear regime, we will use the cosmological perturbation theory to solve the mode functions numerically without relying on the analytical theory that we have developed in \cite{Eroncel:2022vjg}. We do this beacuse in this regime $m/H_{\trap}$ is not very large so the analytical method of \cite{Eroncel:2022vjg} is less reliable. Also, the exact evolution can be obtained without too much difficulty by solving a couple of differential equations numerically. The procedure is described in detail in Section \ref{sec:axion-dens-pert}.
  \item For the benchmark points in the non-linear regime, the use of the cosmological perturbation theory is not possible. Even though precise results will require a non-perturbative analysis such as a lattice simulation, we will rely on the analytical theory in \cite{Eroncel:2022vjg} to calculate the late-time behavior of the mode functions. We then calculate the power spectrum directly from these mode functions using the method in \cite{Enander:2017ogx}. This is the topic of Section \ref{sec:alp-dens-pert-nl}.
\end{itemize}

\section{ALP density perturbations in the incomplete fragmentation regime}
\label{sec:axion-dens-pert}

In this section, we calculate the power spectrum corresponding to the ALP energy density at late times. This is the necessary ingredient in order to study the halo spectrum and the gravitational signatures that we will work out in the next sections. In Section \ref{sec:early-time-evolution} we will calculate the evolution of the mode functions during the fragmentation from which we will obtain the evolution of the density contrast. In Section \ref{sec:evol-at-latet} we describe an effective description of the density contrast at late times. Finally, in Section \ref{sec:comp-with-stand} we will compare the power spectra of KMM and LMM models at late times and discuss the similarities and differences.

\subsection{Early evolution of fluctuations: before and during fragmentation}
\label{sec:early-time-evolution}
Our starting point is the Friedmann-Lemaitre-Robertson-Walker metric including curvature perturbations $\Phi,\Psi$ in the Newtonian gauge\footnote{We adopt the conventions used in \cite{maggiore2008gravitational}.}:
\begin{equation}
  \label{eq:133}
  \dd{s}^2=a^2(\eta)\qty{-\qty[1+2\Psi(\eta,\vb{x})]\dd{\eta}^2+\qty[1+2\Phi(\eta,\vb{x})]\delta_{ij}\dd{x}^i\dd{x}^j},
\end{equation}
The anisotropic stress vanishes for scalar fields at the linear order in perturbation theory which sets $\Psi=-\Phi$. At early times, we assume that the dominant component of the energy density is radiation, so $\Phi$ is an external field for the ALP.

We expand the ALP field as 
\begin{equation}
\theta(\eta,\vb{x})=\Theta(\eta)+\delta\theta(\eta,\vb{x}), 
\end{equation}
and keep only the terms that are linear in fluctuations $\delta\theta$ and $\Phi$. Since the fluctuations are small, the average energy density is given by the energy density of the homogeneous mode:
\begin{equation}
  \label{eq:193}
  \expval{\rho}=\rho_{\Theta}=\frac{\decay^2}{2}\qty(\frac{\Theta'}{a})^2+\decay^2m^2\qty(1-\cos\Theta),
\end{equation}
where $\prime$ denotes derivative with respect to conformal time. The density contrast is
\begin{equation}
  \label{eq:199}
  \delta\equiv \frac{\delta\rho_{\theta}}{\rho_{\Theta}}=\frac{\decay^2}{\rho_{\Theta}}\qty[\frac{\Theta'\delta\theta'}{a^2}+\Phi \frac{\Theta'^2}{a^2}+m^2\sin\Theta \delta\theta].
\end{equation}
The homogeneous mode satisfies the following equation of motion in physical time $t$:
\begin{equation}
  \label{eq:5}
  \ddot{\Theta}+3H\dot{\Theta}+m^{2}\sin\Theta=0,
\end{equation}
while the equation of motion for the mode functions $\theta_k$ reads
\begin{equation}
  \label{eq:201}
  \ddot{\theta}_k+3H\dot{\theta}_k+\qty(\frac{k^2}{a^2}+m^2\cos\Theta)\theta_k=2\Phi_km^2\sin\Theta-4\dot{\Phi}_k\dot{\Theta},
\end{equation}
where $\Phi_{k}$ are the Fourier modes of the curvature perturbations. Following \cite{Arvanitaki:2019rax}, we introduce the following dimensionless quantities:
\begin{equation}
  \label{eq:216}
  t_m\equiv \frac{m}{2H}\simeq mt,\quad t_k\equiv\frac{k/a}{\sqrt{3}H},\quad \tilde{k}^2\equiv \frac{k^2/a^2}{2 m H},
\end{equation}
where the approximate expression for $t_m$ is valid during radiation domination. The momentum variable $\tilde{k}$ is constant during radiation era, therefore it can be used to uniquely identify the momentum mode. The variable $t_k$ is convenient to express the exact solution for the curvature perturbations $\Phi_k$ in radiation era:
\begin{equation}
  \label{eq:218}
  \Phi_k(t_k)=3\Phi_k(0)\qty(\frac{\sin t_k-t_k\cos t_k}{t_k^3}),
\end{equation}
where $\Phi_{k}(0)$ is a stochastic variable that depends on the amplitude of the scalar primordial fluctuations $A_{s}$ by
\begin{equation}
  \label{eq:14}
  \expval{\abs{\Phi_{k}(0)}^{2}}=\qty(\frac{2}{3})^{2}\expval{\abs{\mathcal{R}_{k}(0)}^{2}}=\qty(\frac{2}{3})^{2}\qty(\frac{2\pi^{2}}{k^{3}})A_{s}\qty(\frac{k}{k_{\star}})^{n_{s}-1}.
\end{equation}
Here $\mathcal{R}_{k}$ is the comoving curvature perturbation, $k_{\star}=0.05\,\textrm{Mpc}^{-1}$ is the pivot scale, and $n_{s}$ is the spectral tilt. In this work we take $n_{s}=1$ for simplicity\footnote{In practice, $n_{s}$ is slightly less than unity, so taking a realistic $n_{s}$ will slightly reduce the size of the initial fluctuations, thereby supresses the effects considered in this paper. As we shall see in this section, the amplified modes have typically $k\sim m a_{\trap}$ so the suppression can be estimated by $m a_{\trap}/k_{\star}$. By taking the value measured by Planck 2018 $n_{s}\simeq0.9649$, we checked that this suppression is at most $\sim3$ for all benchmarks models presented in this paper.}, and set $A_{s}=2.1\times 10^{-9}$ consistent with the Planck 2018 measurements (TT,TE,EE+lowE+lensing 68\%) \cite{Aghanim:2018eyx}.

In terms of the variables defined in \eqref{eq:216}, the equations of motion for the homogeneous mode \eqref{eq:5} and the mode functions \eqref{eq:201} during the radiation era take the form \cite{Arvanitaki:2019rax,Zhang:2017dpp,Zhang:2017flu}\footnote{Our convention for the curvature perturbations differs by a sign compared to \cite{Arvanitaki:2019rax}.}:
\begin{framed}
\begin{align}
  \label{eq:6}
  \dv[2]{\Theta}{t_{m}}+\frac{3}{2 t_{m}}\dv{\Theta}{t_{m}}+\sin\Theta&=0,\\
  \label{eq:219}
  \dv[2]{\theta_k}{t_m}+\frac{3}{2 t_m}\dv{\theta_k}{t_m}+\qty(\frac{\tilde{k}^2}{t_m}+\cos\Theta)\theta_k&=2\qty(\Phi_k\sin\Theta-\frac{t_k}{t_m}\dv{\Phi_k}{t_k}\dv{\Theta}{t_m}),
\end{align}
\end{framed}
\noindent where we have used $H=(2t)^{-1}$. After choosing the correct initial conditions for the homogeneous mode $\Theta$ and the mode functions $\theta_{k}$, the evolution of the ALP field and its fluctuations can be determined by numerically solving \eqref{eq:6} and \eqref{eq:219}.


\subsubsection{Initial conditions}

To fix the initial conditions for the homogeneous mode, we used the $\eps$ parameter that we have introduced in \cite{Eroncel:2022vjg}:
\begin{equation}
  \label{eq:224}
  \eps(t_m)\equiv \frac{\rho_{\Theta}(t_{m})}{2m^{2}f^{2}}=\frac{1}{4}\qty(\dv{\Theta}{t_m})^2+\sin^2 \frac{\Theta}{2}.
\end{equation}
The homogeneous mode is rolling when $\eps>1$, and oscillating when $\eps<1$. In the former case, $\eps$ satisfies the following relation \cite{Eroncel:2022vjg}:
\begin{equation}
  \label{eq:225}
  1=\tau^{3/2}\sqrt{\eps(\tau)}\,\ellipticE\qty(\frac{1}{\sqrt{\eps(\tau)}}),\quad \eps>1
\end{equation}
where $\ellipticE$ is the complete elliptic integral of the second kind\footnote{We use the following definition for the elliptic integral:
  \begin{equation*}
    \ellipticE(k)=\int_0^{\pi/2}\dd{\varphi}\sqrt{1-k^2\sin^2\varphi}
  \end{equation*}
  Note that most software packages such as \texttt{Mathematica} and \texttt{scipy} \cite{2020SciPy-NMeth} use $m=k^2$ instead of $k$ as their argument when defining elliptic integrals.
}, and
 \begin{equation}
 \tau\equiv 2 H_{\ast} t=(2 H_{\ast}/m)t_m
 \end{equation} 
 is an another dimensionless time variable defined such that $\tau=1$ coincides with the time of trapping. At any $\tau$, this equation can be solved numerically to get $\eps(\tau)$. At early times when $\eps \gg 1$, we can approximate
\begin{equation}
  \label{eq:226}
  \eps(\tau \ll 1)\approx \frac{4}{\pi^2}\frac{1}{\tau^3}=\frac{4}{\pi^2}\qty(\frac{m}{2 H_{\ast}})^3 \frac{1}{t_m^3}.
\end{equation}
By plugging this result into \eqref{eq:224} we get
\begin{equation}
  \label{eq:227}
  \eval{\dv{\Theta}{t_m}}_{t_{m,i}}=2\sqrt{\frac{4}{\pi^2}\qty(\frac{m}{2 H_{\ast}})^3 \frac{1}{t_m^3} - \sin^2 \frac{\Theta_i}{2}}.
\end{equation}
Here $\Theta_i$ is the initial angle at the start of the simulation which is irrelevant in the case of kinetic misalignment mechanism\footnote{For some choices of $t_{m,i}$ and $\Theta_{i}$ the homogeneous mode can accidently stop on the top of the potential. In this case, all the fragmentation effects we discuss in this section will be amplified. We will neglect such accidental configurations in this work.}. We have chosen it as $\Theta_i=\pi/2$.

In order to fix the initial conditions for the mode functions, we solve \eqref{eq:6} and \eqref{eq:219} analytically at early times when the ALP mass can be neglected by assuming that all the modes are sourced from adiabatic perturbations and were super-horizon when ALP homogeneous mode started to scale as kineation, i.e. $\rho_{\Theta}\propto a^{-6}$. We describe the calculation in Appendix \ref{sec:kinetic-misalignment}, and in much more detail in \cite{Eroncel:2022vjg}. For $t_{k}<1$, we can express the solution as a power series in $t_{k}$ as (see (\ref{eq:9}))
\vspace{0.2cm}
\begin{mdframed}[backgroundcolor=blue!15,innertopmargin=0pt,innerbottommargin=9pt]
\begin{align}
  \label{eq:228}
\text{KMM:}\quad \theta_k(t_m)&\approx 2 t_m\dv{\Theta}{t_m}\Phi_k(0)\sum_{i=0}^4c_i t_k^{2i},
\end{align}
\end{mdframed}
\vspace{0.2cm}
where the coefficients $\qty{c_i}$ are
\begin{equation}
  \label{eq:229}
  c_0=-\frac{1}{2},\quad c_1=\frac{23}{20},\quad c_2=-\frac{491}{1680},\quad c_3=\frac{4427}{\numprint{151200}},\quad c_4=-\frac{\numprint{146131}}{\numprint{93139200}}.
\end{equation}
The conversion between $t_k$ and $t_m$ can be done via
\begin{equation}
  \label{eq:230}
  t_k=\sqrt{\frac{4 t_m}{3}}\tilde{k}.
\end{equation}
The approximation \eqref{eq:228} is quite accurate up to $t_k=1$. Therefore we have chosen our initial time early enough so that for all simulated modes $t_k<1$ initially.
\begin{figure}[tbp]
  \centering
  \includegraphics[width=0.8\textwidth]{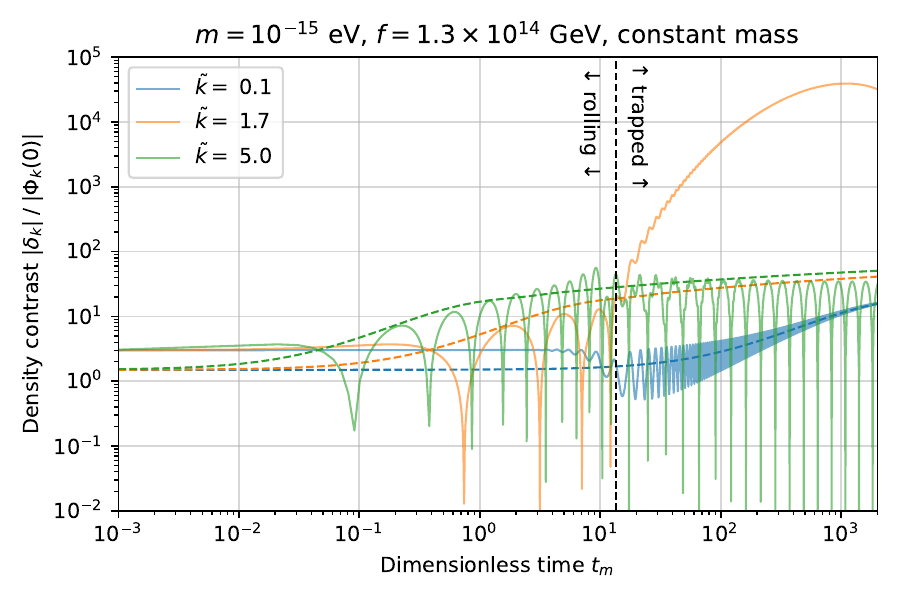}
  \caption{\small \it Evolution of the density contrast, obtained as the numerical solution of \eqref{eq:6} and \eqref{eq:219} together with the initial conditions \eqref{eq:228}. The amplitude is normalized with respect to the value of the primordial curvature perturbations $\abs{\Phi_k(0)}=(2/3)\sqrt{A_s}\approx 3\times 10^{-5}$. The corresponding benchmark point, $m=10^{-15}\,\si{\electronvolt}$, $f=1.3\times 10^{14}\,\si{\giga\electronvolt}$ and $m/H_{\trap}\approx27$, lies in the region where fragmentation is not complete. The dashed lines show the CDM predictions which are calculated using \eqref{eq:232}.}
  \label{fig:cm-mod-function-eval-early}
\end{figure}

\subsubsection{Density contrast evolution}

The evolution of the density contrast is shown for a couple of modes in Figure \ref{fig:cm-mod-function-eval-early}, for the benchmark point $m=10^{-15}\,\si{\electronvolt}$ and $f=1.3\times 10^{14}\,\si{\giga\electronvolt}$, which lies in the second region (incomplete fragmentation) described in Section \ref{sec:brief-review-alp}, but is close to the boundary with the third region. The dashed black vertical line shows the time at which the homogeneous mode gets trapped $t_{m,\trap}=m/2 H_{\trap}$. For comparison we also show the CDM predictions with the colored dashed lines. These can be calculated analytically in the radiation era, and they read \cite{Zhang:2017flu}
\begin{equation}
  \label{eq:232}
  \delta_{\rm{CDM},k}=9\Phi_k(0)\qty[\frac{\sin t_k}{t_k}+\frac{\cos t_k}{t_k^2}-\frac{\sin t_k}{t_k^3}-\cint(t_k)+\ln t_k+\gamma_{\rm E} - \frac{1}{2}],
\end{equation}
where $\cint$ is the cosine integral, and $\gamma_{\rm E}$ is the Euler-Mascheroni constant. Note that all modes approach to $3\Phi_k(0)$ in the super-horizon limit. This is becuase in this limit the adiabatic perturbations satisfy
\begin{equation}
  \label{eq:15}
  \frac{\delta_{i}}{1+w_{i}}=\frac{\delta_{j}}{1+w_{j}},
\end{equation}
for all species $i$ and $j$, where $\delta$ and $w$ are the density contrast and the equation of state respectively. During the radiation era the radiation perturbations are related to the curvature perturbations by $\delta_{r}\approx 2\Phi_{k}(0)$ in the super-horizon limit. Plugging this relation into \eqref{eq:15} and using the fact that the ALP equation of state is $w_{\theta}\approx 1$ before trapping gives $\delta\approx 3\Phi_{k}(0)$ at early times in the super-horizon limit.

Later behavior can be interpreted as follows:
\begin{itemize}
\item $\tilde{k}=0.1$: This mode enters the horizon at $t_m=3/4 \tilde{k}^2=75$ long after the homogeneous mode starts oscillating and redshifts as matter. Therefore, this mode locks on the CDM evolution once the homogeneous mode gets trapped.
\item $\tilde{k}=5.0$: This mode experiences some growth slightly before the homogeneous mode gets trapped, but after that it oscillates with a constant amplitude which is smaller compared to the CDM prediction.
\item $\tilde{k}=1.7$: This mode experiences significant growth due to the parametric resonance effect explained in \cite{Eroncel:2022vjg}, and its amplitude is enhanced significantly compared to CDM.
\end{itemize}

\begin{figure}[tbp]
  \centering
  \includegraphics[width=0.8\textwidth]{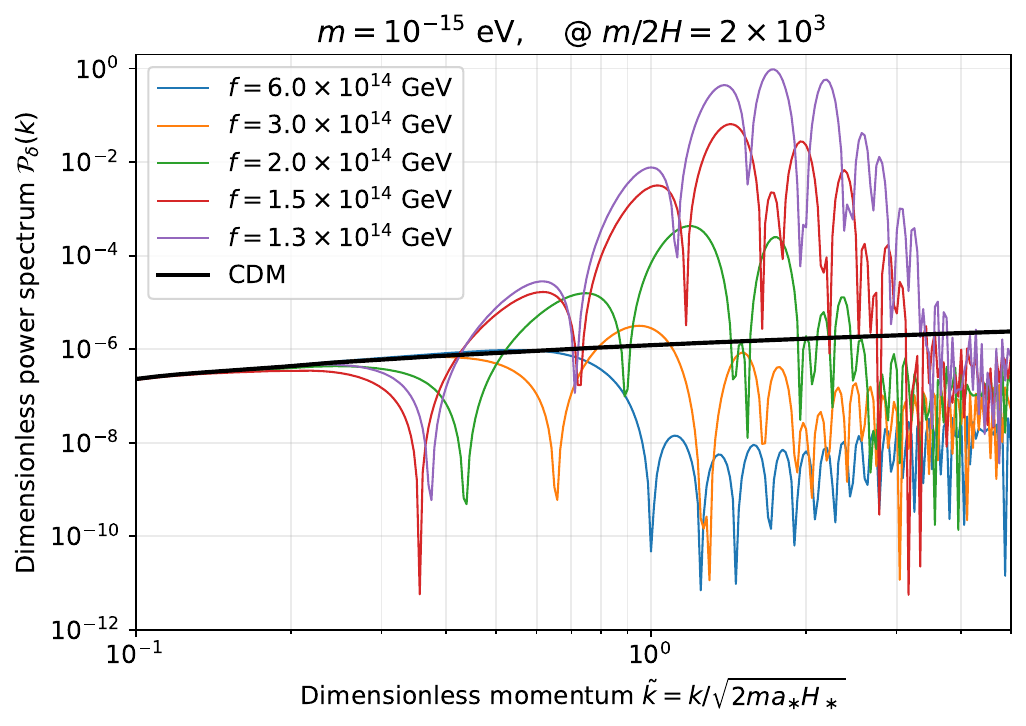}
  \caption{\small \it The dimensionless power spectrum for different benchmarks points (colored lines), and the CDM power spectrum (black line) calculated via \eqref{eq:232} at $t_m=m/2H=2\times 10^3$. For all benchmarks points, the ALP mass is $m=10^{-15}\,\si{\electronvolt}$. We observe that all power spectra converge to the CDM prediction for low momenta. For $f=\qty{6.0, 3.0, 2.0, 1.5, 1.3}\times 10^{14}\,\si{\giga\electronvolt}$, we have $m/H_{\trap}\approx\qty{3, 9, 15, 22, 27}$ respectively. As we see, by decreasing the ALP decay constant $\decay$, we increase the hierarchy between the ALP mass and the Hubble scale at trapping, $m/H_*$, which implies more efficient fragmentation. }
  \label{fig:linear-power-spectrum-cm-early}
\end{figure}

\subsubsection{Power spectrum}

By solving for the mode functions, we obtain the dimensionless power spectrum
\begin{equation}
  \label{eq:22}
  \dimlessps_{\delta}(k)=\frac{k^{3}}{2\pi^{2}}\dimps_{\delta}(k)=A_{s}\qty(\frac{2}{3})^{2}\tilde{\delta}_{k}^{2},
\end{equation}
where $\delta_{k}=\tilde{\delta}_{k}\Phi_{k}(0)$. We show it in Figure \ref{fig:linear-power-spectrum-cm-early} for different $f$ values, and compare to the CDM prediction calculated via \eqref{eq:232}. We set the ALP mass to $m=10^{-15}\,\si{\electronvolt}$. Smaller ALP decay constants correspond to larger hierarchies between the ALP mass and the Hubble scale at trapping which implies more efficient fragmentation. This feature can easily be observed in the plot. We can also see that all power spectra converge to the CDM line at low momenta since these modes enter the horizon long after the homogeneous mode gets trapped and redshifts as matter.

\subsection{Evolution at late times: Well after fragmentation until today}
\label{sec:evol-at-latet}
In order to study the observational signatures such as the halo spectrum, we need to evolve the power spectrum until today. Performing this by numerically solving the mode functions is not feasible due to the rapid oscillations and very long time scales. However, an effective description can be obtained for sub-horizon modes by using the fact that long after the onset of oscillations $a \gg a_{\trap}$, i.e. $t_m\gg 1$, the ALP energy density redshifts like matter with an average equation of state $w\approx 0$. Therefore, one can use the WKB approximation for both the homogeneous mode $\Theta$, and the mode functions $\theta_k(t)$. One makes the ansatz \cite{Park_2012}
\begin{align}
  \label{eq:233}
  \Theta(t)&=a^{-3/2}\qty[\Theta_+\cos(mt)+\Theta_-\sin(mt)],\\
  \label{eq:234}
  \theta_k(t)&=\theta_+(k,t)\cos(mt)+\theta_-(k,t)\sin(mt),
\end{align}
where $\Theta_{\pm}$ are constants. With this approximation, one can directly derive the evolution of the density constrast as \cite{Park_2012}
\begin{equation}
  \label{eq:235}
  \ddot{\delta}_k+2H\dot{\delta}_k+\qty[c_{s,\rm eff}^2\qty(\frac{k^2}{a^2}+16\pi G \rho_r)-4\pi G \rho_{\Theta}]\delta_k=8\pi G\rho_r\delta_{r,k},
\end{equation}
where $\rho_r$ is the energy density in the radiation, $\delta_r$ is the density constrast of the radiation, and $c_{s,\rm eff}^2$ is the effective sound speed of the ALP:
\begin{equation}
  \label{eq:236}
  c_{s,\rm eff}^2=\frac{1}{4}\frac{k^2}{a^2m^2}\qty(1 + \frac{1}{4}\frac{k^2}{a^2m^2})^{-1}\approx \frac{1}{4}\frac{k^2}{a^2m^2},
\end{equation}
where the approximation holds for non-relativistic modes. Even though this expression is derived in the axion comoving gauge, additional terms that arise when converting this result into any other gauge decay on sub-horizon scales \cite{Hlozek:2014lca}. The source term on the RHS of \eqref{eq:235} is proportional to the density constrast of radiation which oscillates rapidly for sub-horizon modes. Therefore, by averaging over these oscillations we can neglect the source term for the sub-horizon modes. The $16\pi G \rho_r$ term on the LHS is negligible during the matter era since $c_{s,\rm eff}^2\rho_r\ll \rho_{\Theta}$\footnote{Here we assume that ALPs make all of dark matter so $\rho_{\Theta}=\rho_{\rm DM}$. }. Also, in radiation era we have $16\pi G \rho_r\approx 6 H^2 \ll k^2/a^2$ for sub-horizon modes. In the end, we find that in both the radiation and matter era, the evolution of the density constrast on sub-horizon scales at late times can be described by
\begin{equation}
\boxed{
  \label{eq:237}
  \ddot{\delta}_k+2H\dot{\delta}_k+\qty(c_{s,\rm eff}^2\frac{k^2}{a^2}-4\pi G \rho_{\Theta})\delta_k=0}
\end{equation}
The terms $4\pi G \rho_{\Theta}$ is responsible for the graviational collapse and the structure formation during the matter era, while $c_{s,\rm eff}^2k^2/a^2$ is the ``pressure'' term. The CDM evolution can be obtained by setting the sound speed to zero. On large scales $c_{s,\rm eff}^2k^2/a^2 \ll 4\pi G \rho_{\Theta}$, density wins over the pressure so the density constrast grows like in the CDM case; logarithmically during the radiation era, and linearly during the matter era with respect to the scale factor. However, at small scales the pressure wins, and the density contrast oscillates with a constant amplitude. The wavenumber at which both become equal is known as the \emph{axion Jeans scale} given by \cite{Marsh:2015xka}
\begin{equation}
  \label{eq:238}
  k_J=\qty(16\pi G a \rho_{\Theta,\present})^{1/4}\sqrt{m}=66.5\times a^{1/4}\qty(\frac{h^2\Omega_{\Theta}}{h^2\Omega_{\rm DM}})^{1/4}\sqrt{\frac{m}{10^{-22}\,\si{\electronvolt}}}\,\rm{Mpc}^{-1}.
\end{equation}
The take-away lesson is for ALPs there is scale-dependent growth, and ALP dark matter differs from CDM on scales below the axion Jeans scale.

\begin{figure}[tbp]
  \centering
  \includegraphics[width=\textwidth]{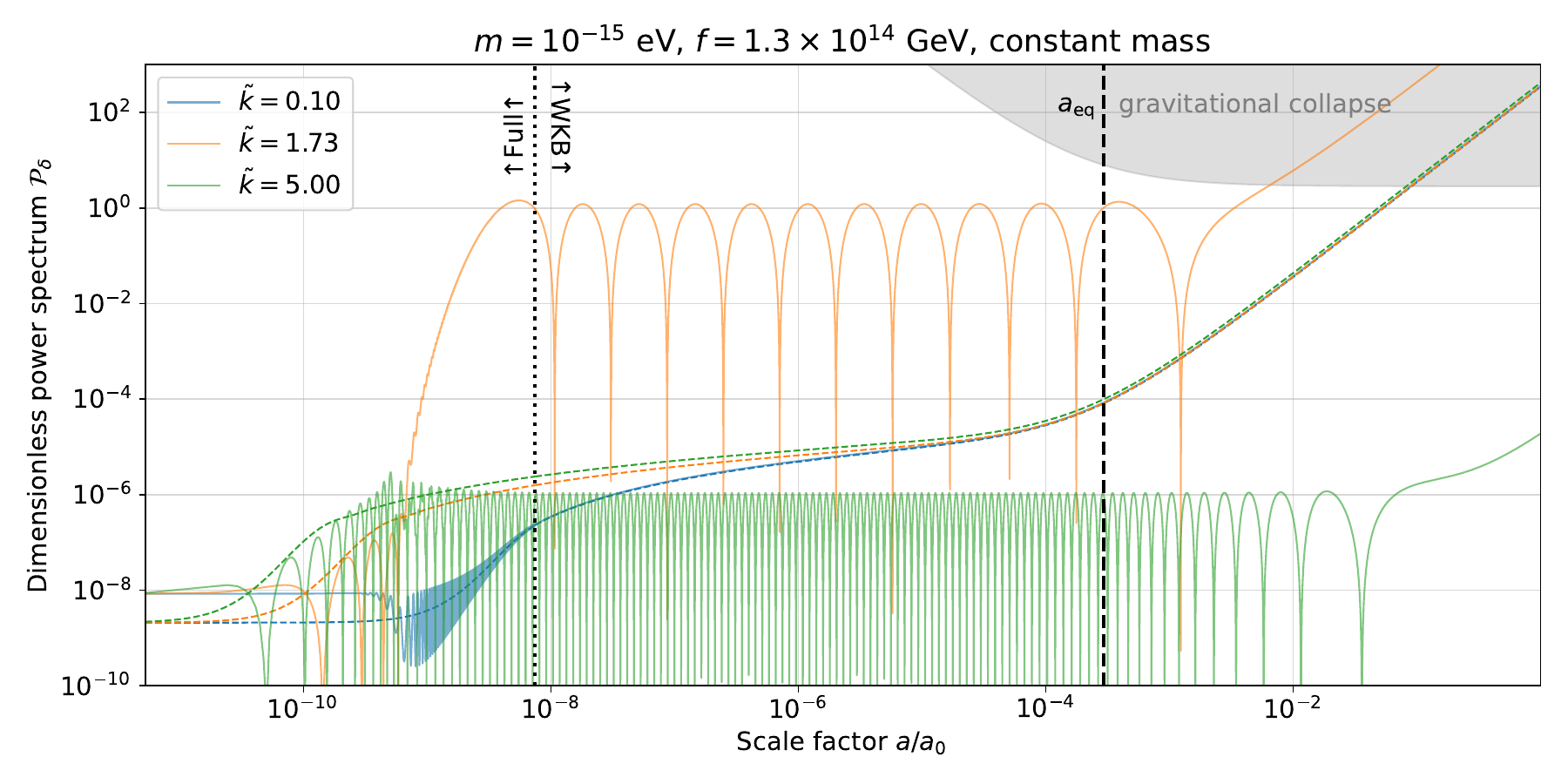}
  \caption{\small \it Evolution of the density contrast (solid lines) for three different momentum modes from the time they enter the horizon until today. Up to the vertical dotted line on the left, the evolution is obtained by solving the full equations of motion \eqref{eq:6} and \eqref{eq:219}, while the WKB approximation \eqref{eq:237} is used for the remaining evolution. The colored dashed lines show the CDM evolution which we have obtained by matching the exact solution \eqref{eq:232} in the ratiation era with the WKB approximation with $c_s^2=0$. The lowest momentum mode shown in blue locks on to the CDM evolution when the homogeneous mode gets trapped, and follows it until today. The intermediate mode shown in orange gets enhanced significantly at trapping, then oscillates with a frequency given by the sound speed until it drops below the Jeans scale in the matter era. After that it starts to grow linearly. The highest momentum mode shown in green does not experience exponential growth at trapping, and oscillates rapidly in its remaining evolution. The grey shaded region on the top right denotes where the overdensities collapse gravitationally. Its expression is derived in Section \ref{sec:crit-dens-coll}. }
  \label{fig:cm-mod-function-eval-full}
\end{figure}

Following \cite{Arvanitaki:2019rax}, we define 
\begin{equation}
y\equiv a/a_{\rm eq} .
 \end{equation}
In terms of this variable \eqref{eq:235} becomes\footnote{In this equation $\tilde{k}$ is constant and defined in radiation era via \eqref{eq:216}.}
\begin{equation}
  \label{eq:239}
  y(1+y)\dv[2]{\delta_k}{y}+\qty(1+\frac{3}{2}y)\dv{\delta_k}{y}+\qty(\frac{\tilde{k}^4}{y}-\frac{3}{2})\delta_k=0.
\end{equation}
Therefore, the momentum mode $\tilde{k}$ drops below the Jeans scale at $y=2\tilde{k}^4/3$. We match the evolution of the density constrast obtained by solving the mode function equation of motion \eqref{eq:219} with the differential equation given above at some $t_m^{\rm match}$ which is much later than the onset of oscillations $t_m^{\rm match}\gg 1$, but long before the matter-radiation equality $y_{\rm match}\ll 1$. By using the fact that $y=2^{1/4}\sqrt{H_{\rm eq} t_m/m}$ in radiation era, we find the matching conditions as
\begin{equation}
  \label{eq:240}
  \eval{\delta}_{t_m^{\rm match}}=\eval{\delta}_{y_{\rm match}}\qand \eval{2 t_m\dv{\delta}{t_m}}_{t_m^{\rm match}}=\eval{y\dv{\delta}{y}}_{y_{\rm match}}.
\end{equation}
The full evolution of the density contrast of three momentum modes $\tilde{k}=0.1$, $\tilde{k}=1.73$, $\tilde{k}=5.0$ is plotted in Figure \ref{fig:cm-mod-function-eval-full} for a benchmark point $m=10^{-15}\,\si{\electronvolt}$ and $f=1.3\times 10^{14}\,\si{\giga\electronvolt}$. The full solution of the equations of motion \eqref{eq:6} and \eqref{eq:219} is matched with the solution of the WKB approximation \eqref{eq:237} by using \eqref{eq:240} at $t_m=2\times 10^3$ or $y\sim 10^{-8}$. The momentum mode $\tilde{k}=0.1$ shown in blue continues to track the CDM evolution shown via colored dashed lines. The mode $\tilde{k}=1.73$ shown in orange oscillate with a constant amplitude while it is above the Jeans scale. Once it drops below the Jeans scale during the matter era, it starts to grow linearly. The evolution of the mode $\tilde{k}=5.0$ is similar, except it does not experience exponential growth during trapping, and drops below the Jeans scale much later.

\subsection{Comparison with the Large Misalignment Mechanism}
\label{sec:comp-with-stand}

In this section, we compute the power spectrum of the ALP fluctuations in the Large Misalignment Mechanism (LMM), and make the comparison with the Kinetic Misalignment. This mechanism \cite{Zhang:2017dpp,Arvanitaki:2019rax} is a special case of the Standard Misalignment Mechanism where the initial angle $\Theta_i$ is very close to the top of the cosine potential, i.e. $\abs{\pi-\Theta_i}\ll 1$.

The calculation is the same as in the KMM case with the exception of initial conditions for the homogeneous mode, and the mode functions. We start by discussing the homogeneous mode evolution. This will also give us the expression for the relic density today, which we will use to determine the model parameters. Then we state the initial conditions for the mode functions, and finally show the comparison.

\subsubsection*{Evolution of the homogeneous mode}

We again use the dimensionless time $t_m=m/2H\simeq mt$ as the dynamical variable. The initial conditions are given by
\begin{equation}
  \label{eq:289}
  \lim_{t_m \rightarrow 0}\Theta(t_m)=\Theta_i,\quad \lim_{t_m \rightarrow 0}\Theta'(t_m)=0.
\end{equation}
If the initial angle is very close to the bottom of the potential $\abs{\Theta_i}\ll 1$, then the cosine potential can be approximated by a quadratic, which implies $\sin\Theta\approx \Theta$. In this case the homogeneous mode equation of motion \eqref{eq:6} with initial conditions \eqref{eq:289} has an exact solution:
\begin{equation}
  \label{eq:290}
  \Theta(t_m)=2^{1/4}\Gamma\qty(\frac{5}{4})\Theta_i \frac{J_{1/4}(t_m)}{t_m^{1/4}},\quad \Theta_i\ll 1.
\end{equation}
In this case, the energy density is given by
\begin{equation}
  \label{eq:291}
  \rho_{\Theta}(t_m)=\frac{m^2f^2}{2}\qty[\Theta'(t_m)^2+\Theta(t_m)^2]=m^2f^2\qty(\Gamma\qty(\frac{5}{4})\Theta_i)^2\qty[\frac{J_{1/4}^2(t_m)+J_{5/4}^2(t_m)}{\sqrt{2 t_m}}].
\end{equation}
In the late time limit $t_m\gg 1$, in other words much after the onset of oscillations, one can use asymptotic forms of the Bessel functions to show that
\begin{equation}
  \label{eq:292}
  \rho_{\Theta}(t_m\gg 1)\approx m^2f^2 \frac{\sqrt{2}}{\pi}\qty(\Gamma\qty(\frac{5}{4})\Theta_i)^2t_m^{-3/2}\propto a^{-3},
\end{equation}
so the ALP field scales as cold dark matter at late times. Redshifting this energy density until today we obtain
\begin{equation}
  \label{eq:293}
  \Omega_{\Theta,0}\approx\qty(5.2\times 10^{-4})\qty(\frac{g_{\rho}^{3/4}(T_m)}{g_s(T_m)})\sqrt{\frac{m}{H_0}}\qty(\frac{f}{\mpl})^2\Theta_i^2,
\end{equation}
where $T_m$ is the temperature at which $t_m=1$, i.e. $2H(T_m)=m$, and $g_{\rho}(T_m),g_s(T_m)$ are the number of effective degrees of freedom in the energy density and entropy respectively.

\begin{figure}[tbp]
  \centering
  \begin{subfigure}[b]{0.49\textwidth}
    \includegraphics[width=\textwidth]{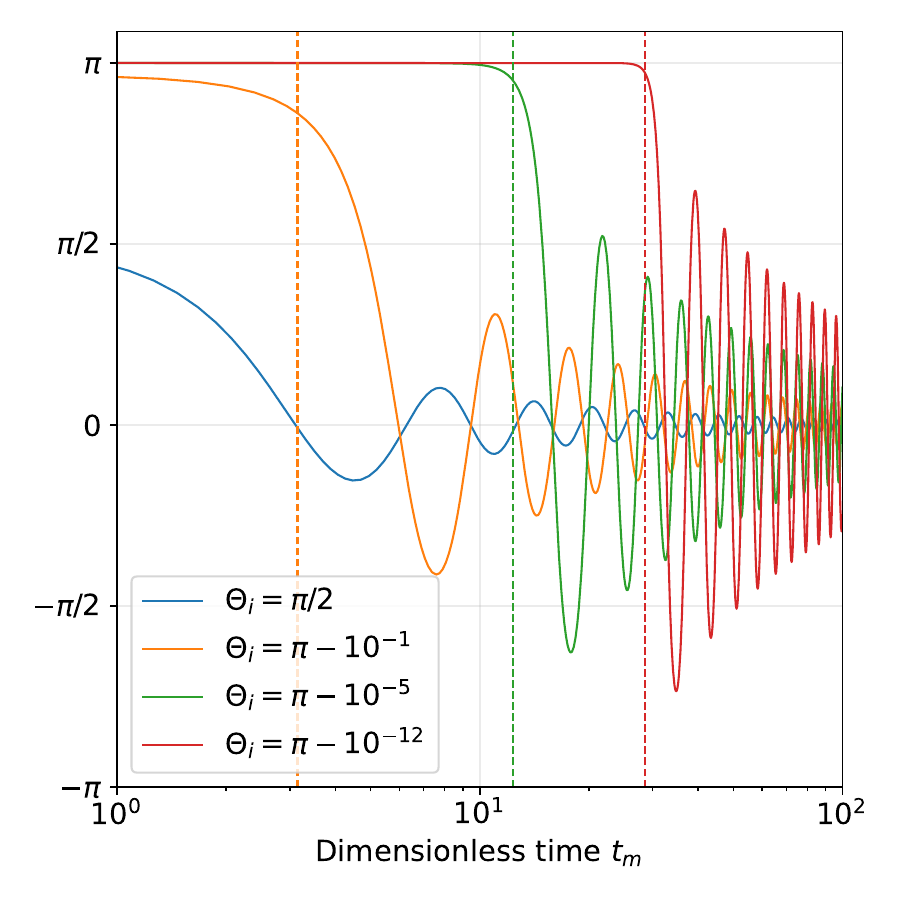}
  \end{subfigure}
  \hfill
  \begin{subfigure}[b]{0.49\textwidth}
    \includegraphics[width=\textwidth]{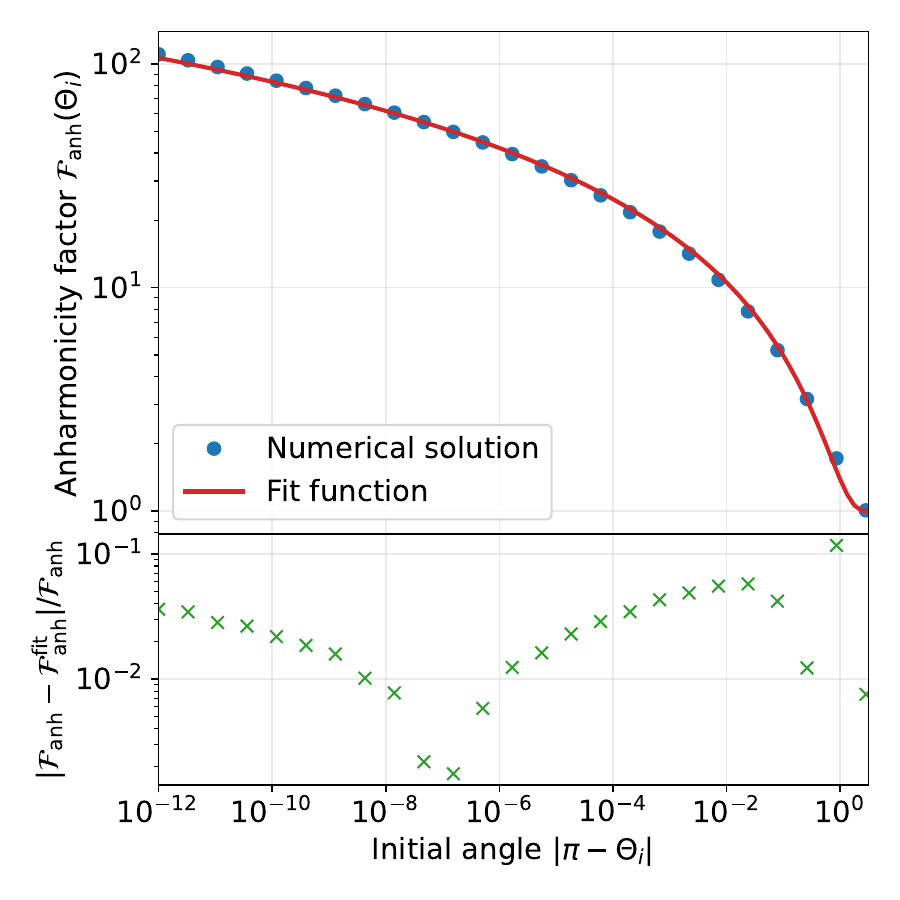}
  \end{subfigure}
  \caption{\small \it \textbf{Left figure:} The evolution of the homogeneous mode in SMM/LMM with different initial conditions obtained by numerically solving \eqref{eq:6}. When the initial angle is very close to the top of the cosine potential, the ALP field starts oscillating much later due to to the tiny potential gradient at the top. The onset of oscillations are shown by vertical dashed lines, and calculated via \eqref{eq:306}. \textbf{Right figure:} Upper plot shows the numerically obtained anharmonicity factor values, and the plot of the fit function \eqref{eq:297}, while the bottom plot show the relative errors of the fit function. In most of the parameter space, the relative error is less than $10\%$. }
  \label{fig:smm-zero-mod}
\end{figure}

If the initial angle $\Theta_i$ is not close to the minimum, the expression \eqref{eq:293} receives corrections from the anharmonicity of the potential. For a general initial angle, the relic density today is given by
\begin{equation}
\boxed{
  \label{eq:296}
 \Omega_{\Theta,0}= \qty(5.2\times 10^{-4})\qty(\frac{g_{\rho}^{3/4}(T_m)}{g_s(T_m)})\sqrt{\frac{m}{H_0}}\qty(\frac{f}{\mpl})^2\Theta_i^2\mathcal{F}_{\rm anh}(\Theta_i),}
\end{equation}
where $\mathcal{F}_{\rm anh}$ is the anharmonicity correction which should behave as $\mathcal{F}_{\rm anh}(\Theta_i)\rightarrow 1$ as $\Theta_i \rightarrow 0$. By solving the equation of motion \eqref{eq:5} numerically with different initial angles, we found that the fit function
\begin{equation}
  \label{eq:297}
  \mathcal{F}_{\rm anh}^{\rm fit}(\Theta_i)=\qty[1-\ln\qty(1-\qty(\frac{\abs{\Theta_i}}{\pi})^{3.829})]^{1.395},
\end{equation}
approximates the numerical solution to an accuracy better than $10\%$ for most of the initial values. This function together with the numerically calculated values and the relative errors are shown on the right plot of Figure \ref{fig:smm-zero-mod}.

As the initial angle becomes closer to the top of the potential, the anharmonicity factor increases, thereby the relic density also increases. This behavior is due to the fact that the tiny potential gradient at the top delays the onset of oscillations as can be seen from the left plot of Figure \ref{fig:smm-zero-mod}. For small displacements from the top $\abs{\pi-\Theta_i}\ll 1$, the dimensionless time at which the oscillations start can be approximated by \cite{Arvanitaki:2019rax}
\begin{equation}
  \label{eq:306}
  t_m^{\rm osc}=\ln\qty[\frac{1}{\abs{\pi-\Theta_i}}\frac{2^{1/4}\pi^{1/2}}{\Gamma(5/4)}],
\end{equation}
which we show via dashed vertical lines in the left plot of Figure \ref{fig:smm-zero-mod}. As a result, the duration of the matter-like scaling decreases which enhances the relic density. Soon we will show that this delay of oscillations also enchances the ALP fluctuations just like in the Kinetic Misalignment case.

\subsubsection*{Initial conditions for the mode functions}

In the case of Standard/Large Misalignment Mechanism, the background ALP field is frozen at early times, and thus behaves as dark energy. At the zeroth order, the adiabatic initial conditions for the ALP perturbations do vanish \cite{Marsh:2015xka}. Higher order corrections can be derived by solving the equations of motion for the homogeneous mode \eqref{eq:6}, and the mode functions \eqref{eq:219} at early times $t_m\ll 1$. 
The calculation is somewhat technical and is presented in Appendix \ref{sec:stand-misal}, leading to 
 equation (\ref{eq:304}), the early-time super-horizon behavior of the mode functions:
\vspace{0.2cm}
\begin{mdframed}[backgroundcolor=blue!15]
\begin{equation}
  \label{eq:307}
  \text{SMM/LMM:}\quad\theta_k(t_m\ll 1)=\frac{2}{5}\Phi_k(0)\sin\Theta_i\qty[t_m^2-\frac{22}{105}\tilde{k}^2t_m^3+\mathcal{O}(t_m^4)].
\end{equation}
\end{mdframed}
\vspace{0.2cm}

\subsubsection*{The results and comparison with KMM}

\begin{figure}[tbp]
  \centering
  \includegraphics[width=0.8\textwidth]{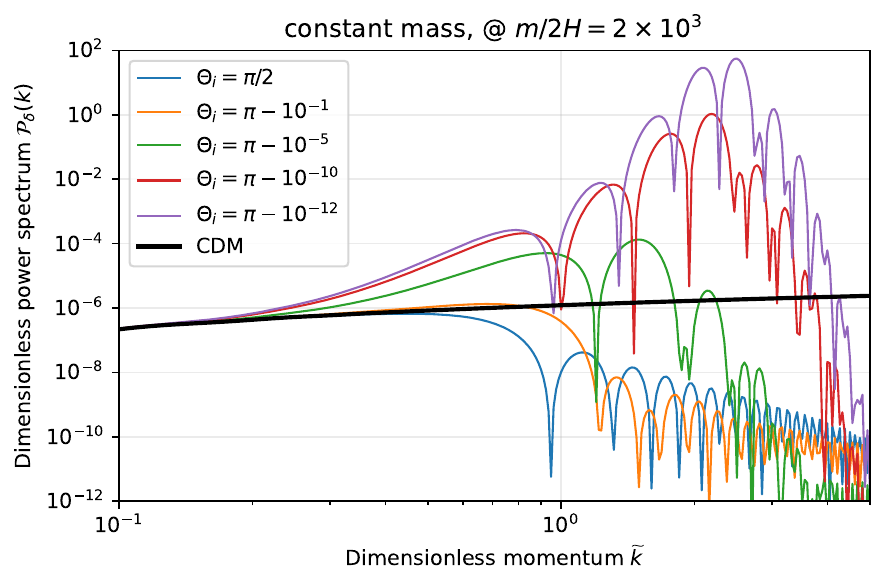}
  \caption{\small \it Power spectra of ALP fluctuations in the Standard/Large Misalignment with different initial conditions. These results are independent of the ALP mass and the decay constant, provided that the oscillations start deep in the radiation era. We observe that if the initial angle is significantly close to the top of the potential $\abs{\pi - \Theta_i}\lesssim 10^{-5}$, there is an enhancement in the power spectrum around the modes $\tilde{k}\sim \mathcal{O}(1)$. }
  \label{fig:smm-linear-power-spectrum-cm-early}
\end{figure}

We first show the dimensionless power spectra at $t_m=m/2H=2\times 10^3$ with different initial conditions in Figure \ref{fig:smm-linear-power-spectrum-cm-early}. These results depend only on the initial conditions, and they are independent of the ALP mass and the ALP decay constant provided that the oscillations start deep in the radiation era. From the plot, we can observe that for a large range of initial values, the power spectra have similar features. The low momentum modes $\tilde{k}\ll 1$ behave as CDM, while larger modes $\tilde{k}\gg 1$ are suppressed with respect to CDM. Only when the initial angle is significantly close to the top of the potential $\abs{\pi-\Theta_i}\lesssim 10^{-5}$, there is an enhancement in the modes $\tilde{k}\sim\mathcal{O}(1)$. The power spectrum reaches to $\mathcal{O}(1)$ values when $\abs{\pi-\Theta_i}\lesssim 10^{-10}$, after which the linear perturbation theory becomes unreliable.

\begin{figure}[tbp]
  \centering
  \includegraphics[width=0.8\textwidth]{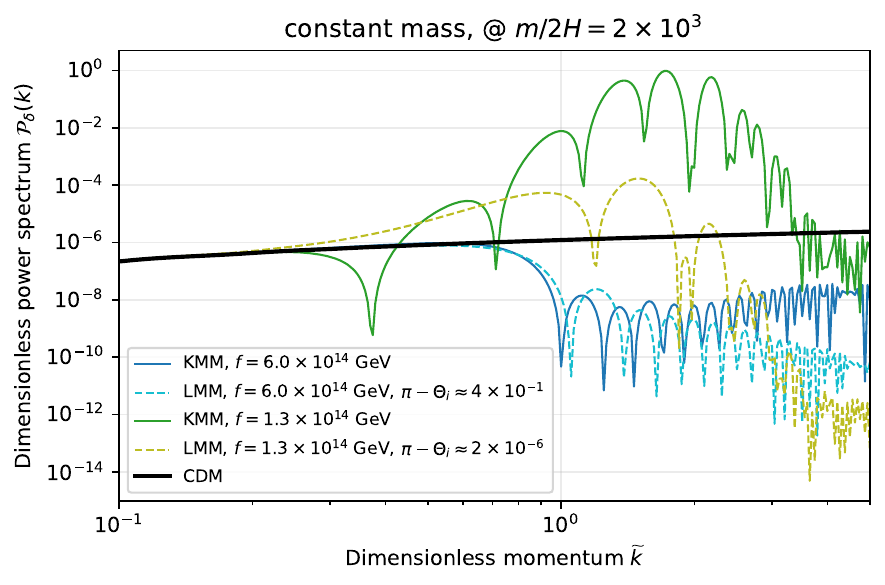}
  \caption{\small \it Comparison of the dimensionless power spectra of ALP fluctuations at $t_m=m/2H=2\times 10^3$ in the Kinetic Misalignment mechanism (solid lines), with Large Misalignment Mechanism (dashed lines). For all curves we assume a constant ALP mass $m=10^{-15}\,\si{\electronvolt}$. For LMM benchmarks, the initial angles are determined via \eqref{eq:296} such that the relic density is the same as the dark matter density. For $f=6.0\times 10^{14}\,\si{\giga\electronvolt}$ (blue lines) both power spectra are similar except there is a suppression of the high-momentum modes in LMM which is absent in KMM. For $f=1.3\times 10^{14}\,\si{\giga\electronvolt}$ (green lines), both models predict an enhancement for the modes around $\tilde{k}\sim \mathcal{O}(1)$, however the growth is much larger in KMM. The reasons behind both of these features are explained in the main text. }
  \label{fig:smm-kmm-ps-compare}
\end{figure}

We now compare the power spectra in the Large and Kinetic misalignment mechanisms. The results are shown in Figure \ref{fig:smm-kmm-ps-compare} which assumes a constant ALP mass equal to $m=10^{-15}\,\si{\electronvolt}$. We have chosen two benchmark values for the ALP decay constant; $f=6.0\times 10^{14}\,\si{\giga\electronvolt}$, and $f=1.3\times 10^{14}\,\si{\giga\electronvolt}$. The former is close to the boundary of the SMM-KMM transition in the language of Section \ref{sec:brief-review-alp}\footnote{In Section \ref{sec:brief-review-alp}, we defined the SMM region as the region where $m_{\trap}/H_{\trap}<3$. This is slightly different what we mean as SMM in this section. In the former, the ALP field has an initial kinetic energy, but it is insufficient to delay the onset of oscillations, while in the latter there is no initial kinetic energy. Despite of this difference, there is no difference from a phenomenological standpoint.}, while the latter is close to the boundary of incomplete-complete fragmentation. To compare the power spectra in these two benchmark points with their LMM counterparts, we kept the ALP decay constant the same, and determined the initial angle $\Theta_i$ via \eqref{eq:296} and \eqref{eq:297} such that all of dark matter is made of ALPs. For $f=6.0\times 10^{14}\,\si{\giga\electronvolt}$ (blue lines), both models predict a similar power spectra except higher momentum modes are suppressed in the case of LMM, while they are not in KMM. On the other hand, for $f=1.3\times 10^{14}\,\si{\giga\electronvolt}$ (green lines) both models predict an enhancement of the modes with $\tilde{k}\sim \mathcal{O}(1)$, however the enhancement in the case of KMM is much larger compared to LMM.

\begin{figure}[tbp]
  \centering
  \includegraphics[width=0.8\textwidth]{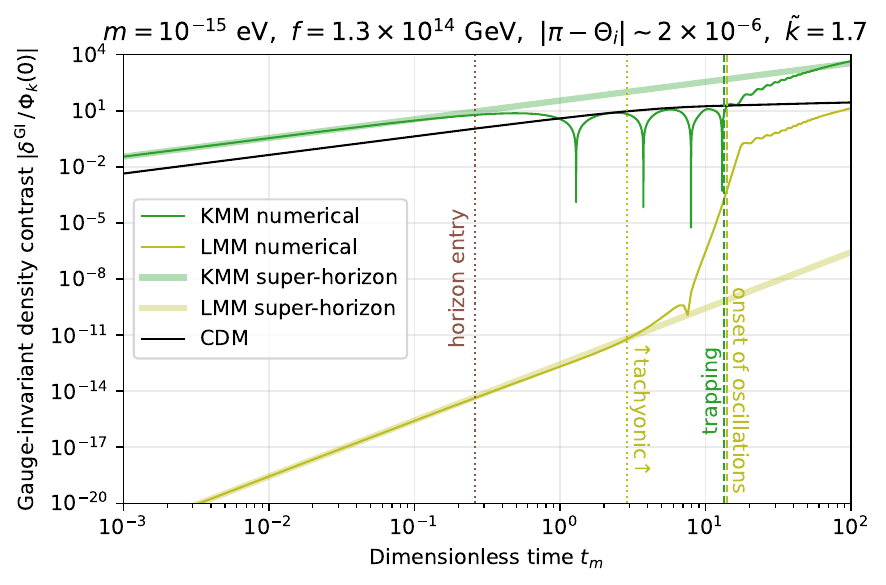}
  \caption{\small \it Comparison of the evolution of the gauge-invariant density contrast of the mode $\tilde{k}=1.7$ in the KMM (green) and LMM (light green) models. The thin lines denote the numerical solution while the thick lines show the analytically obtained super-horizon behavior given in \eqref{eq:318} for KMM and \eqref{eq:319} for LMM. The brown dotted line in the middle shows the time at which the mode enters the horizon. The green line on the right shows the time of trapping in KMM, while the nearby light green dashed line is the onset of oscillations in LMM. The dotted light green line shows the point when $\tilde{k}^2/t_m=1$ indicating the start of the tachyonic instability. We see that even though the enhancement is much larger in LMM when the initial angle is very close to the top, the final value is smaller than its KMM counterpart due to the heavy suppression of the initial fluctuations from small $\abs{\pi - \Theta_i}$, see \eqref{eq:321}. }
  \label{fig:smm-kmm-earlytime-comparison}
\end{figure}

The larger enhancement in KMM compared to LMM deserves an explanation. At first, one might think that this is due to the fact that the onset of oscillations is not delayed by the same amount. However this is not true. Recall that in KMM the oscillations start at $\tau=2 H_{\trap}t$ by definition. In terms of $t_m$ this is
\begin{equation}
  \label{eq:317}
  t_m^{\rm osc}=\frac{m}{2 H_{\trap}},\quad \text{KMM}.
\end{equation}
On the other hand, the onset of oscillations in LMM is given by \eqref{eq:306}. Then for $\decay=1.3\times 10^{14}\,\si{\giga\electronvolt}$ one gets $t_m^{\rm osc}\approx 13.4(12.7)$ for KMM(LMM), so in both cases the onset of oscillations are delayed roughly by the same amount. The real reason behind the difference is imprinted on the evolution of the density contrast before the onset of oscillations, and can be studied analytically. We will make the comparison with the gauge-invariant density contrast
\begin{equation}
  \label{eq:16}
  \delta^{\rm GI}=\delta-3\mathcal{H}(1+w)v,
\end{equation}
where $v$ is the \emph{velocity potential} which for the ALP field takes the form \cite{Eroncel:2022vjg}
\begin{equation}
  \label{eq:17}
  v=-\frac{f^{2}\dot{\Theta}\delta\theta}{a \rho_{\Theta}(1+w)}.
\end{equation}
Thus, in Fourier space the gauge-invariant density contrast is
\begin{equation}
  \label{eq:18}
  \delta_{k}^{\rm GI}=\delta_{k}-3H \frac{f^{2}\dot{\Theta}\theta_{k}}{\rho_{\Theta}}.
\end{equation}
By using the early-time behavior of the KMM mode functions \eqref{eq:228} and the expression for the curvature perturbations \eqref{eq:218}, we can show that at leading order

\begin{equation}
  \label{eq:318}
  \delta^{\rm GI}_{\rm KMM}\approx\Phi_k(0)\times
  \begin{cases}
    12 \,\tilde{k}^2 t_m,& t_k\ll 1,\\
    3 t_k\sin(\sqrt{3}t_k) &t_k\gg 1
  \end{cases}.
\end{equation}
In the SMM/LMM case, we can use \eqref{eq:304} and \eqref{eq:316} to obtain at leading order
\begin{equation}
  \label{eq:319}
  \delta^{\rm GI}_{\rm SMM/LMM}\approx \Phi_k(0)\times
  \begin{cases}
    \dfrac{16}{175}\cos^2\qty(\dfrac{\Theta_i}{2})\tilde{k}^2t_m^3,&t_k\ll 1\\[1.0em]
    \dfrac{81}{100}\qty(\dfrac{\sin^2\Theta_i}{\cos\Theta_i-1})\tilde{k}^{-4} t_k^2\cos(t_k),&t_k\gg 1
  \end{cases}.
\end{equation}
By roughly comparing the density contrast for a mode $\tilde{k}\sim 1$ at $t_m\sim 1$ we find
\vspace{.2cm}
\begin{mdframed}[backgroundcolor=blue!15]
\begin{equation}
  \label{eq:320}
  \frac{\delta_{\rm SMM/LMM}^{\rm GI}}{\delta_{\rm KMM}^{\rm GI}}\sim 8\times 10^{-3}\cos^2\qty(\frac{\Theta_i}{2}).
\end{equation}
\end{mdframed}
\vspace{.1cm}
The enhancement in LMM happens when $\abs{\pi -\Theta_i}\ll 1$. In this case, on top of the two orders of magnitude numerical suppression, there is also a drastic suppression from the cosine term:
\vspace{.2cm}
\begin{mdframed}[backgroundcolor=blue!15]
\begin{equation}
  \label{eq:321}
  \frac{\delta_{\rm LMM}^{\rm GI}}{\delta_{\rm KMM}^{\rm GI}}\sim 8\times 10^{-3}\times \frac{\abs{\pi-\Theta_i}^2}{4},\quad \abs{\pi-\Theta_i}\ll 1.
\end{equation}
\end{mdframed}
\vspace{.2cm}
This behavior can clearly be seen in Figure \ref{fig:smm-kmm-earlytime-comparison}. In this plot we compare the evolution of the gauge-invariant density contrast of the mode $\tilde{k}=1.7$ in KMM and LMM models. The thin lines denote the numerical solution, while the thick lines use super-horizon estimates \eqref{eq:318} and \eqref{eq:319}. We see that these estimates predict the evolution quite accurately until the mode enters the horizon which is shown by the brown dotted line in the middle. Around $t_m=1$, the LMM density constrast is around $12$-orders of magnitude smaller compared to KMM, consistent with the estimate \eqref{eq:321}. We also see that despite the fact that the final value of the density contrast is smaller in the LMM case, the enhancement is significantly larger. The explanation for this is the tachyonic instability. Recall that the mode function equation of motion has an effective frequency term given by $\omega_k^2=\tilde{k}^2/t_m-\cos\Theta$. During the period when the homogeneous mode is stuck at the top we have $\cos\Theta\approx \cos\Theta_i\approx -1$. So the effective frequency of the mode becomes tachyonic when $\tilde{k}^2/t_m>1$. This moment is shown by the vertical dotted light green line in Figure \ref{fig:smm-kmm-earlytime-comparison} labeled as ``tachyonic''. We see that the exponential growth starts shortly after this point.

Before closing this sub-section we also explain briefly why there is a suppression in SMM/LMM power spectra at higher momentum values. These modes are well sub-horizon when the oscillations start so their values slightly before the oscillations can be read from \eqref{eq:319} as $\delta^{\rm GI}\propto \tilde{k}^{-4}t_k^2\propto \tilde{k}^{-2}t_m$. Therefore, higher momentum modes are suppressed with $\tilde{k}^{-2}$ before the onset of oscillations. After the oscillations start their amplitudes are approximately constant which explains the suppression at high momentum modes in SMM/LMM.

\section{ALP density perturbations in the complete fragmentation regime}
\label{sec:alp-dens-pert-nl}

If the fragmentation is complete, the system dynamics is non-linear, and the equation of motion for the mode functions  become coupled to each other. In this case, one cannot use the cosmological perturbation theory as we did in the previous section. To study this regime precisely, one needs to use non-perturbative methods such as lattice simulations which we leave for future work. However, it is still possible to get some semi-analytical estimates by approximating the axion potential at late times by
\begin{equation}
  \label{eq:200}
  V(\theta)\approx\frac{1}{2}\decay^2m^2\theta^2;
\end{equation}
This approximation clearly breaks down during the fragmentation, since it neglects the strong mode mixing due to the non-linearities. But, it will eventually be a good one once the self-interactions become negligible\footnote{Another feature which cannot be studied with this approximation is the formation of \emph{oscillons} which are localized field configurations of a real scalar field sustained only by self-interactions \cite{Bogolyubsky:1976yu,Gleiser:1993pt,Copeland:1995fq}. For more information about the oscillons and their consequences for dark matter see \cite{Olle:2019kbo} and the references therein.}, and this is the best we can do without lattice simulations.
\begin{equation}
  \label{eq:19}
  \delta(t,\vb{x})\equiv \frac{\rho(t,\vb{x})-\expval{\rho}}{\expval{\rho}},
\end{equation}
where $\expval{\rho}$ is the average energy density. For a quadratic potential \eqref{eq:200} it takes the form
\begin{equation}
  \label{eq:241}
  \begin{split}
    \delta=\frac{\decay^2}{\expval{\rho}}\Bigg\lbrace&\frac{1}{2}\qty[\dot{\theta}^{2}\qty(1+2\Phi)-\expval{\dot{\theta}^2}]+\frac{1}{2a^2}\qty[\grad{\theta}\cdot\grad{\theta}\qty(1-2\Phi)-\expval{\grad{\theta}\cdot\grad{\theta}}]\\
    &+m^2\qty(\theta^2-\expval{\theta^2})\Bigg\rbrace.  
  \end{split}
\end{equation}
In deriving this expression we have neglected the $\mathcal{O}(\Phi^{2})$ terms, but the ALP field $\theta(t,\vb{x})$ is kept non-perturbative. We now expand the ALP field as
\begin{equation}
  \label{eq:242}
  \theta(t,\vb{x})=\theta_B(t)+\theta_F(t,\vb{x})=\theta_B(t)+\int_{k>k_{\trap}} \frac{\dd[3]{k}}{(2\pi)^3}\hat{\theta}_{\vb{k}}\theta_k(t)e^{-i\vb{k}\cdot\vb{x}}.
\end{equation}
Here $\hat{\theta}_{\vb{k}}$'s are stochastic variables that carry the statistical properties, while $\theta_{k}$'s are $c$-number functions which depend only on the magnitude of the momentum $k\equiv \vb{k}$. We assume that $\hat{\theta}_{\vb{k}}$'s satisfy
\begin{equation}
  \label{eq:20}
  \expval{\hat{\theta}_{\vb{k}}\hat{\theta}_{\vb{k'}}^{\ast}}=(2\pi)^{3}\delta^{(3)}(\vb{k}-\vb{k}').
\end{equation}
The function $\theta_B$ describes the evolution of the long wavelength modes $k<k_{\trap}$ for which $\grad{\theta}\approx 0$. We took $k_{\trap}=a_{\trap}H_{\trap}$, but this choice does not affect our results as long as $k_{\trap}$ is much smaller than the relevant modes for the fragmentation for which $k\sim m a_{\trap}$ \cite{Eroncel:2022vjg}. The reason why we have made such a split even in the case of complete fragmentation is the following: When calculating the power spectrum in the case of complete fragmentation, we will use the semi-analytical results for the mode functions that we have derived in \cite{Eroncel:2022vjg}. This method does not work for the modes that are super-horizon during fragmentation, i.e. $k_{\trap}<a_{\trap}H_{\trap}$.

By plugging the expansion \eqref{eq:242} into \eqref{eq:241} we find
\begin{equation}
  \label{eq:202}
  \begin{split}
    \delta=&\frac{\decay^2}{\expval{\rho}}\qty[\frac{\theta_B'\theta_F'}{a^2}\qty(1+2\Phi)+m^2\theta_B\theta_F+\Phi \frac{\theta_B^2}{a^2}]\\
    +&\frac{\decay^2}{\expval{\rho}}\Bigg\lbrace\frac{1}{2a^2}\qty[\theta_F'^2\qty(1+2\Phi)-\expval{\theta_F'^2}]+\frac{1}{2a^2}\qty[\grad{\theta_F}\cdot\grad{\theta_F}\qty(1-2\Phi)-\expval{\grad{\theta_F}\cdot\grad{\theta_F}}]\\
    &\qquad+\frac{m^2}{2}\qty(\theta_F^2-\expval{\theta_F^2})\Bigg\rbrace
  \end{split}
\end{equation}
By approximating $1\pm 2\Phi\approx 1$, the density contrast reduces to
\begin{equation}
  \label{eq:215}
  \begin{split}
    \delta=&\frac{\decay^2}{\expval{\rho}}\qty[\frac{\theta_B'\theta_F'}{a^2}+m^2\theta_B\theta_F+\Phi \frac{\theta_B^2}{a^2}]\\
    +&\frac{\decay^2}{\expval{\rho}}\Bigg\lbrace\frac{1}{2a^2}\qty[\theta_F'^2-\expval{\theta_F'^2}]+\frac{1}{2a^2}\qty[\grad{\theta_F}\cdot\grad{\theta_F}-\expval{\grad{\theta_F}\cdot\grad{\theta_F}}]  +\frac{m^2}{2}\qty(\theta_F^2-\expval{\theta_F^2})\Bigg\rbrace
  \end{split}
\end{equation}
We see that the density contrast has two contributions. The first term contains terms which are linear in fluctuations, and it reduces to the linear density constrast \eqref{eq:199} for a quadratic potential if the fragmentation is incomplete since in this case $\theta_B\rightarrow \Theta$, and $\expval{\rho}\approx \rho_{\Theta}$. Therefore, we refer this contribution to the density contrast as $\delta^{\rm lin}$. Since the second term is quadratic in fluctuations, we refer it as $\delta^{\rm quad}$, and write $\delta=\delta^{\rm lin}+\delta^{\rm quad}$. On the scales $k\sim  \mathcal{O}(1)\times m a_{\trap}$ where the fragmentation is efficient, the quadratic contribution $\delta^{\rm quad}$ dominates over the linear contribution $\delta^{\rm lin}$. However, at scales $k\ll m a_{\trap}$ the quadratic contribution is small as we shall show soon, so the linear contribution dominates.

After this splitting, the power spectrum takes the form
\begin{equation}
  \label{eq:266}
  \dimps_{\delta}(k)\equiv \frac{1}{\mathcal{V}}\expval{\abs{\delta_{k}}^{2}}=\frac{1}{\mathcal{V}}\qty[\expval{\abs{\delta_k^{\rm lin}}^2}+\expval{\abs{\delta_k^{\rm quad}}^2}+2\expval{\Re{\qty(\delta_k^{\rm lin})^{\ast}\delta_k^{\rm quad}}}],
\end{equation}
where $\mathcal{V}$ is a volume factor which is much larger than all scales in the problem.  The last term in this expression does contain terms of the form $\expval{\hat{\Phi}\hat{\theta}\hat{\theta}}$ where $\hat{\Phi}$ and $\hat{\theta}$ are the operators which encode the statistical properties of the curvature perturbations and ALP fluctuations respectively. If the latter are sourced by adiabatic perturbations, then $\hat{\theta} \sim \hat{\Phi}$. Therefore, by assuming that the three-point function for the curvature perturbations vanishes $\expval{\hat{\Phi}\hat{\Phi}\hat{\Phi}}=0$, the last term in \eqref{eq:266} becomes zero. Then the power spectrum becomes
\begin{equation}
  \label{eq:267}
  \dimps_{\delta}(k)=P_{\delta}^{\rm lin}(k)+P_{\delta}^{\rm quad}(k),
\end{equation}
where $P_{\delta}^{\rm lin}(k)=\mathcal{V}^{-1}\expval{\abs{\delta_k^{\rm lin}}^2}$ and $P_{\delta}^{\rm quad}(k)=\mathcal{V}^{-1}\expval{\abs{\delta_k^{\rm quad}}^2}$. The linear contribution is given by
\begin{equation}
  \label{eq:268}
  P_{\delta}^{\rm lin}(k)=\qty(\frac{\decay^2}{\expval{\rho}})^2\qty(\dot{\theta}_{B}\dot{\theta}_{k}+m^2\Theta_B\theta_k+\Phi_k \dot{\theta}_{B}^2)^2,
\end{equation}
and it reduces to the power spectrum in the linear case when $\theta_b\rightarrow \Theta$, and $\expval{\rho}\rightarrow \rho_{\Theta}$. The quadratic contribution is derived in \cite{Enander:2017ogx}, and it reads
\begin{equation}
\boxed{
  \label{eq:206}
  \dimps_{\delta}^{\rm quad}(k)=2(2\pi)^3\dfrac{\int\dd[3]{q}\abs{F(q,q-k)}^2}{\qty(\int\dd[3]{q}F(q,q))^2},}
\end{equation}
where
\begin{equation}
\boxed{
  \label{eq:207}
  F(k,k')=\dot{\theta}_{k}\dot{\theta}_{k'}+\qty(\frac{\vb{k}\cdot\vb{k}'}{a^2}+m^2)\theta_k\theta_{k'},}
\end{equation}
The relevant contribution for the late time observables is the quadratic one. So we discuss it first, and then comment on the linear contribution.

\subsection{Quadratic contribution}
\label{sec:quadr-contr}
We are interested in the late time behavior of the power spectrum, when all modes of interest are non-relativistic. In the non-linear regime (complete fragmentation) we will use the following ansatz to express the mode functions \cite{Eroncel:2022vjg}:
\begin{equation}
  \label{eq:209}
  \theta_k(t)=\sqrt{\dimps_{\theta}(k)}A_k(t)N_k(t)\cos(mt +\varphi_k),
\end{equation}
where $\varphi_k$ is a momentum-dependent phase factor, $\dimps_{\theta}(k)$ is the initial power spectrum before the fragmentation
\begin{equation}
  \label{eq:208}
  P_{\theta}(k)=\frac{2\pi^2}{k^3}\qty(\frac{1}{3})^2A_s\qty(\frac{\dot{\Theta}_i^2}{H_i^2}),
\end{equation}
and the redshift factor $A_k(t)$ is given by
\begin{equation}
  \label{eq:210}
  A_k(t)=\frac{\omega_k^{1/2}(t_i)a_i^{3/2}}{\omega_k^{1/2}(t)a^{3/2}}\approx \frac{(k/a_i)^{1/2}a_i^{3/2}}{m^{1/2}a^{3/2}}.
\end{equation}
By definition, the backreaction will be important in the non-linear regime so the amplification factors $N_{k}$ are calculated via the workflow outlined in Section 3.2 of \cite{Eroncel:2022vjg}. By plugging these terms into the expression for the power spectrum \eqref{eq:206} we obtain
\begin{equation}
  \label{eq:211}
  \dimps_{\delta}^{\rm quad}(k)=\pi \qty(\int\dd{q}N_q^2)^{-2}\int\dd[3]{q}\frac{N_q^2N_{\abs{\vb{q}-\vb{k}}}^2\cos^2\qty(\varphi_q-\varphi_{\abs{\vb{q}-\vb{k}}})}{q^2\abs{\vb{q}-\vb{k}}^2},
\end{equation}
where momentum integrals are over the modes which are amplified during the fragmentation, hence they are finite. The amplification factors $N_k$ are interpreted as their asymptotic limits, so they and the whole power spectrum is independent of time\footnote{This is true only when the potential is approximated by a quadratic, in other words when the self-interactions are neglected.}. By averaging over the phase factors we can replace the cosine square term by $1/2.$ Also, without loss of generality, we can choose $\vb{q}$ to be aligned with the $\vu{z}$-direction. Then $\abs{\vb{k}-\vb{q}}=\sqrt{k^2+q^2-2kq u}$ where $u$ is the cosine of the angle between $\vb{k}$ and $\vb{q}$. After integrating over the azimuthal angle we find
\begin{equation}
  \label{eq:212}
  \dimps_{\delta}^{\rm quad}(k)=\pi^2 \qty(\int\dd{q}N_q^2)^{-2}\int\dd{q}N_q^2\int_{-1}^1\dd{u}\frac{N_{\abs{\vb{q}-\vb{k}}}^2}{\abs{\vb{q}-\vb{k}}^2}.
\end{equation}
It is more convenient to make a change of parameter from $u$ to $p\equiv \abs{\vb{q}-\vb{k}}$. This way we obtain
\begin{equation}
  \label{eq:213}
  P_{\delta}^{\rm quad}(k)=\frac{1}{2}\frac{2\pi^2}{k}\qty(\int\dd{q}N_q^2)^{-2}\int \frac{\dd{q}}{q}N_q^2\int_{\abs{q-k}}^{q+k}\frac{\dd{p}}{p}N_p^2.
\end{equation}
The dimensionless power spectrum is
\vspace{0.2cm}
\begin{mdframed}[backgroundcolor=blue!15]
\begin{equation}
  \label{eq:214}
  \dimlessps_{\delta}^{\rm quad}(k)=\frac{k^3}{2\pi^2}\dimps_{\delta}^{\rm quad}(k)=\frac{1}{2}k^2\qty(\int\dd{q}N_q^2)^{-2}\int \frac{\dd{q}}{q}N_q^2\int_{\abs{q-k}}^{q+k}\frac{\dd{p}}{p}N_p^2.
\end{equation}
\end{mdframed}
\vspace{0.2cm}
Let $\mathfrak{q}\equiv q/(a_{\ast}m)$. Then, one can show that at small momenta
\vspace{0.2cm}
\begin{mdframed}[backgroundcolor=blue!15]
\begin{equation}
  \label{eq:269}
  \dimlessps_{\delta}^{\rm quad}(\kappa\ll 1)\approx \kappa^3\qty(\int\dd{\mathfrak{q}}N_{\mathfrak{q}}^2)^{-2}\int \frac{\dd{\mathfrak{q}}}{\mathfrak{q}^2}N_{\mathfrak{q}}^4\propto \kappa^3,
\end{equation}
\end{mdframed}
\vspace{0.2cm}
where $\kappa\equiv k/(a_{\ast}m)$. Therefore we found that the power spectrum approaches to that of a white noise at low momenta. On the other hand, the large momentum behavior is
\vspace{0.2cm}
\begin{mdframed}[backgroundcolor=blue!15]
\begin{equation}
  \label{eq:322}
  \dimlessps_{\delta}^{\rm quad}(\kappa\gg 1)\approx \kappa \qty(\int \dd{\mathfrak{q}}N_{\mathfrak{q}})^{-1}
\end{equation}
\end{mdframed}
\vspace{0.2cm}
These modes collapse too late to contribute to structure formation  and will thus be not too relevant. We note that this result assumes that for all the modes the initial power spectrum is given by \eqref{eq:208} which is true as long as the mode is super-horizon at the onset of the kination-like scaling of the homogeneous mode. Therefore, the initial conditions are not valid for the modes $k>a_{\rm kin}H_{\rm kin}$. As a result, the high-momenta behavior will also be modified. When and how this modification happens depends on the UV completion, and is irrelevant for the gravitational signatures that we will study in the next section. Thus, we will not comment on the high-momenta behavior any further.

\begin{figure}[tbp]
  \centering
  \includegraphics[width=\textwidth]{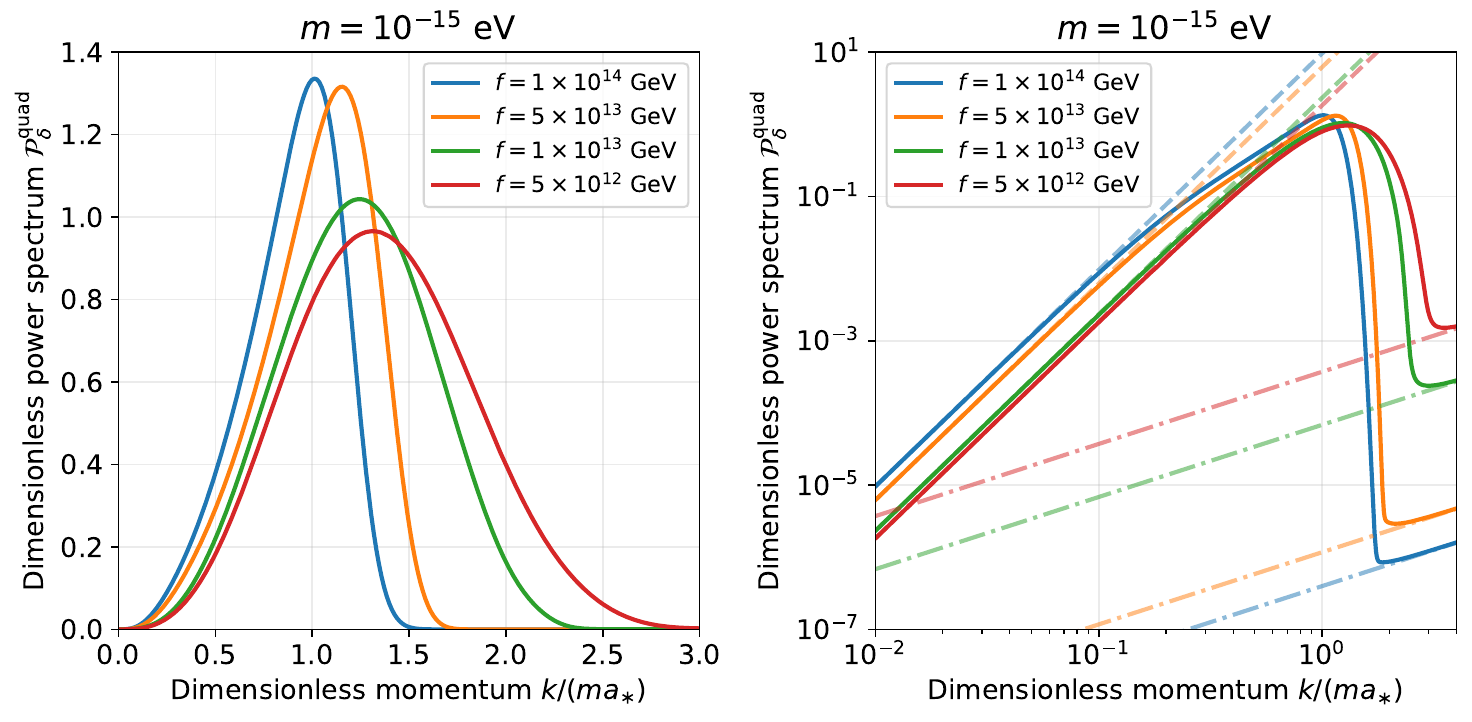}
  \caption{\small \it The dimensionless power spectrum $\dimlessps_{\delta}^{\rm quad}(k)$ for an ALP model with constant mass $m=10^{-15}\;\si{\electronvolt}$ as a function of the dimensionless momentum $\kappa=k/(ma_{\trap})$ for different values of the decay constant. The power spectra is plotted on the linear scale in the left figure, and on the log scale in the right figure. The dashed lines show the low-momenta limit of the power spectrum $\mathcal{P}\propto \kappa^3$ calculated via \eqref{eq:269}, while the dot-dashed lines show the high-momenta limit $\mathcal{P}\propto \kappa$ calculated via \eqref{eq:322}. In all cases, ALP make all of dark matter.}
  \label{fig:dimless-ps-cm-m15}
\end{figure}

We show the plots of the quadratic contribution to the power spectrum in Figure \ref{fig:dimless-ps-cm-m15} for a couple of benchmark points. We assumed a constant ALP mass $m=10^{-15}\,\si{\electronvolt}$, and varied the ALP decay constant $\decay$. For all the curves, the ALP makes all of dark matter. We can see that all power spectra are peaked at $\kappa\sim\mathcal{O}(1)$ while the peak position moves to slightly higher momenta for smaller decay constants. The reason behind this is that the fragmentation starts earlier for smaller decay constants which permits exponential growth of higher modes. However, this also causes amplification of a larger range of modes which results in a more broad spectrum that has a smaller peak value. We can also see that the low/high momentum behavior of all spectra is consistent with the expressions \eqref{eq:269} and \eqref{eq:322} respectively.

\subsection{Linear contribution}
\label{sec:linear-contribution}

For fragmented axions to explain all of dark matter, the resulting matter power spectrum should be consistent with the observations around the CMB scales $k\sim 0.1\,\rm{Mpc}^{-1}$. The relevant modes for fragmentation are those with $k\sim k_{\rm peak}\sim a_{\trap}m$. To express this in dimensionless units, we use the following relation derived in \cite{Eroncel:2022vjg}:
\begin{equation}
  \label{eq:21}
  \frac{m_{\trap}}{m_{0}}\qty(\frac{a_{\trap}}{a_{0}})^{3}=\frac{\pi}{8}\frac{\rho_{\Theta,0}}{\Lambda_{b,0}^{4}},
\end{equation}
where $m_{\trap}$ is the ALP mass at trapping, $\Lambda_{b,0}=\sqrt{m_{0}f}$ is the today's value of the barrier height, and $\rho_{\Theta,0}$ is the ALP energy density today without taking fragmentation into account\footnote{The fragmentation might slightly modify the relic density today. We made a semi-analytical estimate in \cite{Eroncel:2022vjg}, and found that this modification is not significant.}.  From (\ref{eq:21}) and assuming a constant mass we get
\begin{equation}
  \label{eq:270}
  k_{\rm peak}\sim 10^5\,\mathrm{Mpc}^{-1}\times \qty(\frac{m}{10^{-15}\,\si{\electronvolt}})^{1/3}\qty(\frac{\decay}{10^{14}\,\si{\giga\electronvolt}})^{-2/3}.
\end{equation}
Therefore, the fragmentation scales are much shorter compared to the CMB ones\footnote{The fragmentation scales can be close to the CMB scales at the very low ALP mass and very high ALP decay constant region of the parameter space. However, this region is excluded by the BBN constraints \cite{Eroncel:2022vjg}.}. As a result, the quadratic contribution will be very suppressed due to the $k^3$ decay at low momenta. Thus, the matter power spectrum at CMB scales will be determined mainly by the linear contribution.

We cannot evaluate the linear contribution since $\theta_{B}$ is not known. However, from the discussion of the linear case in Section \ref{sec:axion-dens-pert}, we expect that the modes which are super-horizon during fragmentation will lock on the CDM prediction once they enter the horizon. Therefore, we approximate the linear contribution of the low momentum modes by the power spectrum of CDM:
\begin{equation}
  \label{eq:271}
  \dimps_{\delta}^{\rm lin}(k;z)\approx \dimps_{\delta,\rm CDM}(k;z).
\end{equation}
This approximation has been used in \cite{Lee:2020wfn} to study the gravitational signatures of axion miniclusters when Peccei-Quinn breaking happens after inflation, and also in \cite{Graham:2015rva,Alonso-Alvarez:2018tus} where dark matter is generated from inflationary perturbations.

\begin{figure}[tbp]
  \centering
  \includegraphics[width=\textwidth]{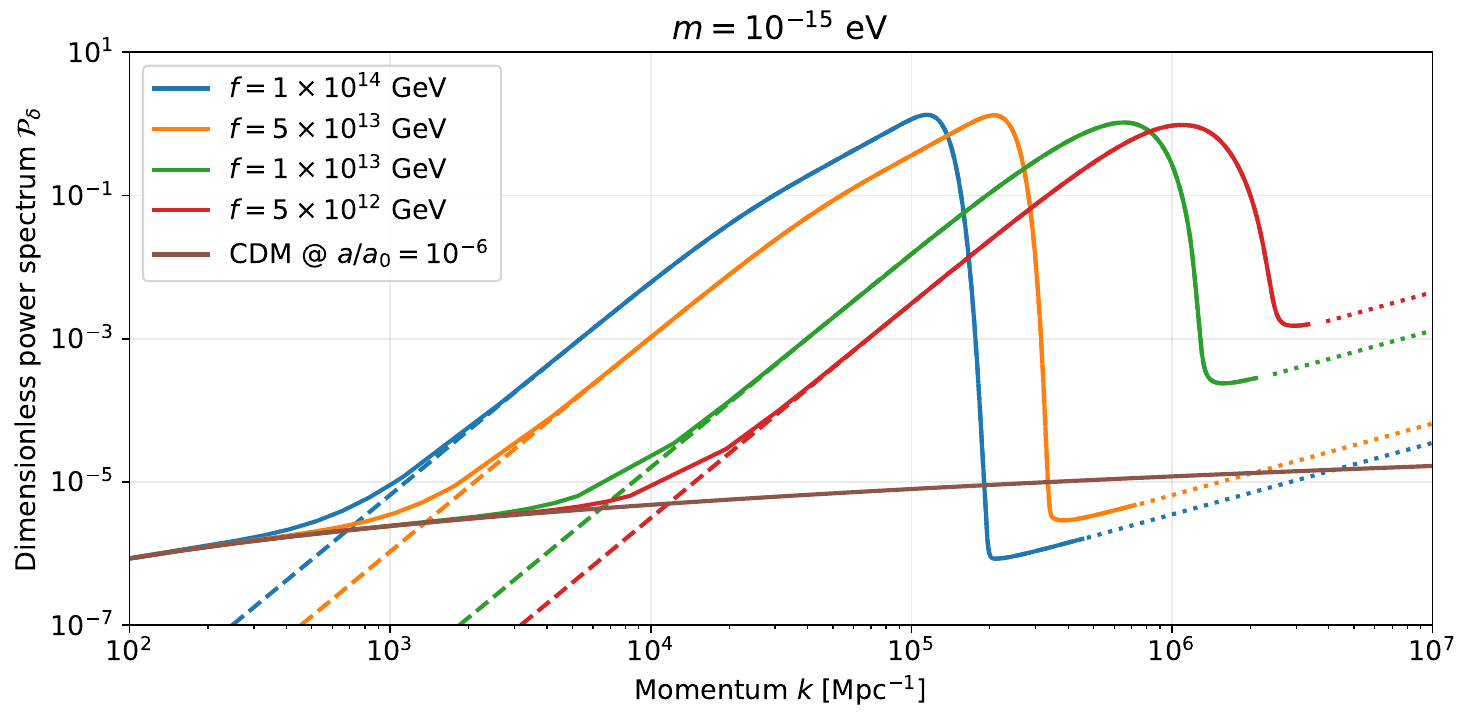}
  \caption{\small \it The combined dimensionless power spectrum $\dimlessps_{\delta}=\dimlessps_{\delta,\rm CDM}+\dimlessps_{\delta}^{\rm quad}$ for an ALP model with constant mass $m=10^{-15}\,\si{\electronvolt}$ as a function of the comoving momentum $k$ for different values of the decay constant. The dashed lines show the low-momenta behavior without the CDM contribution, while the dotted lines show the extrapolation to high momenta using \eqref{eq:322}. Note that the CDM contribution is added only to the low-momenta part of the power spectrum, and we assumed $a/a_0=10^{-6}$ when calculating the CDM spectrum which is much later than the trapping scale factor $a_{\trap}$ for all benchmark points. }
  \label{fig:dimless-ps-m15-cm-combined}
\end{figure}

We show the full power spectrum including the CDM contribution in Figure \ref{fig:dimless-ps-m15-cm-combined}. The benchmark points are the same in Figure \ref{fig:dimless-ps-cm-m15}. Even though the quadratic contribution $\dimlessps_{\delta}^{\rm quad}$ is time-independent, the CDM contribution has a logarithmic time-dependence during the radiation era. We have evaluated the CDM contribution at $a/a_0=10^{-6}$ which is much later than the trapping scale factor $a_{\trap}$ for all benchmark points. The high momenta behaviors are shown by the dotted lines since the position of their cutoff  is model-dependent. We can see that the spectra is peaked at higher comoving momenta for smaller decay constants. This is because the peak momentum scales as $k_{\rm peak}\sim m a_{\ast}$, and the trapping happens later for smaller decay constants. This shift to larger momenta will have a significant effect on the gravitational signatures as we shall see in the next section.


In order to study the gravitational signatures we need to evolve the power spectrum $\dimlessps^{\rm quad}$ until today. For this we will use the linearized evolution equation \eqref{eq:237} even though the fluctuations are clearly non-linear. The reasoning behind this approximation is that the linearized analysis is all we need to have an estimate of the gravitational signatures which becomes more clear when we discuss the Press-Schechter formalism in Section \ref{sec:press-schecht-form}.

\subsection{Comparison with the post-inflationary scenario}
\label{sec:comparison-with-post}

In our discussion so far, we have always considered the \emph{pre-inflationary} scenario where the spontaneous symmetry breaking that creates ALPs as Nambu-Goldstone bosons happens before inflation. In the absence of an initial kinetic energy, the ALP angle $\theta$ takes random values $\theta\in \closedopeninterval{-\pi}{\pi}$ in different patches of the universe after  symmetry breaking. If this process happens before inflation,  each patch is inflated to huge sizes so that the initial angle is the same everywhere in the observable universe which makes the ALP field homogeneous at early times.

However, the symmetry breaking can also happen after inflation. This is the so-called \emph{post-inflationary} scenario. In this case the observable universe will contain many patches all having a different initial angle. This leads to the emergence of topological defects like cosmic strings and domain walls, also known as the string-wall network \cite{Kibble_1976,PhysRevD.26.435,PhysRevLett.48.1867}. As a result, the ALP field becomes very inhomogeneous in contrast to the pre-inflationary scenario.

It is interesting to compare our predictions for the power spectrum with the one predicted in the post-inflationary scenario. A simplified linear analysis of the post-inflationary scenario for the QCD axion is given in \cite{Enander:2017ogx} by assuming a quadratic potential. Full non-linear analysis of ALP models with different temperature dependencies including the constant mass scenario has been recently presented in \cite{OHare:2021zrq}\footnote{An important variable in the simulations of the post-inflationary scenario is the string tension $\kappa\equiv\ln(m_{\phi}/H)$ where $m_{\phi}$ is the mass of the radial mode of the Peccei-Quinn field. Current computational resources permit simulations up to $\kappa\approx8$, however realistic values are $\kappa\sim70$, for example in the case of QCD axion. How to make an extrapolation to larger tension is still an open problem. In particular, \cite{OHare:2021zrq} finds the string density parameter as $\xi\simeq1.2$. Even though this value is in agreement with some extrapolations in the literature \cite{Hindmarsh:2019csc,Hindmarsh:2021vih}, \cite{Gorghetto:2020qws} claims a much larger value $\xi\sim15$ due to the logorithmic growth of $\xi$ with $\kappa$. In this paper we are using the results of \cite{OHare:2021zrq} since it is the only work that simulates the late-time power spectrum for a constant mass ALP.}. In this section, first we calculate the power spectrum of a constant mass ALP model by adapting the method used in \cite{Enander:2017ogx}, and then make a comparison with the numerical results of \cite{OHare:2021zrq}. Finally we compare the predictions in the post-inflationary scenario with the predictions of KMM and SMM/LMM.

\subsubsection*{Linearized analysis}
Following \cite{Enander:2017ogx}, we separate the ALP mode functions as
\begin{equation}
  \label{eq:345}
  \theta_k(t)=\hat{\theta}_{k}f_k(t),
\end{equation}
where the statistical properties and the initial conditions are encoded in $\hat{\theta}_k$, while the functions $f_k(t)$ contain the time evolutions, which are normalized such that $f_k(t_i)=1$. For a quadratic potential, the equations of motion for $f_k$'s are
\begin{equation}
  \label{eq:346}
  \ddot{f}_k+3H\dot{f}_k+\qty(\frac{k^2}{a^2}+m^2)f_k=0,
\end{equation}
where $m$ is the ALP mass which is assumed to be constant. The oscillations start around $t=t_1=(2H_1)^{-1}$, where in the constant mass case $H_1$ is defined by \cite{Blinov:2019rhb,OHare:2021zrq}
\begin{equation}
  \label{eq:347}
  m=\frac{8}{5}H_1.
\end{equation}
We also define $\mathcal{H}_1=a_1H_1=L_1^{-1}$ where $L_1$ is the comoving horizon at $t_1$ \cite{OHare:2021zrq}. Then, in terms of the dimensionless time $t_m=mt$, \eqref{eq:346} takes the form
\begin{equation}
  \label{eq:348}
  f_k''(t_m)+\frac{3}{2t_m}f_k'(t_m)+\qty[\frac{5}{16}\frac{K^2}{t_m}+1]f_k(t_m)=0,
\end{equation}
where we have also defined $K\equiv k L_1$.

The evolution of the mode functions after the onset of oscillations can well be described by the WKB approximation:
\begin{equation}
  \label{eq:349}
  f_k(t_m)=\frac{c_k}{t_m^{3/4}\sqrt{\tilde{\omega}_k(t_m)}}\cos\qty(\int_{t_m^{\rm WKB}}^{t_m}\dd{t_m'\tilde{\omega}_k(t_m')}+\varphi_k),\quad \tilde{\omega}_k(t_m)\equiv\sqrt{\frac{5}{16}\frac{K^2}{t_m}+1},
\end{equation}
where $c_k$ and $\varphi_k$ are numerical factors which we have determined by numerically solving the equation of motion \eqref{eq:348} until $t_m^{\rm WKB} \gg 1$. We have verified that by choosing $t_m^{\rm WKB}=50$, the WKB approximation reproduces the numerical solution very well.

The power spectrum is given by \cite{Enander:2017ogx}
\begin{equation}
  \label{eq:356}
  \dimps(q)=2\qty(2\pi)^3 \frac{\int \dd[3]{k}\dimps_{\theta}(k)\dimps_{\theta}(\abs{\vb{k}-\vb{q}})F(k,k-q)^2}{\qty[\int\dd[3]{k}\dimps_{\theta}(k)F(k,k)]^2},
\end{equation}
where
\begin{equation}
  \label{eq:357}
  F(k,k')=\dot{f}_k(t)\dot{f}_{k'}(t)+\qty(\frac{\vb{k}\cdot\vb{k}'}{a^2}+m^2)f_k(t)f_{k'}(t),
\end{equation}
and $\dimps_{\theta}(k)$ is the field power spectrum at some initial temperature $T_i$ defined such that
\begin{equation}
  \label{eq:358}
  \expval{\hat{\theta}_k\hat{\theta}_{k'}^{\ast}}=\qty(2\pi)^3\delta^{(3)}(\vb{k}-\vb{k}')P_{\theta}(k).
\end{equation}
A reasonable choice for $\dimps_{\theta}(k)$ is \cite{Enander:2017ogx,OHare:2021zrq}
\begin{equation}
\boxed{
  \label{eq:359}
  P_{\theta}(k)=\frac{8\pi^4}{3\sqrt{\pi}k_{\rm cr}^3}\exp(-\frac{k^2}{k_{\rm cr}^2}),}
\end{equation}
where the cutoff is given by $k_{\rm cr}=a_iH_i$, and the normalization is fixed such that $\expval{\theta(\vb{x})^2}=\pi^2/3$. Note that at late times when all the relevant modes become non-relativistic we have $\tilde{\omega}_k\approx 1$, and the power spectrum becomes independent of time\footnote{As we have stated in Section \ref{sec:quadr-contr}, this is true only for the linearized analysis where the potential is approximated by a quadratic, and therefore the self-interactions are neglected. }.

Last thing we need to specify is the initial time. This needs to be before the onset of oscillations but not too early since at those times the string-wall network plays an important role \cite{Vaquero:2018tib,OHare:2021zrq}. Again following \cite{Enander:2017ogx} we have chosen $T_i=3T_1$ as the initial temperature where $T_1$ is defined by $H(T_1)=H_1$. This corresponds to an initial time of $t_{m,i}=4/45$ and a momentum cutoff of $k_{\rm cr}L_1=9$. After this point the calculation is identical to the one we have presented in Section \ref{sec:quadr-contr} so we do not repeat it here.

\subsubsection*{Results and comparison with lattice simulations and other mechanisms}

\begin{figure}[tbp]
  \centering
  \includegraphics[width=\textwidth]{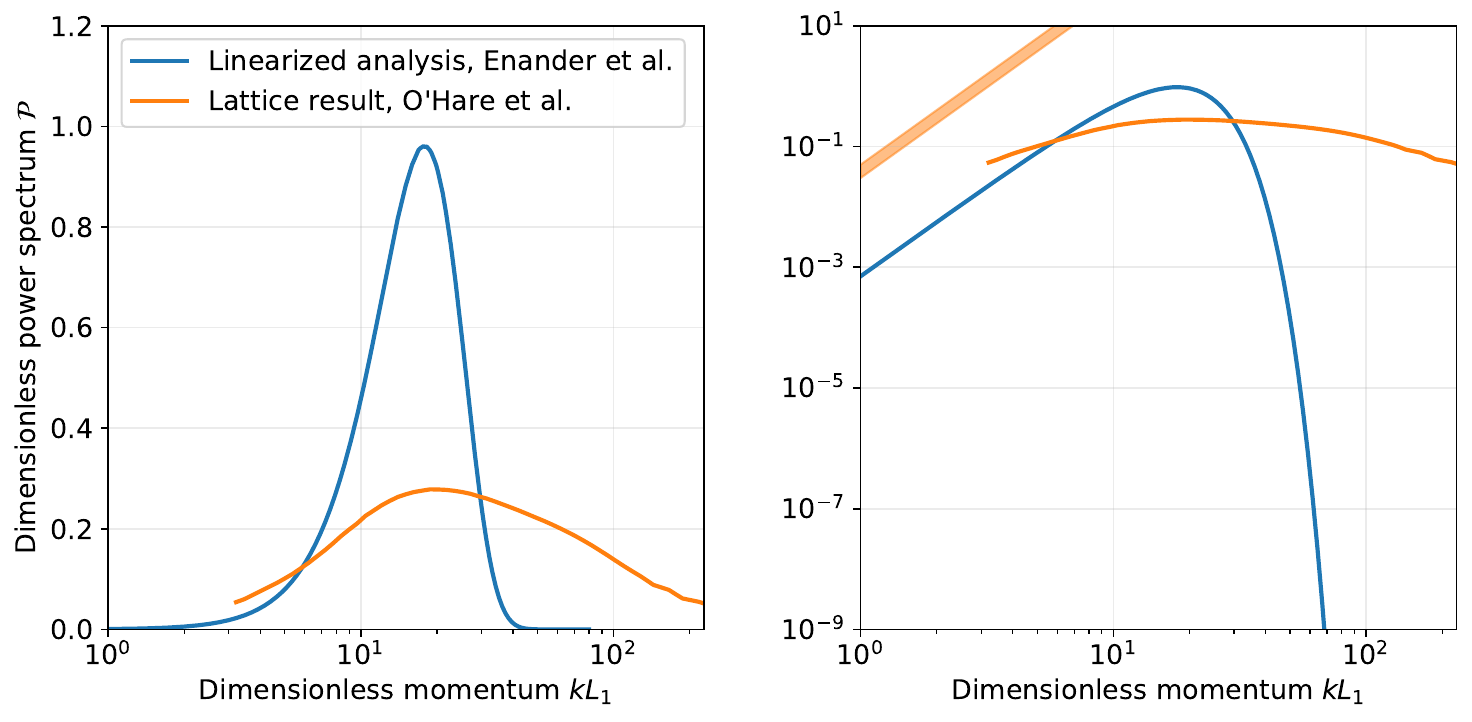}
  \caption{\small \it  Late-time dimensionless power spectrum of the ALP field in the post-inflationary scenario with constant mass obtained via the linearized analysis of \cite{Enander:2017ogx} (blue lines). We compare this spectrum with the lattice result of \cite{OHare:2021zrq} (orange lines). The left and right plots show the same spectra except for the scalings of the vertical axis. The orange band on the right plot show the low-momenta behavior reported in \cite{OHare:2021zrq} such that $\dimlessps(kL_1\ll 1)\approx C\qty(k L_1)^3$ with $C\simeq 0.03\text{ -- }0.05$. The lattice result is obtained at $t_m=2\times 10^3$ in our units, however we have checked that the linearized result does not change significantly after $t_m\sim 10^3$. }
  \label{fig:postinf-linear-lattice-compare-simplified}
\end{figure}

We now present our results and make a comparison with the lattice result obtained in \cite{OHare:2021zrq} in Figure \ref{fig:postinf-linear-lattice-compare-simplified}. The blue lines show the spectra obtained via the linearized analysis, while the orange lines show the lattice results. The orange band on the right plot is given by $\dimlessps(kL_1\ll 1)\approx C\qty(k L_1)^3$ with $C\simeq 0.03\text{ -- }0.05$, which is the low-momentum behavior reported in \cite{OHare:2021zrq}. We can see that the non-linear effects shifts the spectrum to larger momentum values and make the spectrum broader. As a result, the peak value gets reduced by a factor of $\approx 3.5$.

\begin{figure}[tbp]
  \centering
  \begin{subfigure}[b]{0.49\textwidth}
    \includegraphics[width=\textwidth]{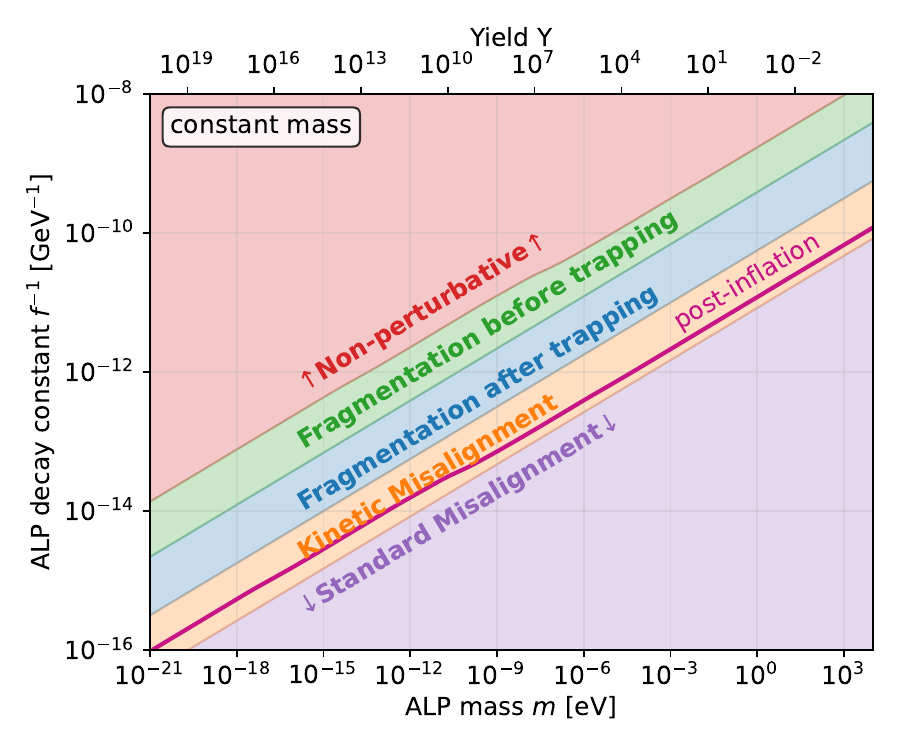}
  \end{subfigure}
  \begin{subfigure}[b]{0.49\textwidth}
    \includegraphics[width=\textwidth]{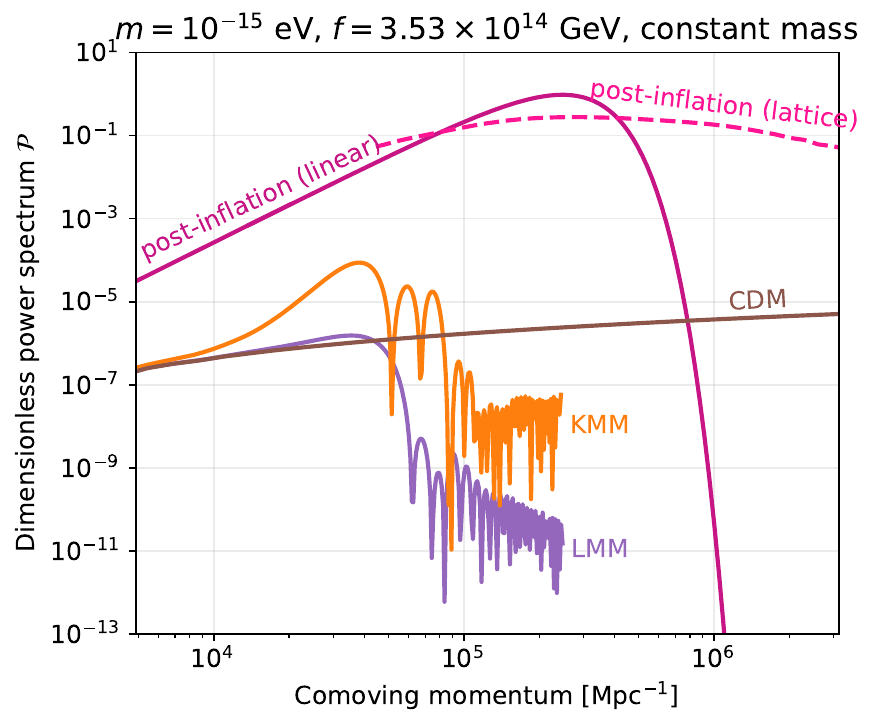}
  \end{subfigure}
  \caption{\small \it\textbf{Left plot: }The line on the ALP parameter space where the correct relic density is obtained in the post-inflationary scenario as determined by \eqref{eq:362}. This line agrees perfectly with the one obtained in the Standard/Large Misalignment Mechanism by setting the initial angle to $\Theta_i\approx 3.06$. \textbf{Right plot: }The comparison of the power spectra in the post-inflationary scenario obtained via the linearized analysis of \cite{Enander:2017ogx}, and the lattice calculation of \cite{OHare:2021zrq} with the spectra obtained in the Kinetic and Large Misalignment Mechanism for the same benchmark point. }
  \label{fig:comparisons}
\end{figure}

Before concluding this section, it would be instructive to compare the power spectra of all the dark matter production mechanisms that we have seen so far on a single plot. For a given ALP mass, the decay constant is fixed in the post-inflationary scenario. The relic density for this scenario is given by \cite{OHare:2021zrq}
\begin{equation}
\boxed{
  \label{eq:362}
  h^2\Omega_{\theta}\simeq 0.019\qty(\frac{g_{\rho}(T_1)}{70})^{3/4}\qty(\frac{g_s(T_{1})}{70})^{-1}\qty(\frac{m}{\si{\micro\electronvolt}})^{1/2}\qty(\frac{f}{10^{12}\,\si{\giga\electronvolt}})^2.}
\end{equation}
One can then solve for the ALP decay constant $\decay$ to get the correct relic density. We show this line on the ALP parameter space in the left plot of Figure \ref{fig:comparisons} with the label ``post-inflation''. This line matches perfectly with the line one gets in the Standard/Large Misalignment Mechanism by setting the initial angle to $\Theta_i\approx 3.06$. However, the power spectrum of the fluctuations is completely different due to the highly inhomogeneous initial conditions. We demonstrate this on the right plot of Figure \ref{fig:comparisons}. In this plot, we set the ALP mass to $m=10^{-15}\,\si{\electronvolt}$, and solve \eqref{eq:362} to get the correct relic density and found $\decay\approx 3.53\times 10^{14}\,\si{\giga\electronvolt}$. We then calculated the power spectra predicted by the Kinetic and the Standard/Large Misalignment Mechanisms and make the comparison with the post-inflationary spectra obtained both with the linearized analysis (solid line), and the lattice results (dashed line). We see that the spectrum is greatly enhanced and broaden in the post-inflationary scenario.

\section{Gravitational collapse of fluctuations and the halo spectrum}
\label{sec:grav-coll-fluct}

Once the matter fluctuations become sufficiently dense, they decouple from the ambient Hubble flow, and form gravitationally bound structures known as \emph{halos}. This process is called \emph{gravitational collapse}. Studying this process precisely is quite difficult, and requires N-body simulations. However, qualitative results can be derived by exploiting the approximate spherical symmetry. This will be our goal in this section. We start by deriving the critical density beyond which a density fluctuation will experience gravitational collapse in Section \ref{sec:crit-dens-coll}. Then we introduce the Press-Schechter formalism in Section \ref{sec:press-schecht-form} which we shall use in Section \ref{sec:mass-distr-halos:} to compute the halo mass function, the number density of halos per logarithmic mass bin. We conclude the section by calculating the halo spectrum in Section \ref{sec:halo-spectrum} that contains information about the mass-density relation of the halos at a given benchmark.

\subsection{Critical density for collapse}
\label{sec:crit-dens-coll}

We consider a universe which can either be matter or radiation dominated, and take a spherical shell of \emph{physical radius} $r$. The time evolution of this radius  obeys \cite{Kolb:1994fi}
\begin{equation}
  \label{eq:203}
  \frac{\ddot{r}}{r}=-\frac{4\pi G}{3}(\rho+3p)=-\frac{4\pi G}{3}\qty(\rho_m+\rho_r+3p_r)=-\frac{4\pi G}{3}\rho_m-\frac{8\pi G}{3}\rho_r,
\end{equation}
where $\rho_m$ is the matter density\footnote{Since we are neglecting the baryons in this work, the matter density is the same as the ALP density, $\rho_m=\rho_{\theta}$. } including the over-densities, and $\rho_r$ is the radiation density which we assume to be homogeneous everywhere. The radiation pressure is $p_r=\rho_r/3$, and we assumed that the matter is pressureless $p_m=0$. The latter assumption is accurate for CDM, however it breaks down for ALPs at small scales due to the quantum pressure discussed in the previous section. We will comment later on the consequences. For now, we continue our discussion by assuming $p_m=0$.

Let $M$ denote the mass inside the spherical region, i.e.
\begin{equation}
  \label{eq:204}
  M=\frac{4\pi}{3}\rho_mr^3=\frac{4\pi}{3}\qty(\overline{\rho}_m+\Delta \rho_m)r^3=\frac{4\pi}{3}\overline{\rho}_m\qty(1+\delta)r^3,
\end{equation}
where $\overline{\rho}_m$ is the background matter density, and $\delta\equiv \Delta \rho_m/\overline{\rho}_m$ is the size of the fluctuation. We will assume that this mass remains constant during the gravitational collapse. In terms of $M$, the equation of motion for $r$ becomes
\begin{equation}
  \label{eq:205}
  \ddot{r}=-\frac{8\pi G}{3}\rho_r r - \frac{MG}{r^2}.
\end{equation}
Let us parametrize $r(t)$ by
\begin{equation}
  \label{eq:243}
  r(t)=a(t)b_R(t)R,
\end{equation}
where $R$ is the \emph{comoving} radial coordinate, and $b_R$ denotes the deviation from the ambient Hubble flow. The $R$-subscript indicates that the evolution is valid for the spherical shell with comoving label $R$. We will drop the subscript below to simplify the notation.

Next, we will switch the conformal time, and make a change of variables to $y\equiv a/a_{\rm eq}$. After some algebra, one can obtain a differential equation for the evolution of $b(y)$:  \cite{Kolb:1994fi}
\begin{equation}
  \label{eq:244}
  y(1+y)\dv[2]{b}{y}+\qty(1+\frac{3}{2}y)\dv{b}{y}+\frac{1}{2}\qty(\frac{1+\delta_i}{b^2}-b)=0,
\end{equation}
where $\delta_i$ is the \emph{initial} over-density in the region $\delta_i=\delta \rho_m(y_i)/\overline{\rho}_m(y_i)$ set at some early time when $b(y_i)=1$. By deriving this result, we have used
\begin{equation}
  \label{eq:245}
  M=\frac{4\pi}{3}\rho_{m,i}r_i^3=\frac{4\pi}{3}\overline{\rho}_{m,i}a_i^3\qty(1+\delta_i)R^3=\frac{4\pi}{3}\frac{\overline{\rho}_{\rm eq}}{2}a_{\rm eq}^3\qty(1+\delta_i)R^3,
\end{equation}
where $\overline{\rho}_{\rm eq}=2\overline{\rho}_{m,\rm eq}=2\overline{\rho}_{r,\rm eq}$ is the \emph{total} average energy density at the matter-radiation equality.

\begin{figure}[tbp]
  \centering
  \includegraphics[width=\textwidth]{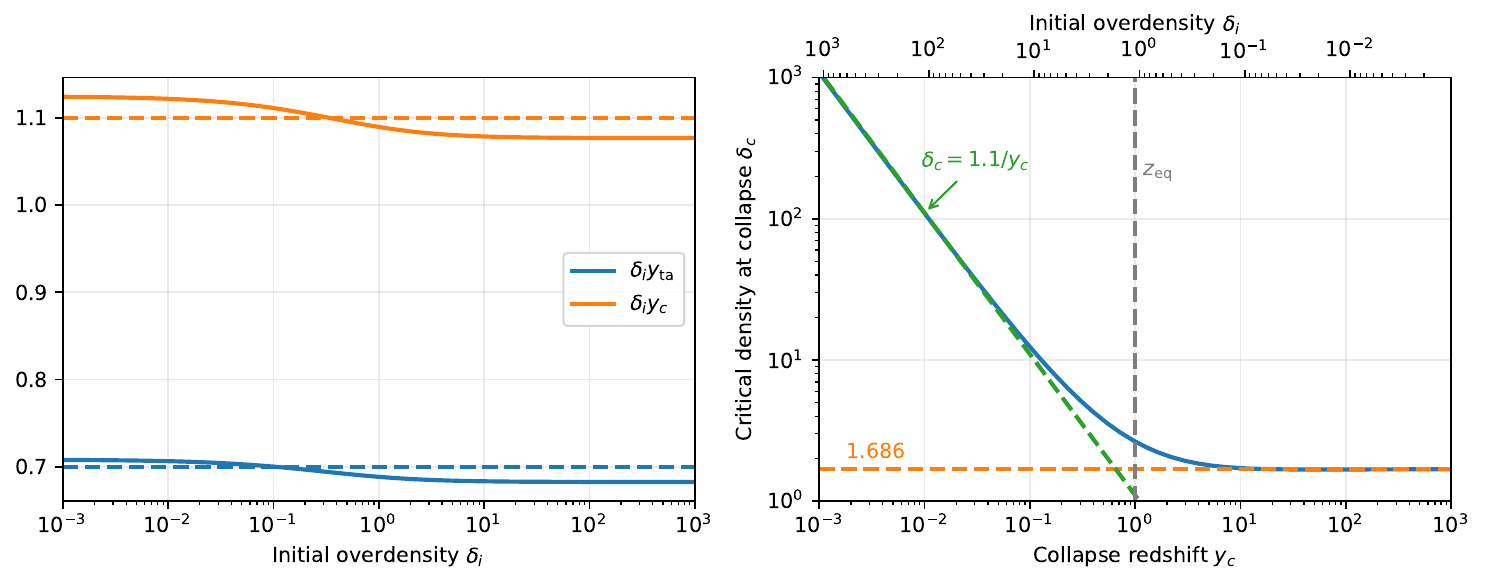}
  \caption{\small \it \textbf{Left: }Plot of $\delta_iy_{\rm ta}$ and $\delta_iy_c$ for various values of the initial density $\delta_i$. The solid lines denote the numerical results, while dashed lines are the mean values for the initial densities considered in the plot. \textbf{Right: }The critical density $\delta_c(x)$ predicted by the linear theory calculated numerically via \eqref{eq:252} together with approximate expressions in the deep matter (dashed orange line), and deep radiation eta (dashed green line). }
  \label{fig:matter-radiation-collapse}
\end{figure}

For a given value of the initial over-density $\delta_i$, one can solve \eqref{eq:244} numerically with the initial conditions $b(y_i)=1$, and $b'(y_i)=0$ where $y_i\ll 1$. Initially, the radius will expand with the Hubble flow. Then it will start to deviate, and at $y_{\rm ta}$ it \emph{turns around}, i.e. the expansion will stop, and the radius will start shrinking. This happens when $\dot{r}=0$, or $b'(y_{\rm ta})+b(y_{\rm ta})/y_{\rm ta}=0$. Numerical solution shows that the turnaround happens at \cite{Enander:2017ogx}
\begin{equation}
  \label{eq:248}
  y_{\rm ta}\approx \frac{0.7}{\delta_i}.
\end{equation}
Without a pressure term, the collapse will continue without a restriction, and the overdensity will eventually collapse to a single point at $y=y_c$ where the overdensity $\delta$ becomes infinite. Again, by numerical solution we observed that
\begin{equation}
  \label{eq:249}
  y_c\approx \frac{1.1}{\delta_i}.
\end{equation}
Numerical calculation of $\delta_iy_{\rm ta}$, and $\delta_iy_c$ for a bunch of initial overdensities is shown on the left plot of Figure \ref{fig:matter-radiation-collapse}.

After having calculated the time of collapse using the \emph{non-linear} theory, we can ask the following question: What would be the prediction of the \emph{linear} theory if we had extrapolated it until the collapse $y_c$?~\cite{Ellis:2020gtq} This is the necessary ingredient for using the Press-Schechter formalism which we will review in the next section. The conservation of the mass $M$ implies
\begin{equation}
  \label{eq:250}
  \qty[1+\delta(y)]b^3(y)=1+\delta_i.
\end{equation}
In the linear approximation we define $\delta(y)=\delta_i + \Delta \delta(y)$, and evaluate all expressions at linear order in $\Delta \delta(y)$. Then \eqref{eq:250} implies $b(y)=1-\Delta \delta(y)/3(1+\delta_i)$, and \eqref{eq:244} becomes
\begin{equation}
  \label{eq:246}
  y(1+y)\dv[2]{\Delta\delta}{y}+\qty(1+\frac{3}{2}y)\dv{\Delta\delta}{y}-\frac{3}{2}\qty[\frac{2}{3}+\frac{1}{3(1+\delta_i)}]\Delta\delta=-\frac{3}{2}\delta_i.
\end{equation}
Note that in the $\delta_i\ll 1$ limit, this equation reduces to the \emph{Meszaros} equation as expected. The solution of this equation with the initial condition $\Delta\delta(y\rightarrow 0)=0$ is
\begin{equation}
  \label{eq:247}
  \Delta\delta(y)=\frac{3\delta_i(1+\delta_i)}{3+2\delta_i}\qty[_2F_1\qty(\delta_-,\delta_+,1,-y)-1],\quad \delta_{\pm}\equiv \frac{1}{4}\pm\frac{\sqrt{17\delta_i^2+42\delta_i+25}}{4+4\delta_i},
\end{equation}
where $_2F_1$ is the hypergeometric function. So in the linear theory, $\delta(y)$ is given by
\begin{equation}
  \label{eq:251}
  \delta^{\rm lin}(y)=\delta_i\qty[\frac{3+3\delta_i}{3+2\delta_i}\,_2F_1\qty(\delta_-,\delta_+,1,-y)-\frac{\delta_i}{3+2\delta_i}].
\end{equation}
Since the non-linear theory predicts that the initial fluctuation $\delta_i$ will collapse at $y_c$, we can calculate the prediction of the linear theory for the overdensity at collapse by replacing $\delta_i$ with the corresponding $y_c$, and setting $y=y_c$ in \eqref{eq:251}. So the prediction of the linear theory at time of the collapse is
\vspace{0.2cm}
\begin{mdframed}[backgroundcolor=blue!15]
\begin{equation}
  \label{eq:252}
  \delta_c(y_c)=\eval{\delta_i\qty[\frac{3+3\delta_i}{3+2\delta_i}\,_2F_1\qty(\delta_-,\delta_+,1,-y_c)-\frac{\delta_i}{3+2\delta_i}]}_{\delta_i\approx 1.1/y_c}.
\end{equation}
\end{mdframed}
\vspace{0.2cm}
The quantity $\delta_c$ is called the \emph{critical density at collapse}. Small fluctuations $\delta_i\ll 1$ collapse during the deep matter era where $y_c\gg 1$. In this case  we recover the well-known result:
\begin{equation}
  \label{eq:253}
  \delta_c(y_c\gg 1)\approx \frac{1.1}{y_c}\qty(1+\frac{3}{2}y_c)\approx \frac{3}{20}\qty(12\pi)^{2/3}\approx 1.686,\quad \delta_i\ll 1.
\end{equation}
On the other hand, for large fluctuations we can consider the $y_c\ll 1$ limit to approximate the hypergeometric function at $1+y_c$. Then we find
\begin{equation}
  \label{eq:254}
  \delta_c(y_c\ll 1)\approx \delta_i\qty(1+\frac{3}{2}y_c)\approx \frac{1.1}{y_c}\qty(1+\frac{3}{2}y_c)\approx \frac{1.1}{y_c},\quad \delta_i\gg 1.
\end{equation}
Suprisingly, the approximate expression for small and large fluctuations are the same. So for all regimes, one can use $\delta_c\approx (1.1/x_c)(1+3 x_c/2)$ as the expression for the critical density at collapse \cite{Ellis:2020gtq}. The plot of the critical density $\delta_c$ as a function of $y_c$ is plotted on the right plot of Figure \ref{fig:matter-radiation-collapse}.

\subsection{Press-Schechter Formalism}
\label{sec:press-schecht-form}

The formation of the dark matter halos can be studied analytically via the Press-Schechter (PS) formalism \cite{Press:1973iz}. It relies on the following postulate: The fraction of matter $\mathscr{F}$ which is inside collapsed structures of comoving size larger than $R$ at any given scale factor $a$ is the same as the probability $\mathscr{P}$ of finding an overdensity $\delta_R(\vb{x},a)>\delta_c(a)$ where $\delta_R(\vb{x},a)$ is the overdensity \emph{smoothed} at the scale $R$, and $\delta_c(a)$ is the critical density for collapse at scale factor $a$. If the initial fluctuations prior to collapse are small, then there is a one-to-one correspondence between the mass $M$ of the halo and the comoving radius $R$. Then the PS postulate can be expressed as
\begin{equation}
  \label{eq:256}
  \mathscr{F}(>M;a)=\mathscr{P}(\delta_R(a)>\delta_c(a)).
\end{equation}
The relation between $M$ and $R$ depends on the choice of the \emph{window function} $W_{R}$ which is used to smooth out the density contrast. For any window function, the smoothed density contrast is given by
\begin{equation}
  \label{eq:257}
  \delta_R(\vb{x},a)=\int\dd[3]{\vb{x}'}W_R(\abs{\vb{x}-\vb{x'}})\delta(\vb{x}',a),\quad\int\dd[3]{\vb{x}}W_R(\vb{x})=1,
\end{equation}
where $\delta$ is the energy density contrast. By a straightforward calculation we can obtain the variance of the smoothed density contrast as
\begin{equation}
  \label{eq:258}
  \sigma_R^2(a)=\expval{\abs{\delta_R(\vb{x},a)}^2}=\int_0^{\infty}\frac{\dd{k}}{k}\abs{\tilde{W}_R(k)}^2\mathcal{P}_{\delta}(k;a),
\end{equation}
where $\tilde{W}$ is the Fourier transform of the window function and $\mathcal{P}_{\delta}$ is the dimensionless power spectrum corresponding to the density contrast.

The most common and natural choice for the window function is the \emph{spherical top-hat (STH)} window function which is defined by
\begin{equation}
  \label{eq:259}
  W_{\rm STH}(\vb{x})=\qty(\frac{4\pi}{3}R^3)^{-1}\times
  \begin{cases}
    1,&\vb{x}\leq R\\
    0,&\vb{x}>R
  \end{cases}.
\end{equation}
Its Fourier transform is
\begin{equation}
  \label{eq:261}
  \tilde{W}_{\rm STH}(kR)=\frac{3}{(kR)^3}\qty[\sin(kR)-kR\cos(kR)].
\end{equation}
With this window function, the smoothed density contrast $\delta_R(\vb{x})$ is just the integral of the density contrast over a spherical region of comoving size $R$ centered at $\vb{x}$. Then the mass enclosed in the region is unambiguously given by
\begin{equation}
  \label{eq:260}
  M(R)=\frac{4\pi}{3}\overline{\rho}a^3R^3\approx\frac{4\pi}{3}\overline{\rho}_{m,0}a_0^3R^3,
\end{equation}
where the approximation holds for small fluctuations. We will use this relation even if the fluctuations are large\footnote{The Press-Schechter Formalism relies on the assumption that there is a one-to-one correspondence between the mass $M$ of the halo and the comoving radius $R$ which breaks down if the fluctuations are large. A modified version of the formalism which does not rely on this one-to-one correspondence is proposed in \cite{Enander:2017ogx}. However, we found that this formalism is not very useful to construct the halo spectrum, which is our main goal in this section. Therefore, we will present our estimates using the original Press-Schechter formalism.}.

Note that in order to calculate the variance via \eqref{eq:258}, one needs to perform an integration over all momentum values, not just the ones we have simulated in the previous sections. The contribution of higher modes can be neglected since the power spectrum is suppressed, however the contribution of the lower momentum modes should be included. Since we have shown that the power spectrum converges to the CDM result for low momentum modes, we calculate the variance as\footnote{In the case of complete fragmentation we took $\dimlessps_{\delta}=\dimlessps_{\delta}^{\rm quad}$ in the second term of \eqref{eq:255} since for all the simulated modes the linear contribution is negligible.}
\begin{equation}
  \label{eq:255}
  \sigma_R^2(a)=\int_0^{k_{\rm min}}\frac{\dd{k}}{k}\abs{\tilde{W}_R(k)}^2\mathcal{P}_{\delta,\rm CDM}(k;a)+\int_{k_{\rm min}}^{k_{\rm max}}\frac{\dd{k}}{k}\abs{\tilde{W}_R(k)}^2\mathcal{P}_{\delta}(k;a),
\end{equation}
where $k_{\rm min}$ and $k_{\rm max}$ are the minimum and the maximum modes which are simulated. To calculate the CDM contribution during the radiation era, we have used the \emph{gauge-invariant} density contrast $\delta_{\rm CDM}^{\rm GI}$ given by \cite{Zhang:2017flu}
\begin{equation}
  \label{eq:262}
  \delta_{\rm CDM}^{\rm GI}(k;a)=9\Phi_k(0)\qty[\frac{\sin t_k}{t_k}+\frac{\cos t_k - 1}{t_k^2}-\cint(t_k)+\ln t_{k}+\gamma_E-\frac{1}{2}],\quad t_k=\frac{k/a}{\sqrt{3}H}
\end{equation}
so that the dimensionless power spectrum becomes\footnote{The difference between the density contrast calculated in the Newtonian gauge and the gauge-invariant one decays rapidly on sub-horizon scales. However, on super-horizon scales the Newtonian one becomes a constant while the gauge-invariant one scales with $t_k^2$. We are using the gauge-invariant one in this calculation in order to have a well-defined integral for the variance.  }
\begin{equation}
\boxed{
  \label{eq:263}
  \mathcal{P}_{\delta,\rm CDM}^{\rm rad}(k;a)=\frac{4}{9}A_s\times 81\qty[\frac{\sin t_k}{t_k}+\frac{\cos t_k - 1}{t_k^2}-\cint(t_k)+\ln t_{k}+\gamma_E-\frac{1}{2}]^2.}
\end{equation}
During the matter era, we use
\begin{equation}
\boxed{
  \label{eq:264}
  \mathcal{P}_{\delta,\rm CDM}^{\rm mat}(k;z)=A_s\qty(\frac{k_{\rm eq}^2}{5 \Omega_m H_0^2})^2 \frac{K^4 \mathcal{T}_{\rm BBKS}^2(K)}{(1+z)^2},}
\end{equation}
where $\Omega_m\approx 0.32$, $K\equiv \sqrt{2}k/k_{\rm eq}$ with $k_{\eq}$ is the wavenumber which enters the horizon at matter-radiation equality, and $\mathcal{T}_{\rm BBKS}$ is the BBKS transfer function \cite{Bardeen:1985tr}:
\begin{equation}
  \label{eq:265}
  \mathcal{T}_{\rm BBKS}(K)=\frac{\ln (1+0.121K)}{0.121K}\qty[1 + 0.201K + \qty(0.831K)^2+\qty(0.282K)^3 +\qty(0.346K)^4]^{-1/4},
\end{equation}
We review the derivation of \eqref{eq:264} in Appendix \ref{sec:derivation-cdm-power}.

\begin{figure}[tbp]
  \centering
  \includegraphics[width=0.8\textwidth]{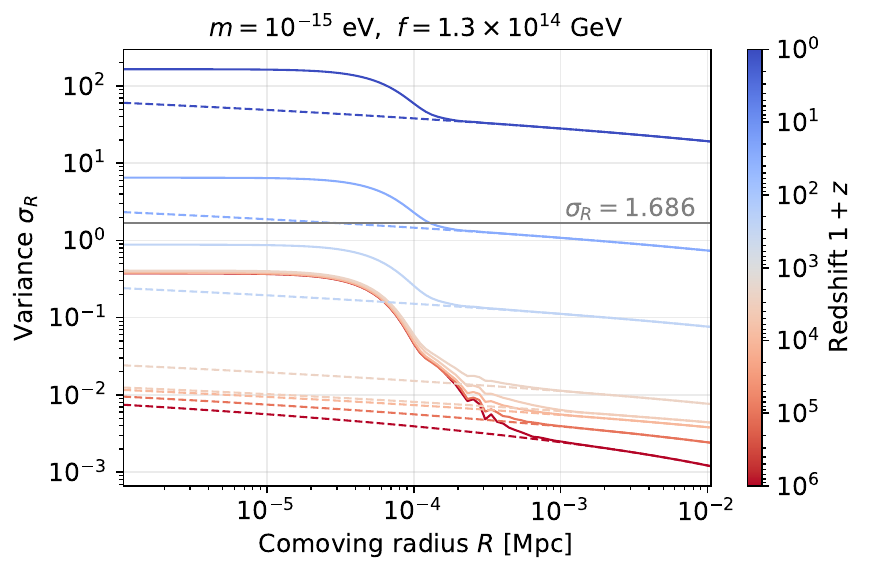}
  \caption{\small \it Evolution of the variance $\sigma_R$ for the benchmark point  $m=10^{-15}\,\si{\electronvolt}$ and $f=1.3\times 10^{14}\,\si{\giga\electronvolt}$ shown with solid lines. The red and blue lines denote redshifts in radiation and matter eras respectively. The dashed lines show the corresponding spectra for CDM. We see that at large scales, i.e. large comoving radius, the variance converges to the CDM prediction at all times. During the radiation era, the variance grows logarithmically on large scales  like CDM, while at small scales it is frozen since the variance integral \eqref{eq:255} is dominated by the modes that are above the Jeans scale. Slightly after the matter-radiation equality, all relevant modes drop below the Jeans scale, so the variance grows linearly with the scale factor at all scales. We also show a horizontal gray line at which the variance becomes equal to the critical density at collapse during the matter era. }
  \label{fig:variance-evolution}
\end{figure}

In Figure \ref{fig:variance-evolution}, we show the plots of the variance calculated at different redshifts for a benchmark ALP model with $m=10^{-15}\,\si{\electronvolt}$ and $f=1.3\times 10^{14}\,\si{\giga\electronvolt}$ with solid lines, while the dashed lines show the CDM predictions. The lines with red and blue color tones denote the redshifts in radiation and matter eras respectively. We can observe that the variance evolution differs between small scales (smaller comoving radius) and large scales (larger comoving radius). On large scales, all lines do converge to the CDM predictions which grow logarithmically/linearly with the scale factor during the radiation/matter eras. On the other hand, the small scales are frozen initially since the variance integral \eqref{eq:255} at those scales are dominated by the modes which are above the Jeans scale. Only slightly after the matter-radiation equality these scales drop below the Jeans scale, and as a result the variance grows linearly at all scales.

\begin{figure}[tbp]
  \centering
  \includegraphics[width=0.8\textwidth]{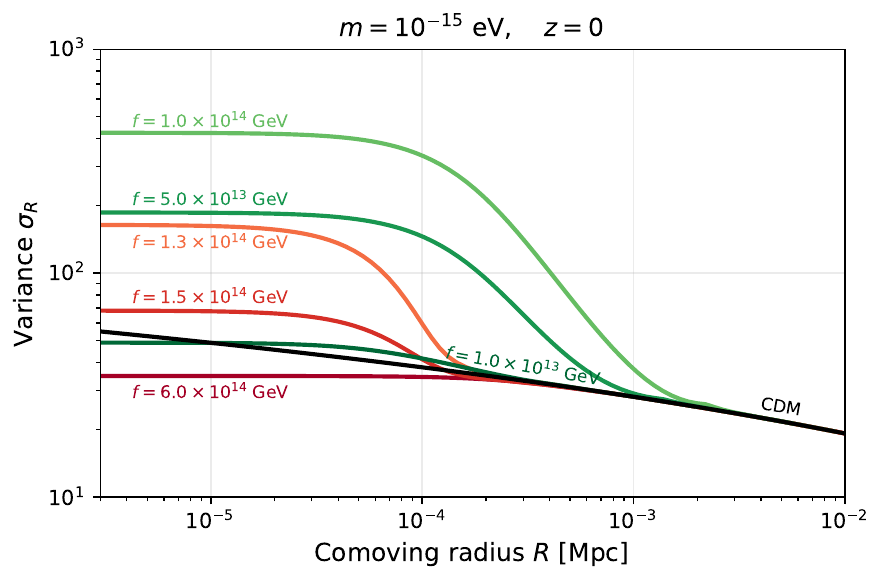}
  \caption{\small \it The variance of the density power spectum today \eqref{eq:255} calculated with the spherical top-hat window function \eqref{eq:259} for different ALP benchmarks with the mass $m=10^{-15}\,\si{\electronvolt}$, and CDM. The curves with the red tones denote the benchmarks with incomplete fragmentation (linear regime), while the ones with the green tones show the benchmarks where the fragmentation is complete (non-linear regime).}
  \label{fig:variance-today}
\end{figure}

We compare the variance evaluated at today for different ALP benchmarks, and CDM in Figure \ref{fig:variance-today}. We fixed the ALP mass to $m=10^{-15}\,\si{\electronvolt}$, and gradually decreased the ALP decay constant starting from a value which is close to the Kinetic-Standard Misalignment boundary. The curves with the red and green tones denote the benchmarks where the fragmentation is incomplete and complete respectively, corresponding to linear and non-linear regimes. In the linear regime, decreasing the axion decay constant yields larger enhancement of the small scales. The variance gets peaked for the benchmark which is closest to the incomplete-complete fragmentation boundary. Decreasing the ALP decay constant further reverses the trend, yielding to less enhancement. This feature is closely related to the shape of the power spectrum that we show in Figure \ref{fig:dimless-ps-cm-m15}. As the ALP decay constant decreases in the non-linear regime, the power spectrum becomes broader, and its peak is also shifted to higher momentum modes. These modes experience less growth in the matter era due to the Jeans suppression, and as a result the variance gets suppressed.

\subsection{Mass distribution of the halos}
\label{sec:mass-distr-halos:}

We now continue our discussion of the PS formalism in order to obtain the mass distribution of the halos. We have seen that the relevant quantity to calculate is $\mathscr{P}(\delta_R(a)>\delta_c(a))$. The easiest approach is to assume that $\delta(\vb{x},a)$, hence $\delta_R(\vb{x},a)$ is a Gaussian random field. Then the probability density function (PDF) for $\delta_R$ is given by
\begin{equation}
  \label{eq:323}
  f(\delta_R;a)=\frac{1}{\sqrt{2\pi \sigma_R^2(a)}}\exp(-\frac{\delta_R^2(a)}{2\sigma_R^2(a)}).
\end{equation}
Then by using the PS postulate we can write
\begin{equation}
  \label{eq:324}
  \mathscr{F}\qty(>M;a)=\mathscr{P}(\delta_R(a)>\delta_c(a))=\int_{\delta_c}^{\infty}\dd\delta f(\delta;a)=\frac{1}{2}\operatorname{erfc}\qty(\frac{\delta_c(a)}{\sqrt{2 \sigma_R^2(a)}}).
\end{equation}
In the case of CDM, the variance $\sigma_R^2$ goes to infinity as $R \rightarrow 0$. This implies that all the matter in the universe must finally be in virialized objects so we expect $\mathscr{F}(>M;a)\rightarrow 1$ as $M\rightarrow 0$ \cite{Maggiore:2009rv}. However, the PS result \eqref{eq:324} has the limit $\mathscr{F}(>M;a)\rightarrow 1/2$ as $M\rightarrow 0$, meaning that the PS formalism predicts that only half of the matter reside in collapsed structures. The reason behind this disagreement is that in PS formalism only overdense regions do collapse. In reality, underdense regions can also collapse if they are enclosed within a larger overdense region \cite{mo2010}. This effect can be included by using the \emph{excursion set formalism} \cite{Bond:1990iw}, for a nice review see \cite{Zentner:2006vw}. The correct expression is\footnote{This result is derived under the assumption that the variance is smoothed via a \emph{Fourier-space top-hat (FTH)} window function $\tilde{W}_{\rm FTH}(kR)=\Theta(1-kR)$, with $\Theta$ being the Heaviside function. This is simply the analog of the spherical top-hat in momentum space. The major drawback of this window function is that one cannot define a mass for the overdensity as we have done in \eqref{eq:260} since its volume is not well-defined \cite{Maggiore:2009rv}. The reason for choosing the FTH window function is that only with this choice the excursion set problem has an analytical solution \cite{Zentner:2006vw}.}
\begin{equation}
  \label{eq:325}
  \mathscr{F}\qty(>M;a)=\operatorname{erfc}\qty(\frac{\delta_c(a)}{\sqrt{2 \sigma_R^2(a)}}),
\end{equation}
which has the correct limit as $M \rightarrow 0$.

With this final result we can write the comoving number density of collapsed objects within the mass range $(M,M + \dd{M})$ as
\begin{equation}
  \label{eq:327}
  \begin{split}
    \dv{n(M;a)}{M}\dd{M}&=\frac{\overline{\rho}_{m,0}}{M}\abs{\dv{\mathscr{F}(>M;a)}{M}}\dd{M}\\
    &=\sqrt{\frac{2}{\pi}}\frac{\overline{\rho}_{m,0}}{M^2}\frac{\delta_c(a)}{\sigma_M(a)}\abs{\dv{\ln \sigma_M(a)}{\ln M}}\exp\qty(-\frac{\delta_c^2(a)}{2\sigma_M^2(a)})\dd{M}.
  \end{split}
\end{equation}
This is known as the \emph{Press-Schechter mass function}. In the literature, one commonly uses the so-called \emph{halo mass function (HMF)} which gives the number density of halos per logarithmic mass bin:
\begin{equation}
\boxed{
  \label{eq:328}
  \dv{n(M;a)}{\ln M}=\frac{1}{2}\frac{\overline{\rho}_{m,0}}{M}f_{\rm PS}(\nu)\abs{\dv{\ln \sigma_M^2(a)}{\ln M}},}
\end{equation}
where $\nu=\delta_c(a)/\sigma_M(a)$, and
\begin{equation}
  \label{eq:329}
  f_{\rm PS}(\nu)=\sqrt{\frac{2}{\pi}}\,\nu\exp(-\frac{\nu^2}{2}).
\end{equation}
The advantage of using this form is that one can replace $f_{\rm PS}(\nu)$ by other fit functions so that the HMF agrees more accurately with the results of the N-body simulations. An often adopted fit function is the Sheth-Tormen fit \cite{Sheth:1999su} which uses an ellipsoidal collapse model rather than a spherical one as in Press-Schechter. However, this fit function contains numerical factors which are obtained by solving the excursion set problem explicitly for CDM. So the numerical factors need to be recalculated for each model \cite{Du:2016zcv}. In this work we will use the PS mass function \eqref{eq:329} since it is more general and simpler.

\begin{figure}[tbp]
  \centering
  \includegraphics[width=\textwidth]{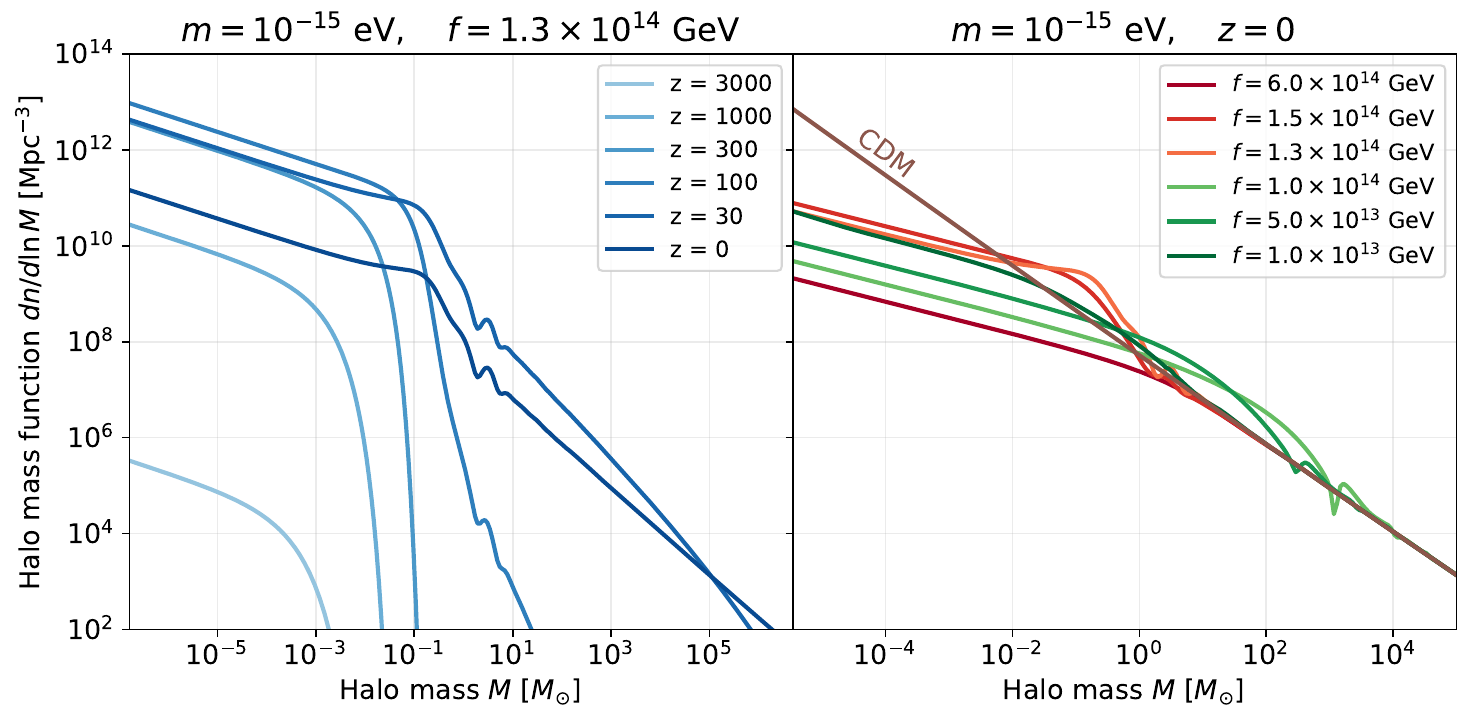}
  \caption{\small \it \textbf{Left plot: }The evolution of the halo mass function \eqref{eq:328} for the benchmark point with $m=10^{-15}\,\si{\electronvolt}$ and $f=1.3\times 10^{14}\,\si{\giga\electronvolt}$. We see that the number density of lighter halos decrease with redshift, while the number density of heavier ones increase. This is a consequence of merging of smaller halos with time forming larger halos. \textbf{Right plot: } The halo mass function today for different benchmark points and CDM. The ALP mass is set to $m=10^{-15}\,\si{\electronvolt}$. The color codes are the same as in Figure \ref{fig:variance-today}. }
  \label{fig:halo-mass-function-m15}
\end{figure}

We show our results about the halo mass function in Figure \ref{fig:halo-mass-function-m15}. On the left plot we show the time evolution of the HMF for the benchmark point $m=10^{-15}\,\si{\electronvolt}$ and $f=1.3\times 10^{14}\,\si{\giga\electronvolt}$. We see that the structure formation starts slightly after the matter-radiation equality with light halos, and with time heavier halos begin to form. The number density of light halos increases initially, however they start to decrease after some time. Meanwhile, the number density of heavy halos is always increasing meaning that the number density shifts from lighter halos to heavier halos. This is in agreement with what we expect from the structure formation where light halos are formed first, and later these halos merge with each other forming heavier halos \cite{mo2010}\footnote{Even though there is no gravitational evolution in the PS formalism, the Excursion Set description of the PS formalism can be used to study halo mergers, for example by constructing halo merger trees \cite{Zentner:2006vw}.}.


The halo mass function today for different benchmark points is presented on the right plot of Figure \ref{fig:halo-mass-function-m15}. The number density of heavy halos are not affected by the fragmentation, while the number density of light halos are suppressed compared to CDM due to the existence of the Jeans scale. On intermediate scales, there is an enhancement compared to CDM due to the fragmentation

\begin{figure}[tbp]
  \centering
  \includegraphics[width=\textwidth]{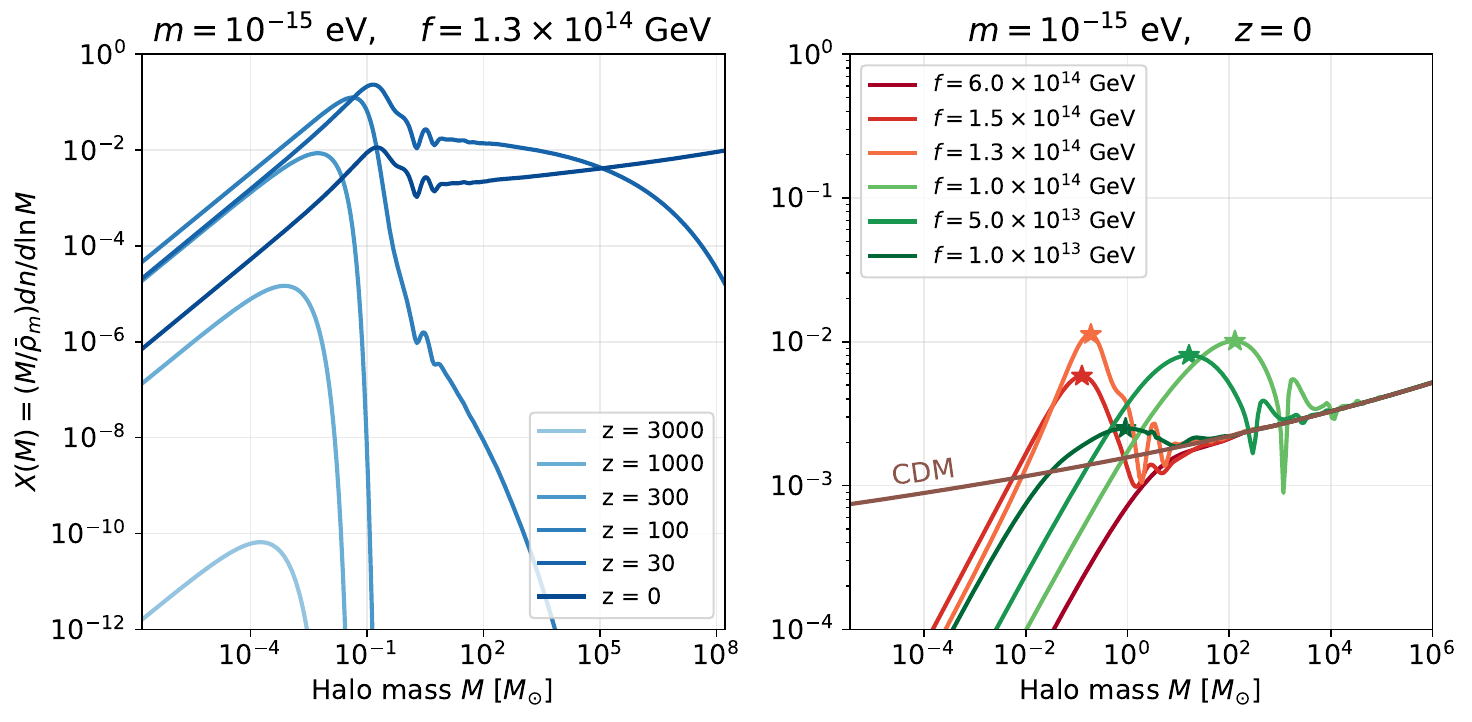}
  \caption{\small \it Analog of Figure \ref{fig:halo-mass-function-m15}, except that the vertical axis is in terms of the dimensionless halo mass function defined in \eqref{eq:326}. This dimensionless quantity gives roughly the fraction of dark matter that resides in halos with masses in the logarithmic mass bin $\qty[\ln M,\ln M+\dd{\ln M}]$. The stars on the right plot show the local maxima of $X(M)$ which approximately gives the mass of the halos that have the largest fraction of dark matter, excluding very heavy halos.}
  \label{fig:xm-function-m15}
\end{figure}

\begin{figure}[tbp]
  \centering
  \includegraphics[width=\textwidth]{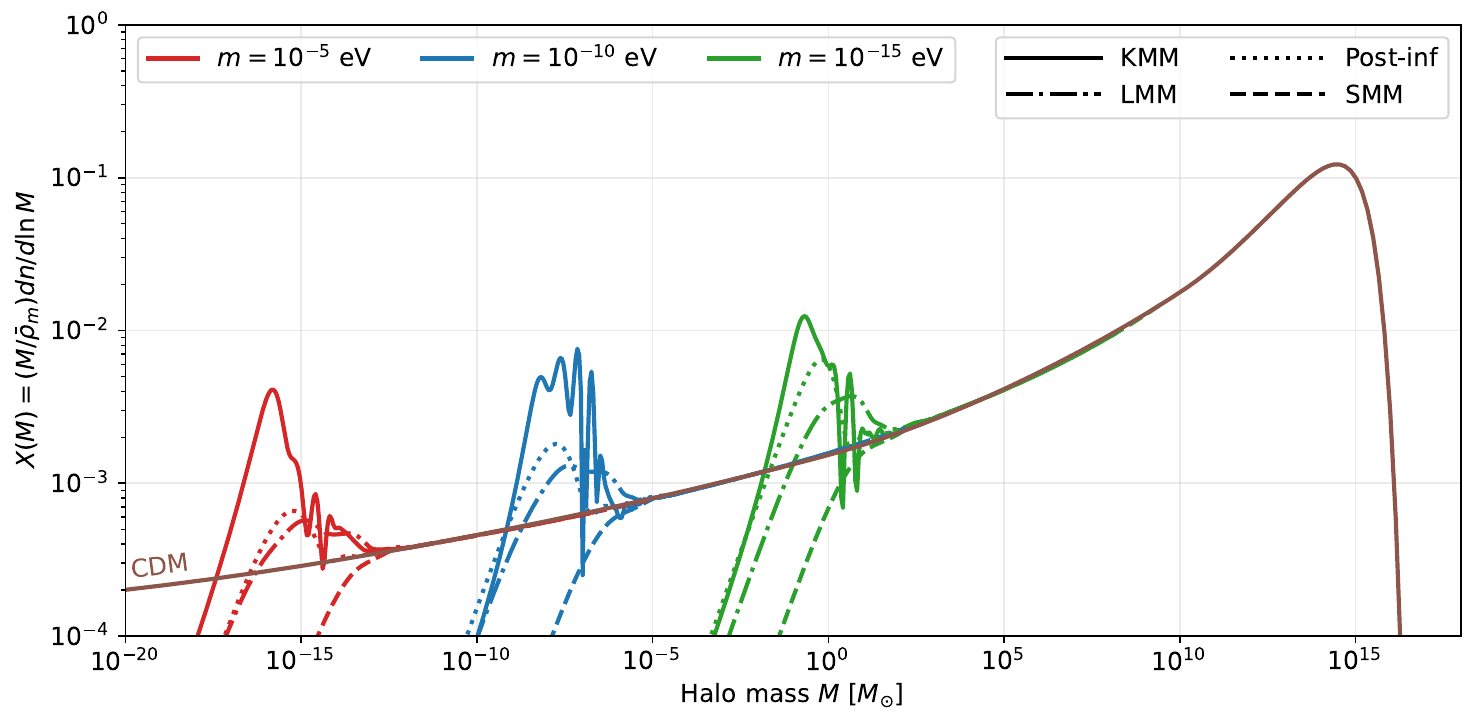}
  \caption{\small \it Dimensionless halo mass functions predicted by the production mechanisms Kinetic Misalignment (solid), Large Misalignment (dot-dashed), post-inflationary scenario (dotted), and Standard Misalignment (dashed). Different colors denote different ALP masses; $m=10^{-5}\,\si{\electronvolt}$ (red, left), $m=10^{-10}\,\si{\electronvolt}$ (blue, middle), $m=10^{-15}\,\si{\electronvolt}$ (green, right). The ALP decay constants are given in Table \ref{tab:grand-money-benchmarks}. The CDM prediction is shown via the brown line, and the curves do converge to the CDM prediction for larger halo masses. }
  \label{fig:grand-money-plot-hmf}
\end{figure}

We also present our results of the halo mass function in terms of the dimensionless variable \cite{Enander:2017ogx}
\begin{equation}
\boxed{
  \label{eq:326}
  X(M)\equiv \frac{M}{\overline{\rho}_{m,0}}\dv{n(M;a)}{\ln M}.}
\end{equation}
in Figure \ref{fig:xm-function-m15}. This approximately gives the fraction of dark matter that reside in halos with masses between $\qty[\ln M,\ln M+\dd{\ln M}]$. We also denote the local maxima of these curves with stars on the right plot.  Note that for large $M$, the curves for all benchmarks converge to the CDM prediction for which $X(M)$ grows with $M$\footnote{The CDM prediction for $X(M)$ has a global maximum around $M\sim 10^{14}M_{\odot}$, see Figure \ref{fig:grand-money-plot-hmf}.}. Therefore heavier halos contain a larger fraction of dark matter even with fragmentation. However, these halos collapse very late, so they are not sufficiently dense to be interesting for observational prospects as we shall see in the next section.

\begin{table}[tbp]
  \centering
  \begin{tabular}[tbp]{c | c | c | c}
    \toprule
    & \makecell{$m=10^{-5}\,\si{\electronvolt}$ \\ red, left} & \makecell{$m=10^{-10}\,\si{\electronvolt}$ \\ blue, middle} & \makecell{$m=10^{-15}\,\si{\electronvolt}$ \\ green, right} \\
    \cmidrule{1-4}\morecmidrules \cmidrule{1-4}
    \makecell{KMM \\ solid} & $f=4.0\times 10^{11}\,\si{\giga\electronvolt}$ & $f=6.7\times 10^{12}\,\si{\giga\electronvolt}$ & $f=1.2\times 10^{14}\,\si{\giga\electronvolt}$\\
    \cmidrule(lr){1-4}
    \makecell{LMM \\ dot-dashed} & \makecell{$f=4.0\times 10^{11}\,\si{\giga\electronvolt}$\\$\abs{\pi-\Theta_i}\approx 1.2\times 10^{-9}$}  & \makecell{$f=6.7\times 10^{12}\,\si{\giga\electronvolt}$\\$\abs{\pi-\Theta_i}\approx 7.0\times 10^{-9}$} & \makecell{$f=1.2\times 10^{14}\,\si{\giga\electronvolt}$\\$\abs{\pi-\Theta_i}\approx 3.7\times 10^{-7}$ }\\
    \cmidrule(lr){1-4}
    \makecell{Post-inf\\ dotted} & $f=1.5\times 10^{12}\,\si{\giga\electronvolt}$ & $f=2.3\times 10^{13}\,\si{\giga\electronvolt}$ & $f=3.5\times 10^{14}\,\si{\giga\electronvolt}$ \\
    \cmidrule(lr){1-4}
    \makecell{SMM \\ dashed} & \makecell{$f=3.0\times 10^{12}\,\si{\giga\electronvolt}$\\$\abs{\pi-\Theta_i}\approx 0.6$}  & \makecell{$f=4.0\times 10^{13}\,\si{\giga\electronvolt}$\\$\abs{\pi-\Theta_i}\approx 0.4$} & \makecell{$f=6.0\times 10^{14}\,\si{\giga\electronvolt}$\\$\abs{\pi-\Theta_i}\approx 0.4$ }\\
    \bottomrule
  \end{tabular}
  \caption{\small \it The list of benchmarks used in constructing the halo mass functions in Figure \ref{fig:grand-money-plot-hmf} and the halo spectra shown in Figure \ref{fig:grand-money-plot}. }
  \label{tab:grand-money-benchmarks}
\end{table}

In Figure \ref{fig:grand-money-plot-hmf}, we show a comparison of the dimensionless halo mass functions corresponding to different ALP masses and production mechanisms. The masses are $m=10^{-5}\,\si{\electronvolt}$ (red, left), $m=10^{-10}\,\si{\electronvolt}$ (blue, middle), $m=10^{-15}\,\si{\electronvolt}$ (green, right). Different linestyles denote different production mechanisms: Kinetic Misalignment (solid), Large Misalignment (dash-dotted), post-inflationary scenario (dotted), and Standard Misalignment (dashed). When comparing Kinetic and Large Misalignment mechanisms we have used the same value for the axion decay constant. For the post-inflationary scenario we set the decay constant such that the correct relic density is obtained, and used the power spectrum obtained via the linear analysis in Section \ref{sec:comparison-with-post}. The list of benchmarks that we used to produce this plot are tabulated in Table \ref{tab:grand-money-benchmarks}. The CDM prediction is shown by the brown line, and all the curves converge to it at large halo masses.

We end this section with a short discussion of the low-mass behavior of the halo mass function. As we can see from the right plot of Figure \ref{fig:halo-mass-function-m15}, even though the halo mass function for ALP is suppressed compared to CDM it still continues to grow with decreasing mass. However, this is not physical since the existence of the Jeans scale should heavily suppress the structure formation below a certain mass. This is a well-known problem of the Press-Schechter formalism applied to ALP dark matter. In \cite{Arvanitaki:2019rax} and \cite{Kulkarni:2020pnb}, it has been argued that this isssue stems from the fact that even for a small comoving radius $R$, the spherical top-hat window function still collects contributions from larger scales, i.e. small $k$-modes. Other works claim that the issue is not the spherical top-hat window function, but the critical density for collapse that we have computed in Section \ref{sec:crit-dens-coll} should be modified to account for the Jeans scale \cite{Marsh:2013ywa,Bozek:2014uqa,Du:2016zcv}. The precise calculation of the halo mass function for ALP dark matter using the N-body simulations has become possible only recently \cite{May:2021wwp,May:2022gus}, and more work is needed to correctly modify the Press-Schechter formalism such that it can also be applied to ALP dark matter. We will not investigate this issue further in this work, and continue our calculations by using the spherical top-hat window function, and scale-independent critical density. 

\subsection{Halo spectrum}
\label{sec:halo-spectrum}
In this section we  derive the halo spectrum predicted by the ALP fragmentation, compare it with the predictions of Standard Misalignment and the post-inflationary scenario. For this we need to know the density profile of the halos. We start by discussing the profiles of the CDM halos, and later comment on the modifications due to the wavelike nature of ALPs.

\subsubsection*{CDM Halos}

The N-body simulations of CDM show that for a halo identified, or measured, at redshift $z$, the density profile is well-approximated by the \emph{Navarro-Frenk-White (NFW)} profile given by \cite{Navarro:1996gj}
\begin{equation}
  \label{eq:330}
  \rho(r;z)=\frac{\delta_{\rm char}\rho_c(z)}{(r/r_s)(1+r/r_s)^2},
\end{equation}
where $r_s$ is called the \emph{scale radius} which is defined by the condition
\begin{equation}
  \label{eq:295}
  \eval{\dv{\ln \rho(r)}{\ln r}}_{r_s}=-2,
\end{equation}
$\delta_{\rm char}$ is the dimensionless \emph{characteristic density}, and $\rho_c(z)$ is the critical density of the universe at redshift $z$. An alternative but equivalent definition is
\begin{equation}
  \label{eq:331}
  \rho(r)=\frac{4\rho_s}{(r/r_s)(1+r/r_s)^2},
\end{equation}
where $\rho_s=\rho(r=r_s)$ is called the \emph{scale density}. From this defition, one can also define the \emph{scale mass} $M_s$ as the mass enclosed inside the scale radius:
\begin{equation}
  \label{eq:332}
  M_s=4\pi\int_0^{r_s}\dd{r}r^2\rho(r)=16\pi \rho_sr_s^3f(1),\quad f(x)\equiv \ln(1+x)-\frac{x}{1+x}.
\end{equation}
It is well known that the energy density inside a virialized halo is approximately $200$ times the mean energy density when the halo was formed. This motivates the definition of a quantity called the \emph{virialized radius} $r_{200}$ such that the mean energy density enclosed in a spherical region of radius $r_{200}$ is equal to $200$ times the critical density:
\begin{equation}
  \label{eq:333}
  200\rho_c=\qty(\frac{4\pi}{3}r_{200}^3)^{-1}4\pi\int_0^{r_{200}}\dd{r}r^2\rho(r).
\end{equation}
One also defines the \emph{concentration parameter} $c_{200}$ by
\begin{equation}
  \label{eq:334}
  c_{200}\equiv \frac{r_{200}}{r_s}.
\end{equation}
Then we can use \eqref{eq:333} to relate the characteristic density $\delta_{\rm char}$ to the concentration parameter $c_{200}$:
\begin{equation}
  \label{eq:335}
  \delta_{\rm char}=\frac{200}{3}\frac{c_{200}^3}{\ln(1+c_{200})-c_{200}/(1+c_{200})}.
\end{equation}
Even though we have not written down the redshift dependence explicitly, both $\delta_{\rm char}$ and $c_{200}$ are redshift-dependent.

The quantities $\delta_{\rm char}$ and $c_{200}$ need to be determined by the N-body simulations. However there is an emprical method originally developed by NFW for CDM \cite{Navarro:1996gj}, which can also be used for axion miniclusters \cite{Ellis:2020gtq}. Consider a halo identified, or measured, at redshift $z=z_{\rm ind}$. Assign this halo a mass $M_{200}$ which is equal to the mass enclosed in the spherical region of radius $r_{200}$. By definition we have
\begin{equation}
  \label{eq:336}
  M_{200}=\frac{4\pi}{3}\qty(200\times \rho_c(z_{\rm ind}))r_{200}^3.
\end{equation}
This mass can be identified with the mass $M$ of the collapsing region defined in \eqref{eq:260}, i.e. $M \leftrightarrow M_{200}$\footnote{The reasoning behing this identification is the assumption that the mass $M$ inside the collapsing region as defined by \eqref{eq:260} remains approximately constant during the collapse and the virialization \cite{Blinov:2021axd}.}. Then assign this halo a \emph{collapse redshift} $z_{\rm col}(M_{200},\mathfrak{f})$\footnote{The parameter $z_{\rm col}$ has nothing to do with the redshift of the linear collapse $y_c = (1 + z_{\rm eq})/(1 + z_{\rm c})$ that we have introduced in Section \ref{sec:crit-dens-coll}. } defined as the redshift at which half the mass of the halo was first contained in progenitors more massive than some fraction $\mathfrak{f}<1$ of the mass $M_{200}$. This can be estimated using the Press-Schechter formalism: \cite{Lacey:1993iv,Lacey:1994su}
\begin{equation}
\boxed{
  \label{eq:337}
  \operatorname{erfc}\qty[X(z_{\rm col})-X(z_{\rm ind})]=\frac{1}{2},}
\end{equation}
where
\begin{equation}
\boxed{
  \label{eq:338}
  X(z)=\frac{\delta_c(z)}{\sqrt{2\qty[\sigma^2(\mathfrak{f}M_{\rm 200};z)-\sigma^2(M_{\rm 200};z)]}}.}
\end{equation}
In this equation, $\delta_c(z)$ is the critical density for gravitational collapse, and $\sigma^2(M_{200};z)$ is the variance at the redshift $z$ evaluated with a spherical tophat window function with the smoothing scale
\begin{equation}
  \label{eq:339}
  R=\qty(\frac{3 M_{\rm 200}}{4\pi \rho_{m,0}a_0^3})^{1/3},
\end{equation}
where $\rho_{m,0}$ is the CDM/axion density today\footnote{We are assuming that CDM/axion make all of matter.}, and $a_0$ is the scale factor today.

Now assume that $\delta_{\rm char}(z_{\rm ind})\rho_c(z_{\rm ind})$, and hence $\rho_s$, is proportional to the energy density of the universe at the time of collapse:
\begin{equation}
  \label{eq:340}
  \delta_{\rm char}(z_{\rm ind})\rho_c(z_{\rm ind})=4\rho_s(z_{\rm ind})=\mathfrak{C}(z_{\rm col})\rho_c(z_{\rm col}),
\end{equation}
where $\mathfrak{C}(z_{\rm col})$ is in general $z_{\rm col}$-dependent proportionality factor. Both $\mathfrak{C}$ and $\mathfrak{f}$ are free parameters which need to be fit with the N-body simulations. For both CDM \cite{Navarro:1996gj} and axion miniclusters \cite{Ellis:2020gtq}, it is found that $\mathfrak{f}=10^{-2}$ provides a good fit to the simulations. Hence, we will also use this value in our calculations. To determine $\mathfrak{C}(z_{\rm col})$ we have used the CDM simulations compiled in \cite{Sanchez-Conde:2013yxa}, and axion minicluster simulations in \cite{Eggemeier:2019khm}. The details of the fitting procedure is explained in the Appendix \ref{sec:fitt-proc-obta}. Once this fitting is done, all the halo properties, most importantly the scale mass $M_s$ and the scale density $\rho_s$, can be calculated.

\subsubsection*{ALP Halos}
At small scales, the halo density profile for ALP dark matter cannot be described by the NFW profile due to the wavelike nature of ALPs. For recent reviews, see for example \cite{Marsh:2015xka} and \cite{Niemeyer:2019aqm}. Numerical simulations show that the central profiles of ALP halos do not obey the $r^{-1}$ scaling, the so-called ``cusp'', as in the NFW halos \cite{Hu:2000ke,Schive:2014dra,Schive:2014hza}. Instead, the density profiles are described by \emph{solitons}\footnote{Precisely, these are \emph{pseudo-solitons} since the ALP field is time-dependent, and they are not absolutely stable \cite{Marsh:2015xka}. True soliton solutions require a conserved Noether charge \cite{Diez-Tejedor:2013sza}. Therefore they occur for complex scalar field dark matter where the conserved $U(1)$ charge guarantees stability. These are known as the \emph{boson stars} in the literature \cite{Liddle:1992fmk}.} which are the eigenstates of the time-independent Schrödinger-Poisson equations \cite{PhysRev.187.1767,PhysRevLett.66.1659}. The soliton profile can be fitted by \cite{Schive:2014dra}
\begin{equation}
  \label{eq:353}
  \rho_{\rm sol}(r)=\rho_{\rm sol}(0)\qty[1+\qty(2^{1/8}-1)\qty(\frac{r}{r_{1/2}})^2]^{-8},
\end{equation}
where
\begin{equation}
  \label{eq:354}
  \rho_{\rm sol}(0)\approx 1.9\times\qty(\frac{m}{10^{-17}\,\si{\electronvolt}})^{-2}\qty(\frac{r_{1/2}}{\rm{pc}})^{-4}M_{\odot}\rm{pc}^{-3}
\end{equation}
is the density at the core, and $r_{1/2}$ is the radius at which the density is half of the core density. By using this form of the profile one can calculate all the scale parameters $r_s, \rho_s$ and $M_s$. Then one can show that the scale density and the scale mass are related to each other by
\begin{equation}
\boxed{
  \label{eq:355}
  \rho_s^{\rm sol}\approx 1.2\times 10^{-4}\qty(\frac{m}{10^{-17}\,\si{\electronvolt}})^6\qty(\frac{M_s^{\rm sol}}{M_{\odot}})^4\,M_{\odot}\rm{pc}^{-3}.}
\end{equation}
This ground state configuration is independent of the power spectrum, and is reached rapidly by a process called \emph{gravitational cooling} \cite{Seidel:1993zk,Guzman:2006yc}.

\subsubsection*{Results for the halo spectrum}

Far away from the center of the halo, the density profile of the ALP halos can again be explained by the NFW profile. Therefore, a realistic description requires a profile which smoothly interpolates between the soliton profile \eqref{eq:353} at small scales, and the NFW profile \eqref{eq:331} on large scales. A precise way to obtain this matching is currently not known. In this work, we calculated the halo spectrum by assuming the NFW profile at all scales, however used the $M_s^{\rm sol}$--$\rho_s^{\rm sol}$ relation \eqref{eq:355} as an upper bound on the scale density for a given scale mass.

\begin{figure}[tbp]
  \centering
  \includegraphics[width=\textwidth]{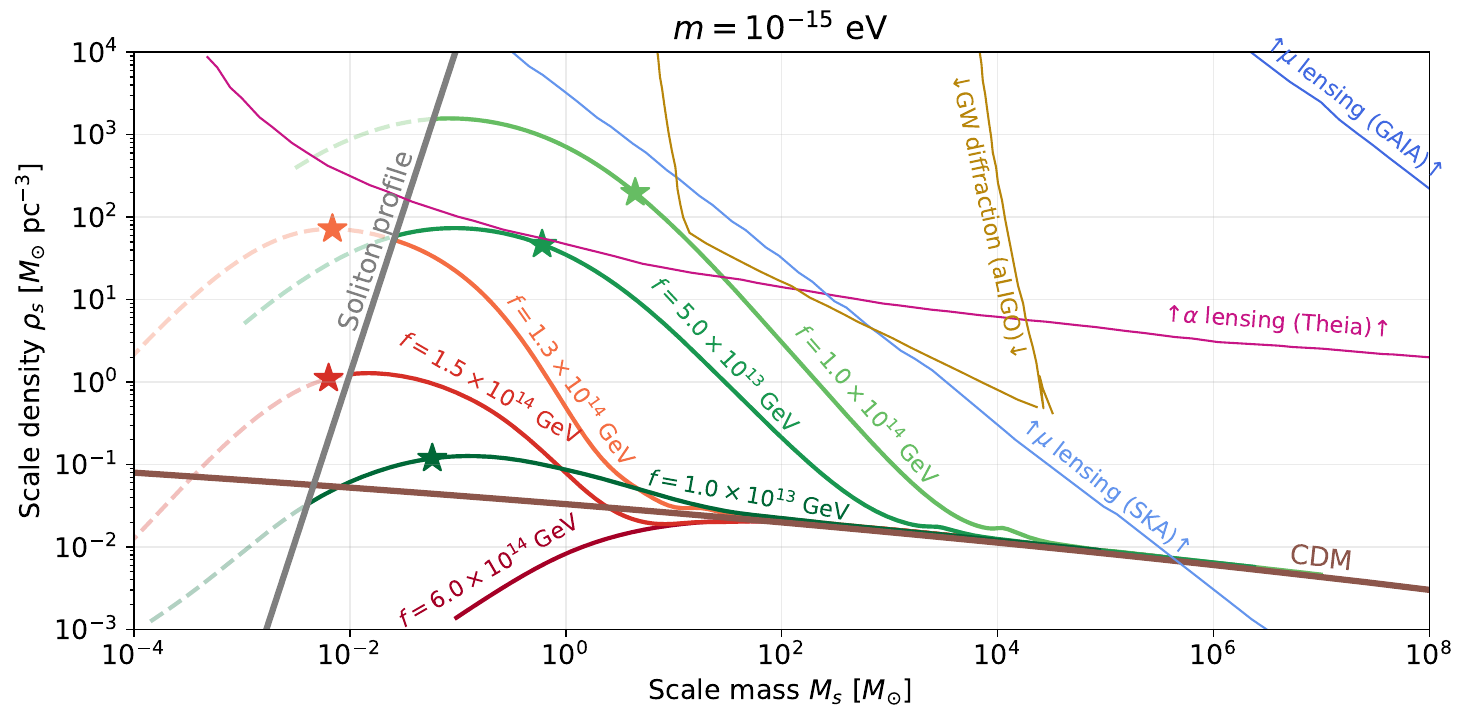}
  \caption{\small \it Halo spectrum for some benchmark points predicted by kinetic fragmentation assuming the NFW profile for the halos. The ALP mass is assumed to be constant and is fixed to $m=10^{-15}\,\si{\electronvolt}$. Labels on top of the curves denote the ALP decay constants. The gray line labeled as the ``soliton profile'' is the spectrum if all ALP halos have the soliton profile \eqref{eq:353}, and it is given by \eqref{eq:355}. The curves with the red tones denote the benchmarks with incomplete fragmentation (linear regime), while the ones with the green tones are the benchmarks with complete fragmentation (non-linear regime). The dashed lines on the left of the soliton line are the halo spectra if the ALP halos are described by the NFW profile at all scales. The stars denote the halos for which $X(M)$ as defined by \eqref{eq:326} has a local maxima, meaning that the probability of these halos is larger compared to nearby halos. Above the light/dark thin blue lines, weak gravitational lensing of compact halos can induce localized distortions in the proper motion $\mu$ of the background sources that can be observed by SKA/GAIA. Above the thin pink line, the weak lensing can cause correlations in stellar proper accelerations $\alpha$ that can be probable by Theia. In the region enclosed by the dark yellow line, the compact halos can cause distortions in the waveform of the gravitational waves emitted from a merger of two black holes that can be observed with aLIGO. See Section \ref{sec:probes-halo-spectrum} for more details.}
  \label{fig:m15-halo-spectrum}
\end{figure}

We show the results of the halo spectrum in Figure \ref{fig:m15-halo-spectrum}. We assumed a constant ALP mass, fixed it to $m=10^{-15}\,\si{\electronvolt}$, and varied the ALP decay constant. The color codes of the curves are the same as in Figure \ref{fig:variance-today}; the ones with the red tones denote the benchmarks with incomplete fragmentation, and the ones with the green tones are the cases with complete fragmentation. By comparing the results with the Figure \ref{fig:m15-halo-spectrum} we can clearly see that there is a direct relationship between the variance today and the peak scale density. The benchmark which has the largest variance today, is also the benchmark which has the largest scale density. This is a direct consequence of the fact that a larger variance implies larger fluctuations, and they collapse earlier yielding to denser halos. We also show the soliton relation \eqref{eq:355} by the gray line on the left. The dashed lines on the left of it are the spectra if the ALP halos have the NFW profile at all scales.

\begin{figure}[tbp]
  \centering
  \includegraphics[width=\textwidth]{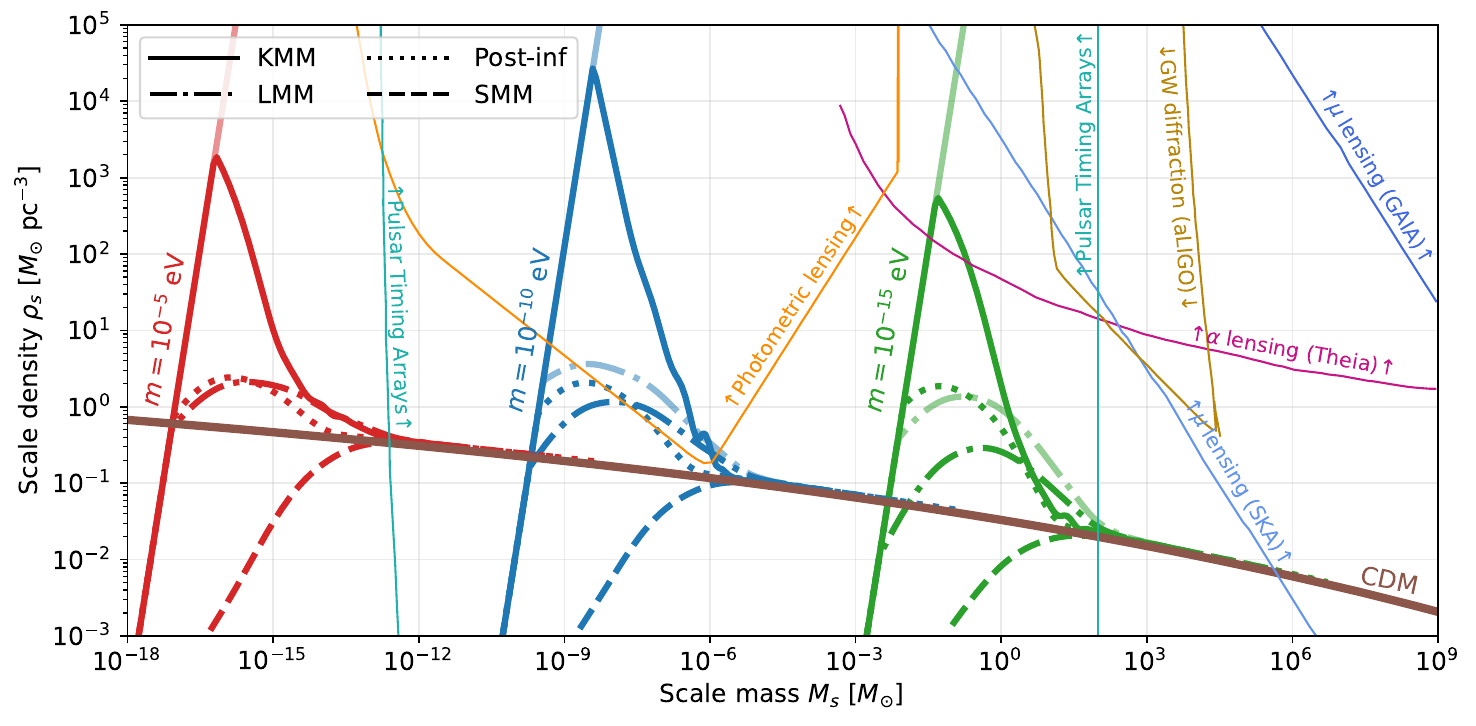}
  \caption{\small \it The halo spectra corresponding to the benchmarks listed in Table \ref{tab:grand-money-benchmarks} together with the regions observable by future lensing probes that we briefly summarized in Section \ref{sec:probes-halo-spectrum}. Different colors show the different ALP masses; $m=10^{-5}\,\si{\electronvolt}$ (red, left), $m=10^{-10}\,\si{\electronvolt}$ (blue, middle), $m=10^{-15}\,\si{\electronvolt}$ (green, right). Different linestyles show different production mechanisms; Kinetic misalignment with fragmentation (solid), Large misalignment (dot-dashed), post-inflationary scenario (dotted), and Standard misalignment (dashed). The straight faint lines labeled via the ALP mass show the soliton spectrum corresponding to the given ALP mass.}
  \label{fig:grand-money-plot}
\end{figure}

Finally, we compare the halo spectra corresponding to the different ALP masses and production mechanisms in Figure \ref{fig:grand-money-plot}. The ALP masses are $m=10^{-5}\,\si{\electronvolt}$ (red, left), $m=10^{-10}\,\si{\electronvolt}$ (blue, middle), and $m=10^{-15}\,\si{\electronvolt}$ (green, right). For all masses we compare the production mechanisms Kinetic misalignment with fragmentation (solid lines), Large misalignment (dash-dotted lines), post-inflationary scenario (dotted lines), and Standard misalignment (dashed lines)\footnote{Kinetic misalingment  without fragmentation line is indistinguishable from the Standard misalignment line so we don't show in the plot. }. The straight faint lines labeled via the ALP mass show the soliton spectrum corresponding to the given ALP mass. When comparing Kinetic and Large misalignment mechanisms we have used the same value for the ALP decay constant. For the post-inflationary scenario we set the decay constant such that ALPs make up all of the dark matter, and used the power spectrum obtained via the linear analysis in Section \ref{sec:comparison-with-post}. The list of benchmarks that we used when constructing the halo spectra can be found in Table \ref{tab:grand-money-benchmarks}. We also show the region of the $M_s$--$\rho_s$ plane which can be probed by future experiments by thin lines. We see that low-mass axions provide much more optimistic discovery prospects since the halo spectra are peaked at larger masses.

Before concluding this section, we briefly comment on why the scale density of the KMM halos is much larger than the other scenarios, namely LMM and the post-inflationary scenario. The KMM wins over the LMM due to the reasons explained in Section \ref{sec:comp-with-stand}. In summary, the size of the fluctuations before parametric resonance in KMM are many orders of magnitude larger compared to their LMM counterpart. Therefore, the size of the density contrast after parametric resonance is larger in KMM, despite the fact that the amplification is stronger in LMM, see Figure \ref{fig:smm-kmm-earlytime-comparison}. The reason that the KMM beats the post-inflationary scenario is the fact that in the latter scenario, the power spectrum is peaked at a larger comoving momenta; see the right plot of Figure \ref{fig:comparisons}. For a fixed ALP mass, the mode with a larger comoving momentum drops below the Jeans scale later, thus it collapses later. Also, the scale factor at the Jeans scale crossing scales as $a\propto k^{4}$, i.e. \eqref{eq:238}, which explains the large difference between the halo spectra in KMM and the post-inflationary scenario.

\section{Observational prospects}
\label{sec:observ-prosp}

In this section, we briefly comment on the phenomenological consequences of the halo spectra that we derived in the previous section. In Section \ref{sec:probes-halo-spectrum} we discuss various experiments that have a potential to probe the halo spectrum at small scales. In Section \ref{sec:cons-terr-exper} we discuss the consequences of the compact ALP halos for the terrestrial ALP detection experiments, such as holoscopes.

\subsection{Probes of the halo spectrum}
\label{sec:probes-halo-spectrum}

\begin{figure}[tbp]
  \centering
  \includegraphics[width=0.8\textwidth]{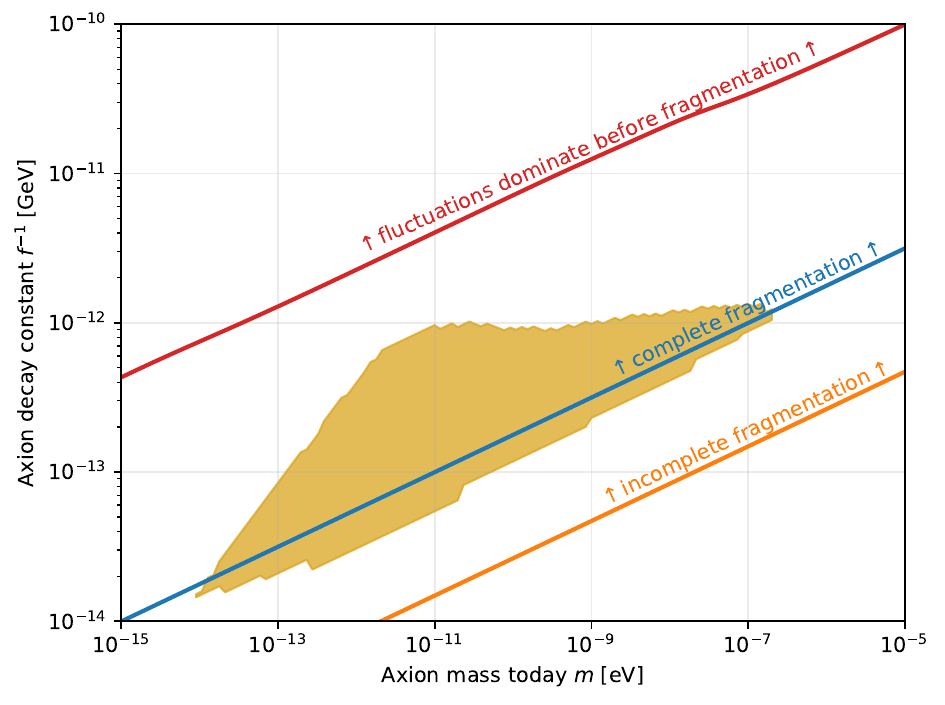}
  \caption{\small \it In this plot, we show the region on the ALP parameter space which has the potential to be probed via photometric lensing. To check the observability, we used the thin orange line in Figure \ref{fig:grand-money-plot} obtained in \cite{Arvanitaki:2019rax}. This means that we assumed a monochromatic mass distribution for the Milky-Way sub-halos, and 30\% of dark matter resides in compact halos.}
  \label{fig:photometric-prospects-constant-mass-single}
\end{figure}

The compact dark matter halos which are denser than the CDM ones can be probed by future gravitational surveys via their direct gravitational interactions. A detailed study of the discovery prospects is beyond the scope of this paper. Here we give a quick overview by using the sensitivity curves in the $M_s$--$\rho_s$ plane presented in \cite{Arvanitaki:2019rax}.
\begin{itemize}
  \item \textbf{Weak gravitational lensing: }When a compact DM halo passes near the line of sight of a background light source such as a star or any other luminuous object, it can induce apparent motions of these light sources through gravitational lensing. In the case of \emph{strong} lensing, these distortions are easily visible and form Einstein rings, arcs, and multiple images. If the distortions are weak so that they can only be detected by analyzing a large number of sources statistically, then one speaks of \emph{weak} gravitational lensing. The discovery prospects of compact DM halos via weak gravitational lensing has been studied in \cite{VanTilburg:2018ykj}. The weak lensing of a compact halo can induce localized distortions in the proper motion $\mu$ of the background light sources that can be observed by GAIA \cite{2018A&A...616A...1G} and SKA \cite{2004NewAR..48.1473F}. We call this \emph{$\mu$ lensing}, and show the observable region via thin blue lines in Figures \ref{fig:m15-halo-spectrum} and \ref{fig:grand-money-plot}. Similarly, the weak lensing can cause correlations in stellar proper accelerations $\alpha$ that can be probable by Theia \cite{Theia:2017xtk}. We call this \emph{$\alpha$ lensing} and show the observable region via thin pink line in Figures \ref{fig:m15-halo-spectrum} and \ref{fig:grand-money-plot}.
  \item \textbf{Photometric microlensing: }Another way of probing the halo spectrum is the \emph{photometric microlensing} where the brightness of a background light source changes due to the passage of a compact halo \cite{1986ApJ...304....1P,Dai:2019lud}. The potentially observable region as obtained in \cite{Arvanitaki:2019rax} is shown by the thin orange line in Figure \ref{fig:grand-money-plot}. We also show in Figure \ref{fig:photometric-prospects-constant-mass-single} the region on the ALP parameter space where the halo spectrum lies in the observable region.
\item \textbf{Pulsar Timing Arrays: } The Pulsar Timing Arrays (PTAs) are powerful probes of DM substructure at small scales \cite{Dror:2019twh,Ramani:2020hdo,Lee:2020wfn}. A transit of a compact object near the timing system can modify the frequency of the pulsar in two ways. First, it can modify the photon geodesic along the line of sight causing a \emph{Shapiro} time delay \cite{Siegel:2007fz}. Second, the compact objects can lead to an acceleration of the pulsar or the Earth modifying the period of the pulsar via the \emph{Doppler} effect \cite{Seto:2007kj}. We show the potentially observable region obtained in \cite{Ramani:2020hdo} via the thin green line in Figure \ref{fig:grand-money-plot}.
\item \textbf{Diffraction of Gravitational Waves: }Final observable that we comment on is the diffraction of gravitational waves. A distribution of compact DM halos can also induce distortions in the amplitude and the phase of the gravitational waves emitted from a merger of two black holes \cite{Dai:2018enj}. The potentially observable region via aLIGO as obtained in \cite{Arvanitaki:2019rax} is shown via dark yellow lines in Figures \ref{fig:m15-halo-spectrum} and \ref{fig:grand-money-plot}.
\end{itemize}

All the observable prospects mentioned above assume a monochromatic halo mass distribution, in other words $X(M)$ is taken to be proportional to a $\delta$-function. This means that, for a given $M_s$, the lines are derived under the assumption that some fraction of dark matter resides in halos with scale mass $M_s$ and scale density $\rho_s$. In \cite{Arvanitaki:2019rax} this fraction is chosen to be $0.3$ which we also use. The monochromatic assumption is necessary to draw the observable regions on the $M_s$--$\rho_s$ plane. The observable prospects will be modified with more realistic mass distributions like we have derived in Section \ref{sec:mass-distr-halos:}. For the halometry observations, such as $\mu$- and $\alpha$-lensing it is expected that this will not modify the qualitative features too much \cite{VanTilburg:2018ykj}. However, taking into account realistic sub-halo mass distributions reduces the prospects of the PTAs significantly, at least for CDM \cite{Ramani:2020hdo,Lee:2020wfn}.

\subsection{Consequences for terrestrial experiments}
\label{sec:cons-terr-exper}

If the ALPs have couplings to the Standard Model, they can be detected via direct detection terrestrial experiments, most notably via Haloscopes\footnote{For a detailed list of haloscope experiments and the corresponding references we refer the reader to our companion paper \cite{Eroncel:2022vjg}.}. These experiments search for a monochromatic signal with frequency set by the ALP mass $m_{0}$, and amplitude set by the local DM density $\rho^{\odot}_{\rm DM}\approx 1.1\times 10^{-2}\,M_{\odot}\,\textrm{pc}^{-3}$. The coherence time of the DM signal is given by $\tau\sim \frac{2\pi}{m_{0}v^{2}}$ where $v\sim 10^{-3}$ is the galactic virial velocity of dark matter \cite{Graham:2013gfa}.

If a significant fraction of DM in the Milky Way is bound to minihalos that are much denser that the local DM density, they can have a big impact on the terrestrial DM searches if they pass through the Earth \cite{Arvanitaki:2019rax,Blinov:2021axd}. During the crossing of a minihalo, the DM signal is boosted by a factor of $\mathcal{B}\sim \rho_{s}/\rho_{\rm DM}^{\odot}$, and is coherent for a much longer time. However, detecting such a signal would also require some modifications in the experimental setups. See \cite{Arvanitaki:2019rax} for more details on this point.

\begin{figure}[tbp]
  \centering
  \includegraphics[width=\textwidth]{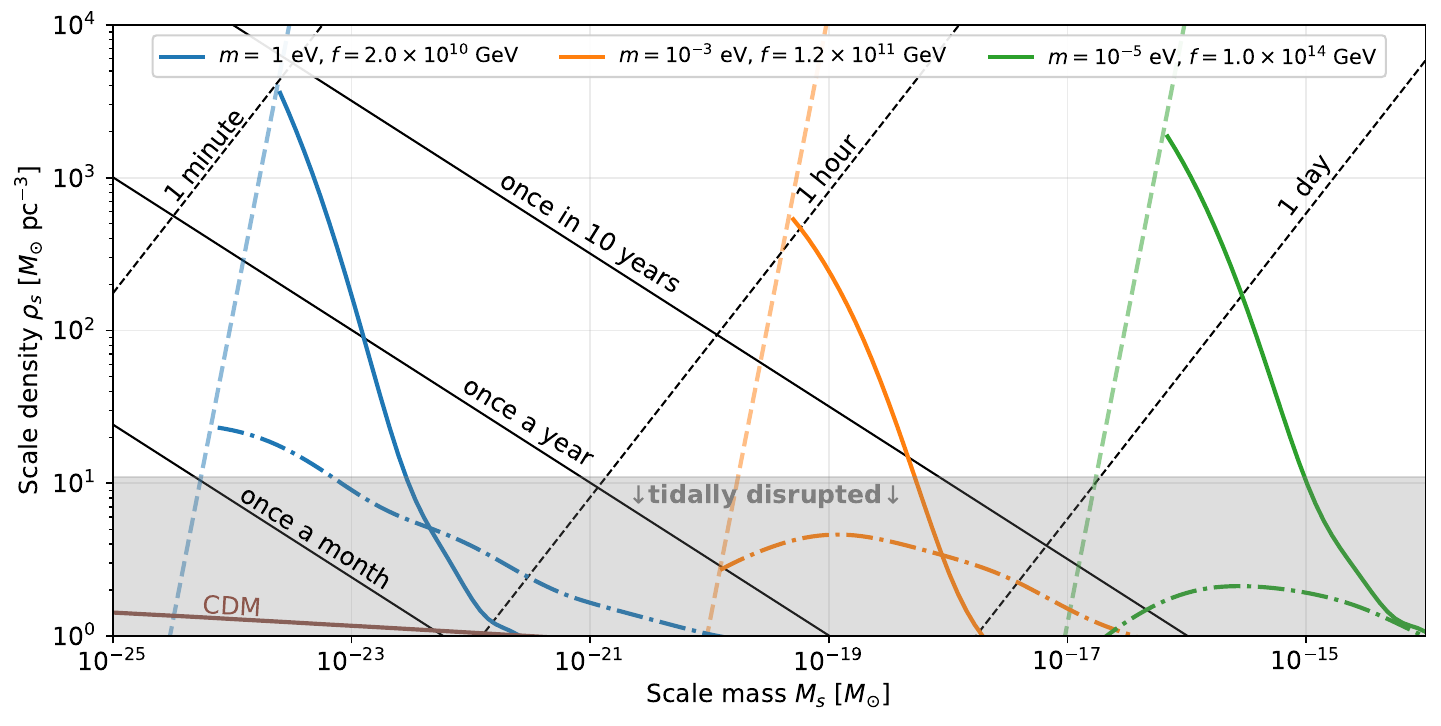}
  \caption{\small \it Contours of the encounter rate (solid black lines) calculated via \eqref{eq:381} and halo crossing times (dashed black lines) calculated via \eqref{eq:382} on the $M_{s}$--$\rho_{s}$ plane. We assume a monochromatic sub-halo mass function, and take $\Omega_{\rm sub}=0.3$ as the fraction of DM bound to sub-halos. For comparison, we also show the halo spectra for three benchmark points indicated in the top caption. The solid and dot-dashed lines show the spectra in the case of Kinetic and Large Misalignment respectively with the same $m_{0}$ and $f$ values. The dashed lines are the soliton profiles. The halos that lie inside the gray region are expected to be tidally distrupted according to the estimate in \cite{Arvanitaki:2019rax}.}
  \label{fig:encounter-rate}
\end{figure}

To assess the relevance of such an effect, we need to quantify how likely a microhalo will pass through the Earth. This information is encoded in the \emph{differential encounter rate} given by
\begin{equation}
  \label{eq:272}
  \dv{\Gamma_{\rm enc}}{\ln M}=\dv{\tilde{n}}{\ln M}\sigma_{\rm enc}v_{\rm vir}\equiv \frac{\rho_{\rm DM}^{\odot}}{M} \tilde{X}(M)\sigma_{\rm enc}v_{\rm vir},
\end{equation}
where $\sigma_{\rm enc}\sim \pi r_{s}^{2}$ is the encounter cross-section, $v_{\rm vir}\sim 200\,\si{\kilo\meter}/\si{\second}$ is the virial velocity of DM inside the Milky-Way, and $\dd \tilde{n}/\dd \ln M$ is the sub-halo mass function (sHMF). The latter tells us the distribution of sub-halos of the Milky Way and it is different than the HMF that we have calculated in Section \ref{sec:mass-distr-halos:} that tells us about the global distribution of halos in the whole universe. Therefore, a precise calculation of the encounter rate requires the calculation of the sHMF which we have left for future work. Here we assume a monochromatic sHMF by taking
\begin{equation}
  \label{eq:380}
  \tilde{X}(M)=\Omega_{\rm sub}\delta\qty(\ln M - \ln M_{s}),
\end{equation}
where $\Omega_{\rm sub}$ is the fraction of DM in the Milky Way that is bound to minihalos. With this assumption we can estimate the encounter rate as
\begin{equation}
  \label{eq:381}
  \Gamma_{\rm enc}\sim \frac{10\pi \Omega_{\rm sub}}{\rm year}\qty(\frac{\rho_{\rm DM}^{\odot}}{\rho_{s}})^{2/3}\qty(\frac{10^{-18}M_{\odot}}{M_{s}})^{1/3},
\end{equation}
where we have assumed that the minihalos have the NFW profile and used \eqref{eq:332}. We see that this effect is relevant mostly for very light sub-halos, and therefore for heavy ALP masses $m_{0}\gtrsim 10^{-5}\,\si{\electronvolt}$. The crossing time of a minihalo can be estimated as
\begin{equation}
  \label{eq:382}
  t_{\rm cross}\sim \frac{r_{s}}{v_{\rm vir}}\sim 4\,\textrm{days}\times\qty(\frac{\rho_{\rm DM}^{\odot}}{\rho_{s}})^{1/3}\qty(\frac{M_{s}}{10^{-18}M_{\odot}})^{1/3}.
\end{equation}

We show our results in Figure \ref{fig:encounter-rate}. The solid and dashed black contours show the encounter rate and crossing time respectively. We took $\Omega_{\rm sub}=0.3$ as the fraction of DM bound to sub-halos, consistent with the previous section. We also show the halo spectra for three benchmark points given in the legend. The solid and dot-dashed lines show the spectra in the case of Kinetic and Large Misalignment respectively with the same $m_{0}$ and $f$ values. The dashed lines show the soliton profiles. The halos that have densities $\rho_{s}\lesssim 10^{3}\rho_{\rm DM}^{\odot}$ are expected to be tidally distrupted before the encounter according to the estimate in \cite{Arvanitaki:2019rax}.

We see that the probability of a minihalo encounter during the lifetime of an experiment is rather low, even with highly optimistic assumptions. Repeating the calculation with a more realistic sHMF will likely give a lower encounter rate according to the analysis done in \cite{Blinov:2021axd} for a different model. We plan to revisit this in a future work.

\section{Conclusion}
\label{sec:conclusion-1}
In summary, we have analysed the implications of ALP fragmentation in the Kinetic Misalignment Mechanism (KMM) for the formation of ALP mini-clusters, and the observational consequences. KMM is particularly appealing as it makes the low $f$/large coupling region, which is accessible to terrestrial experiments,  viable for dark matter. This work is a follow-up to our companion article \cite{Eroncel:2022vjg}. The main logic and results of this paper are summarised in Figure \ref{fig:pipeline}, quoting the key equations. We solved the equations for the mode functions of the ALP fluctuations (equations \eqref{eq:6} and \eqref{eq:219}), obtained the initial conditions \eqref{eq:228}, and solved for  the late time evolution \eqref{eq:237}. A key result is the comparison of the density contrasts in KMM versus the Standard Misalignement Mechanism (SMM) and Large Misalignment Mechanism (LMM)  before fragmentation starts (equations \eqref{eq:320} and \eqref{eq:321}). This shows the large suppression in SMM and LMM in comparison with KMM, initial fluctuations being much larger in KMM.

An important figure is Figure \ref{fig:cm-mod-function-eval-full}, showing the full evolution of the modes, comparing CDM with KMM. The period just before fragmentation is illustrated in Figure \ref{fig:smm-kmm-earlytime-comparison}, showing clearly that the large initial fluctuations in KMM is what makes the difference.

Our two main final predictions are the halo mass functions (Figure \ref{fig:grand-money-plot-hmf}) and the halo spectra (Figure \ref{fig:grand-money-plot}).
This leads to promising prospects for experiments such as GAIA, SKA, Theia, PTA, aLIGO.
On the other hand, we do not expect that the formation of dense ALP halos will impact  direct detection experiments such as haloscopes as we found that  the probability of a minihalo encounter during the lifetime of an experiment is rather low.

We will provide a more precise study of the predictions for the QCD axion, relying on lattice simulations, in a future work. A key question for the relevance of this work is the motivation for the kinetic misalignment mechanism. For this, the general class of UV completions realising this framework should be scrutinised. This is what is done  in \cite{Eroncel:2022a} where we analyse in detail the parameter space in the $(m,f)$ plane.
 
In conclusion, there are great experimental prospects for probing the early evolution of the ALP field and thus access high energy scales through a variety of detection approaches, combining  haloscopes, helioscopes, light-shining-through-the-wall experiments as well as astronomical observations as the ones we have discussed in this paper.

\appendix

\acknowledgments

Special thanks go to Aleksandr Chatrchyan and Matthias Koschnitzke for important discussions. We  acknowledge as well contributions from Ryosuke Sato and Philip S\o{}rensen at the earlier stages of this work. We also thank Hyungjin Kim, Andrea Mitridate and Marcello Musso for useful discussions and feedback. This work is supported by the Deutsche Forschungsgemeinschaft under Germany Excellence Strategy - EXC 2121 ``Quantum Universe'' - 390833306.

\section{Derivation of the initial conditions for the mode functions}
\label{sec:deriv-adiab-init-1}

In this Appendix we derive the initial conditions for the mode functions assuming they originate from adiabatic perturbations. For the kinetic misalignment, we did this calculation in detail in our companion paper \cite{Eroncel:2022vjg}. In Appendix \ref{sec:kinetic-misalignment} we review the key results. In Appendix \ref{sec:stand-misal}, we present the calculation for the standard/large misalignment case, by generalizing the calculation in \cite{Zhang:2017flu} that has been done for a quadratic potential. 

\subsection{Kinetic misalignment}
\label{sec:kinetic-misalignment}

At early times the ALP mass can be neglected, so the equation of motions for the homogeneous mode, and the mode functions can be approximated respectively as
\begin{align}
  \label{eq:7}
  \ddot{\Theta}+3H\ddot{\Theta}&\approx 0,\\
  \ddot{\theta}_{k}+3H\dot{\theta}_{k}+\frac{k^{2}}{a^{2}}\theta_{k}&\approx -4\dot{\Phi}_{k}\dot{\Theta}.  
\end{align}
From these equations, the full solution for the mode functions can be derived as
\begin{equation}
  \label{eq:8}
  \theta_{k}(t_{k})=\frac{1}{t_{k}}\qty{\cos(\sqrt{3}t_{k})\qty[c_{1}-\mathcal{I}_{s}(t_{k})+\mathcal{I}_{s}(t_{k,i})]+\sin(\sqrt{3}t_{k})\qty[c_{2}+\mathcal{I}_{c}(t_{k})-\mathcal{I}_{c}(t_{k,i})]},
\end{equation}
where $\mathcal{I}_{c,s}$ are the indefinite integrals
\begin{align}
  \label{eq:10}
  \mathcal{I}_{c}(t_{k})&=-4\sqrt{3}\Phi_{k}(0)\frac{\dot{\Theta}_{i}}{H_{i}}t_{k,i}\int^{t_{k}}\dd{t_{k}'}\cos(\sqrt{3}t_{k}')\qty(\frac{\sin t_{k}'}{t_{k}'^{3}}+\frac{3\cos t_{k}'}{t_{k}'^{4}}-\frac{3\sin t_{k}'}{t_{k}'^{5}}),\\
 \label{eq:12} \mathcal{I}_{s}(t_{k})&=-4\sqrt{3}\Phi_{k}(0)\frac{\dot{\Theta}_{i}}{H_{i}}t_{k,i}\int^{t_{k}}\dd{t_{k}'}\sin(\sqrt{3}t_{k}')\qty(\frac{\sin t_{k}'}{t_{k}'^{3}}+\frac{3\cos t_{k}'}{t_{k}'^{4}}-\frac{3\sin t_{k}'}{t_{k}'^{5}}).
\end{align}
Here $\dot{\Theta}_{i}$ and $H_{i}$ are respectively the ALP velocity, and the Hubble scale at some early time $t_{i}$, and $t_{k,i}\equiv (k/a_{i})/\sqrt{3}H_{i}$. The coefficients $c_{1}$ and $c_{2}$ are determined by the initial conditions at $t_{k}=t_{k,i}$. By assuming that all the relevant modes are deep super-horizon at $t_{i}$, i.e. $t_{k,i}\ll 1$, and sourced by adiabatic perturbations fix these coefficients uniquely as
\begin{equation}
  \label{eq:11}
  c_{1}=-\frac{1}{2}t_{k,i}\frac{\dot{\Theta}_{i}}{H_{i}}\Phi_{k}(0),\quad c_{2}=0.
\end{equation}
Therefore, the full solution can be obtained by evaluating the integrals \eqref{eq:10} and \eqref{eq:12}.

We will set the initial conditions for all the simulated modes when they are super-horizon. Therefore we can expand the full solution as a Taylor series around $t_{k}=0$. Moreover, since we assume that all the relevant modes are super-horizon at $t_{i}$ we can also take the $t_{k,i}\rightarrow 0$ limit in the end. This way we obtain
\vspace{0.2cm}
\begin{mdframed}[backgroundcolor=blue!15,innertopmargin=0pt,innerbottommargin=9pt]
\begin{equation}
  \label{eq:9}
  \theta_{k}(t_{k}<1)\approx\frac{\dot{\Theta}(t)}{H(t)}\Phi_{k}(0)\qty(-\frac{1}{2}+\frac{23}{20}t_{k}^{2}-\frac{491}{1680}t_{k}^{4}+\frac{4427}{\numprint{151200}}t_{k}^{6}-\frac{\numprint{146131}}{\numprint{93139200}}t_{k}^{8})
\end{equation}
\end{mdframed}
\vspace{0.2cm}
When deriving this expression we have used
\begin{equation}
  \label{eq:13}
  \frac{t_{k,i}}{t_{k}}\frac{\dot{\Theta}_{i}}{H_{i}}=\frac{a_{i}}{a}\frac{\dot{\Theta}_{i}}{H_{i}}=\frac{\dot{\Theta}(t)}{H(t)}.
\end{equation}
Changing to dimensionless time variable $t_{m}=m/2H\simeq mt$ gives \eqref{eq:228}.
\subsection{Standard/Large misalignment}
\label{sec:stand-misal}

We now repeat the calculation of the previous section in the case of Standard/Large Misalignment. That is, we want to obtain an approximate solution of the mode function equation \eqref{eq:219} at early times $t_m\ll 1$ for super-horizon modes. This calculation has been done in \cite{Zhang:2017flu} for a quadratic potential, but the generelization to the full cosine potential is straightforward. In order to do this, we first need to get an approximate solution for the homogeneous mode equation of motion which is given by
\begin{equation}
  \label{eq:294}
  \Theta''(t_m)+\frac{3}{2 t_m}\Theta'(t_m)+\sin\Theta(t_m)=0.
\end{equation}
At early times, the ALP angle $\Theta$ is very close to its initial value, so we can approximate
\begin{equation}
  \label{eq:298}
  \Theta(t_m)\approx \Theta_i + \delta\Theta(t_m),
\end{equation}
where $\delta\Theta \ll \Theta_i$. Then at leading order $\delta\Theta$ obeys the following differential equation:
\begin{equation}
  \label{eq:299}
  \delta\Theta''(t_m)+\frac{3}{2t_m}\delta\Theta'(t_m)+\qty[\sin\Theta_i+\cos\Theta_i\delta\Theta(t_m)]\approx 0.
\end{equation}
By solving this equation with the initial condition $\delta\Theta(t_m)\rightarrow 0$ as $t_m\rightarrow 0$, and then expanding the result around $t_m=0$ one finds the asymptotic behavior of $\Theta(t_m)$ as
\begin{equation}
  \label{eq:300}
  \Theta(t_m)=\Theta_i-\frac{1}{5}\sin\Theta_i t_m^2+\mathcal{O}(t_m^4).
\end{equation}
This is in agreement with the results of \cite{Zhang:2017flu} after replacing $\sin\Theta_i\rightarrow \Theta_i$ in the case of quadratic potential.

At leading order in $t_{m}$, the mode function equation of motion \eqref{eq:219} becomes
\begin{equation}
  \label{eq:301}
  \begin{split}
    \dv[2]{\theta_k}{t_m}+\frac{3}{2 t_m}\dv{\theta_k}{t_m}+\qty(\frac{\tilde{k}^2}{t_m}+\cos\Theta_i)\theta_k&\approx 2\qty[\Phi_k(t_k)\sin\Theta_i+\frac{2}{5}\sin\Theta_i \,t_k\dv{\Phi_k}{t_k}]\\
    &\equiv \mathcal{S}(\tilde{k},t_m).
  \end{split}
\end{equation}
The homogeneous solutions are given by the Kummer's function of the first kind\footnote{Also known as the Kummer's confluent hypergeometric function, or the hypergeometric function $_1 F_1$. }: \cite{Zhang:2017flu}
\begin{align}
  \label{eq:302}
  \theta_1(t_m)&=\exp(-i \sqrt{\cos\Theta_i} t_m)M\qty(\frac{3}{4}+\frac{i \tilde{k}^2}{2\sqrt{\cos\Theta_i}},\; \frac{3}{2}, \;2i\sqrt{\cos\Theta_i}t_m)\\
  \theta_2(t_m)&=\frac{\exp(-i \sqrt{\cos\Theta_i} t_m)}{\sqrt{t_m}}M\qty(\frac{1}{4}+\frac{i \tilde{k}^2}{2\sqrt{\cos\Theta_i}},\; \frac{1}{2}, \;2i\sqrt{\cos\Theta_i}t_m)
\end{align}
The particular solution is
\begin{equation}
  \label{eq:303}
  \theta_k^{\rm part}(t_m)=\theta_2(t_m)\int_0^{t_m}\dd{t_m'}\frac{\theta_1(t_m')\mathcal{S}(\tilde{k},t_m')}{W[\theta_1(t_m'),\theta_2(t_m')]}-\theta_1(t_m)\int_0^{t_m}\dd{t_m'}\frac{\theta_2(t_m')\mathcal{S}(\tilde{k},t_m')}{W[\theta_1(t_m'),\theta_2(t_m')]},
\end{equation}
where $W$ is the Wronskian. Again, by expanding all the terms around $t_m=0$ we can obtain the asymptotic behavior of the particular solution as\footnote{Note that here we are also assuming that the modes are super-horizon, otherwise the second term can dominate the first term.}
\vspace{0.2cm}
\begin{mdframed}[backgroundcolor=blue!15]
\begin{equation}
  \label{eq:304}
  \theta_k^{\rm part}(t_m)=\frac{2}{5}\Phi_k(0)\sin\Theta_i\qty[t_m^2-\frac{22}{105}\tilde{k}^2t_m^3+\mathcal{O}(t_m^4)].
\end{equation}
\end{mdframed}
\vspace{0.2cm}
Again this is in agreement with \cite{Zhang:2017flu} after the replacement $\sin\Theta_i \rightarrow \Theta_i$.

The full solution is
\begin{equation}
  \label{eq:305}
  \theta_k(t_m)=c_1\theta_1(t_m)+c_2\theta_2(t_m)+\theta_k^{\rm part}(t_m),
\end{equation}
where we have omitted the momentum subscripts from the homogeneous solutions in order not to clutter the notation. As $t_m \rightarrow 0$, the first solution has the limit $\theta_1 \rightarrow 1$, while the second solution $\theta_2$ diverges. In the case of SMM, the adiabatic initial conditions for the ALP mode functions should vanish in the super-horizon limit; see in particular \eqref{eq:15}. Therefore, both of the integrations constants $c_{1,2}$ need to be set to zero, and the solution is given solely by the particular solution.

The sub-horizon behavior of the mode functions at early time can also be derived analytically. For these modes we have
\begin{equation}
  \label{eq:312}
  \frac{\tilde{k}^2}{t_m}=\frac{3 t_k^2}{4 t_m^2}\gg \cos\Theta_i
\end{equation}
at early times, so we can neglect the $\cos\Theta_i$ term on the RHS of \eqref{eq:301}. In this regime, it is more convenient to use $t_k$ as the dynamical variable. Then \eqref{eq:301} becomes
\begin{equation}
  \label{eq:311}
  \dv[2]{\theta_k}{t_k}+\frac{2}{t_k}\dv{\theta_k}{t_k}+3\theta_k\approx \frac{9}{4}\frac{t_k^2}{\tilde{k}^4}\mathcal{S}(t_k),
\end{equation}
where the source term $\mathcal{S}$ is still defined by \eqref{eq:301}. Now the homogenous solutions are
\begin{equation}
  \label{eq:313}
  \theta_{1}(t_k)=\frac{\cos(\sqrt{3}t_k)}{t_k}\qand\theta_2(t_k)=\frac{\sin(\sqrt{3}t_k)}{t_k}.
\end{equation}
The particular solution is
\begin{equation}
  \label{eq:314}
  \theta_k^{\rm part}(t_k)=\frac{9 \tilde{k}^{-4}}{4 \sqrt{3}}\qty[\frac{\sin(\sqrt{3}t_k)}{t_k}\mathcal{I}_c(t_k)-\frac{\cos(\sqrt{3}t_k)}{t_k}\mathcal{I}_s(t_k)],
\end{equation}
where the integrals are
\begin{align}
  \label{eq:315}
  \mathcal{I}_c(t_k)&=\int_{t_{k,i}}^{t_k}\dd{t_k'}t_k'^3\cos(\sqrt{3}t_k)\mathcal{S}(t_k'),\\
  \mathcal{I}_s(t_k)&=\int_{t_{k,i}}^{t_k}\dd{t_k'}t_k'^3\sin(\sqrt{3}t_k)\mathcal{S}(t_k'),
\end{align}
In the sub-horizon limit $t_k\gg 1$, both integrands behave as $\propto t_k'^2$, so the integrals are dominated by their upper cutoffs $t_k$. Then, the particular solution becomes
\begin{equation}
  \label{eq:316}
  \theta_k^{\rm part}(t_k)=\frac{27}{10}\Phi_k(0)\sin\Theta_i \tilde{k}^{-4}\qty[t_k\sin t_k-\frac{3}{2}\cos t_k-3 \frac{\sin t_k}{t_k}].
\end{equation}
Since the particular solution grows with $t_k$, it quickly dominates the homogeneous solution so again the full solution is equal to the particular solution.

\section{Derivation of the CDM power spectrum during the matter era}
\label{sec:derivation-cdm-power}

It is well known that the metric perturbations are constant during the matter domination. This constant value is commonly expressed in terms of a \emph{transfer function} $\mathcal{T}$ defined by
\begin{equation}
  \label{eq:360}
  \Phi_k(a_{\rm MD})=\frac{9}{10}\mathcal{T}(k)\Phi_k(a_{\rm in}).
\end{equation}
Here the scale factors $a_{\rm MD}$ and $a_{\rm in}$ are irrelevant as long as they are chosen at deep matter and radiation domination respectively. The normalization is chosen such that $\mathcal{T}(0)=1$. The main idea for introducing a transfer function is to separate the evolution of the modes from their initial conditions.

The evolution at later times $a>a_{\rm MD}$ can be described by
\begin{equation}
  \label{eq:361}
  \Phi_k(z)=\frac{9}{10}\mathcal{T}(k)\mathcal{D}(z)(1+z)\Phi_k(0),
\end{equation}
where $\mathcal{D}(z)$ is called the \emph{growth function}. It is defined such that $\mathcal{D}(z)=(1+z)^{-1}$ in a matter-dominated universe without a cosmological constant $\Omega_{\Lambda}=0$. A non-zero $\Omega_{\Lambda}$ suppresses the growth at very late times so that todays value of the growth factor is $\mathcal{D}(0)\approx 0.79$ instead of $1$. Since this suppression happens very late, and most of the structures that we are interested in collapse deep in the matter era, we neglect the suppression from the vacuum energy and assume $\mathcal{D}(z)=(1+z)^{-1}$.

To get the density constrast, we make use of the Poisson equation given by
\begin{equation}
  \label{eq:363}
  \nabla^2\Phi=-4\pi G a^2\overline{\rho}_m\delta_{\rm m}^{\rm GI},
\end{equation}
which is valid for all modes not just the sub-horizon ones. From this we can solve for the Fourier modes of the density constrast as
\begin{equation}
  \label{eq:364}
  \delta_{m,k}=\frac{3}{5}\frac{k^2\mathcal{T}(k)}{\Omega_mH_0^2(1+z)}\Phi_k(0)=\frac{2}{5}\frac{k^2\mathcal{T}(k)}{\Omega_mH_0^2(1+z)}\mathcal{R}_k(0).
\end{equation}
The power spectrum is
\begin{equation}
  \label{eq:365}
  \dimps_m(k)=\frac{4}{25}\frac{k^4\mathcal{T}(k)}{\Omega_m^2H_0^4}\qty(1+z)^{-2}\dimps_{\mathcal{R}}^{\rm in}(k)=\frac{8\pi^2}{25}\frac{k \mathcal{T}^2(k)}{\Omega_m^2H_0^4}\qty(1+z)^{-2}A_s\qty(\frac{k}{k_{\star}})^{n_s-1}.
\end{equation}
In this work we neglect the spectral tilt by taking $n_s=1$, and we also neglect the baryons so the matter power spectrum is the same as the CDM power spectrum. Finally by changing to the dimensionless power spectrum $\dimlessps=k^3 P(k)/2\pi^2$, and defining $K\equiv \sqrt{2}k/k_{\rm eq}$ we get \eqref{eq:264}.

\section{Fitting procedure to obtain $\mathfrak{C}(z_{\rm col})$}
\label{sec:fitt-proc-obta}

\begin{figure}[tbp]
  \centering
  \includegraphics[width=0.8\textwidth]{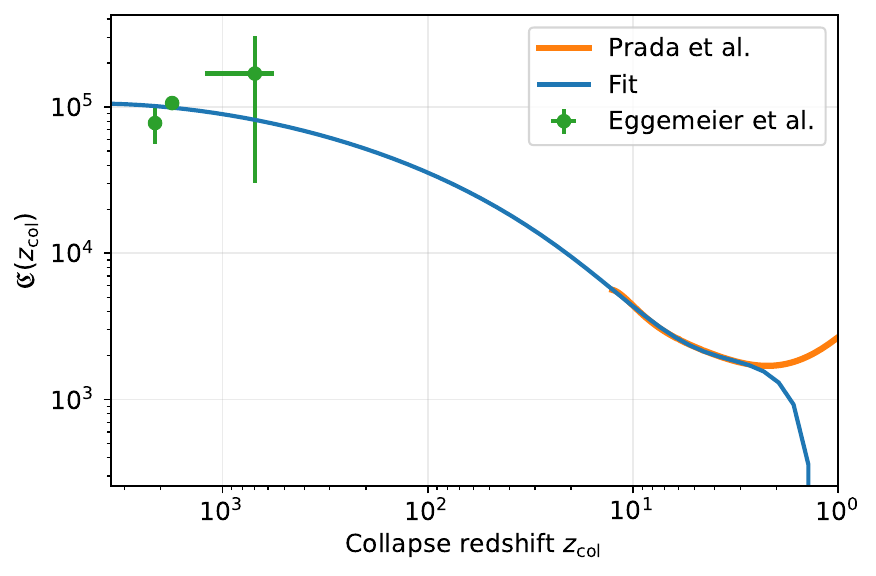}
  \caption{\small \it Plot of the $\mathfrak{C}(z_{\rm col})$ fit function \eqref{eq:351}, together with the N-body results of Conde and Prada \cite{Sanchez-Conde:2013yxa} for CDM, and Eggemeier et al. \cite{Eggemeier:2019khm} for axion miniclusters. }
  \label{fig:zcol-cfrak-fun}
\end{figure}

To determine $\mathfrak{C}(z_{\rm col})$ we have used two sets of simulations:
\begin{itemize}
\item A set of CDM simulations compiled by Conde and Prada \cite{Sanchez-Conde:2013yxa}.
\item An axion minicluster simulation by Eggemeier et al. \cite{Eggemeier:2019khm} based on the lattice calculation of the post-inflationary scenario for the QCD axion that is performed in \cite{Vaquero:2018tib}.
\end{itemize}
Both sets of simulations give the results for the concentration parameter $c_{200}$ as a function of the halo mass $M_{200}$. To perform the matching, first we did calculate the collapse redshifts for the tabulated halo masses. For CDM halos we use the power spectrum \eqref{eq:264}, while for the axion miniclusters we took the latetime power spectrum obtained in \cite{Vaquero:2018tib}, calculated the variance using the spherical top-hat window function, and finally evaluated until today as \cite{Ellis:2020gtq,Lee:2020wfn}
\begin{equation}
  \label{eq:350}
  \sigma^2(M;z)=\sigma_i^2(M)\qty(1+\frac{3}{2}\frac{1+z_{\rm eq}}{1+z})^2,
\end{equation}
where $\sigma_i$ is the initial variance obtained from the power spectrum. The reason that the Jeans scale plays no role in this evolution is because all relevant modes are well below the Jeans scale at the matter-radiation equality.

After the collapse redshifts are calculated, it becomes a trivial task to determine $\mathfrak{C}(z_{\rm col})$ using the given concentration parameter data. For CDM halos that collapse around $z\sim 10$ we found that $\mathfrak{C}\sim \mathcal{O}(1)\times 10^3$, and it is an increasing function of $z_{\rm col}$. On the other hand, axion miniclusters collapse around $z_{\rm col}\sim 10^3$ for which $\mathfrak{C}\approx 9.76\times 10^4$ did provide the best fit. Finally by combining the CDM and the axion minicluster data we obtained the following fit function for $\mathfrak{C}(z_{\rm col})$:
\begin{equation}
  \label{eq:351}
  \mathfrak{C}(z_{\rm col})=\sum_{i=0}^5c_i\qty[\ln z_{\rm col}]^i,\quad 3\lesssim z_{\rm col}\lesssim 3000,
\end{equation}
with the coefficients:
\begin{equation}
  \label{eq:352}
  \qty{c_i}_{i=0}^5=\qty{-1.55\times 10^3, 7.67\times 10^3, -6.56\times 10^3, 2.39\times 10^3,-2.33\times 10^2, 5.98}
\end{equation}
We show this fit function together with data from CDM and minicluster simulations in Figure \ref{fig:zcol-cfrak-fun}. The observationally relevant halos collapse well before $z_{\rm col}\sim 3$ so the disagreement between the Prada et al. results and our fit is not important.

\bibliographystyle{JHEP}
\bibliography{miniclusters_v3}
\end{document}